\documentclass[final,5p,times,twocolumn]{elsarticle}

\usepackage[utf8]{inputenc}
\usepackage[T1]{fontenc}

\graphicspath{{graphics/}}

\DeclareGraphicsExtensions{.pdf,.jpeg,.png}

\usepackage{amsmath}
\usepackage{amsfonts}
\usepackage{amssymb}
\usepackage{float}

\usepackage[normalem]{ulem}

\usepackage{booktabs}
\usepackage{tabularx}
\usepackage{threeparttable}
\usepackage{siunitx}

\usepackage{url}
\usepackage[colorlinks=true, citecolor=blue, linkcolor=blue, filecolor=blue,urlcolor=blue]{hyperref}

\usepackage[gen]{eurosym}

\newcommand{\ra}[1]{\renewcommand{\arraystretch}{#1}}

\def\co{CO${}_2$}
\def\el{${}_{\textrm{el}}$}
\def\th{${}_{\textrm{th}}$}

\newcommand{\ubar}[1]{\text{\b{$#1$}}}

\newcommand*\rot{\rotatebox{90}}
\newcommand*\OK{\ding{51}}

\usepackage{tikz}

\usepackage[europeanresistors,americaninductors]{circuitikz}
\usepackage{adjustbox}

\usepackage{fixltx2e}

\journal{Energy}

\begin{document}
\begin{frontmatter}



\title{Synergies of sector coupling and transmission reinforcement in a cost-optimised, highly renewable European energy system}


\author[kit,fias]{T.~Brown\corref{cor1}}
\ead{tom.brown@kit.edu}
\author[fias]{D.~Schlachtberger}
\author[fias]{A.~Kies}
\author[fias]{S.~Schramm}
\author[aarhus]{M.~Greiner}

\cortext[cor1]{Corresponding author}
\address[kit]{Institute for Automation and Applied Informatics, Karlsruhe Institute of Technology, Hermann-von-Helmholtz-Platz 1, 76344 Eggenstein-Leopoldshafen, Germany}
\address[fias]{Frankfurt Institute for Advanced Studies, Ruth-Moufang-Straße 1, 60438 Frankfurt am Main, Germany}
\address[aarhus]{Department of Engineering, Aarhus University, 8000 Aarhus C, Denmark}

\begin{abstract}

  There are two competing concepts in the literature for the
  integration of high shares of renewable energy: the coupling of
  electricity to other energy sectors, such as transport and heating,
  and the reinforcement of continent-wide transmission networks. In
  this paper both cross-sector and cross-border integration are
  considered in the model PyPSA-Eur-Sec-30, the first open,
  spatially-resolved, temporally-resolved and sector-coupled energy
  model of Europe. Using a simplified network with one node per
  country, the cost-optimal system is calculated for a 95\% reduction
  in carbon dioxide emissions compared to 1990, incorporating
  electricity, transport and heat demand. Flexibility from battery
  electric vehicles (BEV), power-to-gas units (P2G) and long-term
  thermal energy storage (LTES) make a significant contribution to the
  smoothing of variability from wind and solar and to the reduction of
  total system costs. The cost-minimising integration of BEV pairs
  well with the daily variations of solar power, while P2G and LTES
  balance the synoptic and seasonal variations of demand and
  renewables.  In all scenarios, an expansion of cross-border
  transmission reduces system costs, but the more tightly the energy
  sectors are coupled, the weaker the benefit of transmission reinforcement becomes.
\end{abstract}


\end{frontmatter}

\section{Introduction}

It has been established in many studies that the integration of high
shares of renewable energy in the European electricity sector is both
technically feasible and affordable
\cite{Czisch,Schaber,Schaber2,Scholz,Rodriguez2013,Bussar201440,OptHet,Breyer2017} (see also
the review \cite{burdenresponse}). Typically, these studies show that
the most cost-effective solutions are dominated by wind generation and
require the expansion of a pan-continental transmission network, which
enables the exploitation of the best renewable production sites and
smooths out the variations from weather systems on the synoptic scale
($\sim$ 600--1000~km) as they pass over the continent. Without an
expansion of the transmission network, more expensive electricity
storage solutions are needed to balance the variability of renewables
in time \cite{HALLER2012282,Hagspiel,GILS2017173,Schlachtberger2017,CEBULLA2017211}.


However, focussing on the electricity sector means not only neglecting
the significant greenhouse gas emissions from other energy demand sectors,
such as heating and transport, but also ignoring important sources of
flexibility in these sectors. In what some authors term `smart
energy systems' \cite{LUND2017556}, demand from, for example, battery
electric vehicles or intelligent heating systems can be brought
forward or delayed to reduce system costs, and low-cost long-term
storage can be provided either chemically, using power-to-gas units to
produce synthetic fuels such as hydrogen and methane (so called
`electrofuels'), or thermally \cite{Lund2010}. Long-term storage can smooth out both
the seasonal variations of renewables and the synoptic variations
($\sim$ 3-10 days in the time dimension).

Modelling all energy sectors in high spatial and temporal detail is
computationally demanding. In order to maintain computational
tractability, previous sector coupling studies have either focused on
just a few demand sectors, or sacrificed spatial or temporal
resolution.

Studies of a few sectors have either considered just electricity and
heat, electricity and transport, or electricity and gas. For example,
in \cite{Meibom2007,PENSINI2014530} the possibility of using excess
renewable electricity in the heating sector was considered, but no
requirements were set to defossilise all heating, or to couple to
other demand sectors.  In another set of studies, a simplified
investment and dispatch scheme was used for a one-node-per-country model of
Europe to study electricity-heat coupling
\cite{ASHFAQ2017363}. Interactions between the electricity sector and
transport were studied for electric vehicles in
\cite{KEMPTON1997157,KIVILUOMA20111758,SCHILL2015185} and including
fuel cell electric vehicles in \cite{en10070956,en10070957}. More
general coupling of electricity to gas for use in either heating or
transport was considered in \cite{SCHIEBAHN20154285,ROBINIUS2018182}.


Studies that include multiple sectors, often encompassing all energy
usage, but that sacrifice spatial resolution have typically either
considered single countries (e.g. Germany
\cite{Henning20141003,PALZER20141019,IEESWV,Quaschning}, Denmark
\cite{LUND2009524,mathiesen2014smart,Lund201296}, Ireland
\cite{CONNOLLY2011502,Deane2012303}) or considered the whole continent
of Europe without any spatial differentiation \cite{Connolly20161634}
so that international network bottlenecks are not visible. In one
study two countries, Denmark to represent Northern Europe and Italy to
represent Southern Europe, were coupled to compare cross-border with
cross-sectoral coupling \cite{THELLUFSEN2017492}; while both
strategies demonstrated benefits, cross-sectoral coupling gave the
best performance.

Another option to reduce computation times is to include multiple
sectors and/or multiple countries, but reduce the number of
representative demand and weather situations to several typical days
\cite{Primes,REMIND,CAPROS2014231,SIMOES2017353,en10101468}. A lower
intra-annual resolution allows optimisation of investment paths over
multiple decades, but does not allow enough resolution to assess the
variability and flexibility requirements for high shares of wind and
solar power \cite{LUDIG20116674,KOTZUR2018474}.



In this paper both sector coupling and international grid integration
are considered in the model PyPSA-Eur-Sec-30, the first open, hourly,
country-resolved, sector-coupled investment model of the European
energy system. Generation, storage and transmission investment are
optimized so that demand for electricity, space and water heating, and
land transport is met under the condition that carbon dioxide emissions
are reduced by 95\% compared to 1990 levels, in line with European
Union targets \cite{Council2009}.\footnote{1990 is the standard reference year both for the European Union targets and for the Kyoto Protocol \cite{kyoto}.} It is assumed that both heating and
transport can be electrified, using for example heat pumps to meet
heating demand and electric vehicles for transport, both of which
leverage significantly higher efficiencies than their fossil-fuelled
counterparts.

The novelty in the model presented here is that the combination of
pan-continental integration and sector coupling in one model with
hourly time resolution over a full year allows a full consideration of
which competing concept is more cost-effective: smoothing of renewable
fluctuations in space with networks or in time with demand-side
management and low-cost long-term storage. Compared to
electricity-only models it has heating and transport demand for more
energy coverage and enhanced flexibility; compared to the
sector-coupled models with low spatial resolution, a full
consideration of cross-border exchange is possible; compared to the
models with low temporal resolution, the model can distinguish between
flexibility options at different time scales, and thus account for all
the cross-correlations of weather patterns over time.

The model is further distinguished by being fully open, in the sense
that all the input data, processing code and output data is freely
available online \cite{zenodo,zenodo-full} and may be re-used by
anyone, thereby enhancing transparency and reproduceability \cite{PFENNINGER2017211,PFENNINGER201863}.

In Section \ref{sec:model} the model framework is
described, before the input data which defines the model instance
PyPSA-Eur-Sec-30 is documented in Section \ref{sec:data}. In Section
\ref{sec:results} the results are presented and analysed; in Section
\ref{sec:discussion} the results are compared to the literature and
the limitations of the study are discussed. Conclusions are drawn in
Section \ref{sec:conclusions}.

\section{Model}\label{sec:model}

In this section the equations of the model are described, as
implemented in the modelling framework PyPSA \cite{PyPSA}.

The model uses linear optimisation to minimise annual operational and
investment costs subject to technical and physical constraints,
assuming perfect competition and perfect foresight. Market prices are
derived in the model that guarantee that each asset owner recovers
their costs from the market.

\begin{figure}[!t]
  \begin{adjustbox}{scale=0.60,trim=5 6.8cm 0 0}
  \begin{circuitikz}
  \draw (1.5,14.5) to [short,i^=grid connection] (1.5,13);
  \draw [ultra thick] (-5,13) node[anchor=south]{electric bus} -- (6,13);
  \draw(2.5,13) |- +(0,0.5) to [short,i^=$$] +(2,0.5);
  \draw (0,-0.5) ;
  \draw (0.5,13) -- +(0,-0.5) node[sground]{};
  \draw (2.5,12) node[vsourcesinshape, rotate=270](V2){}
  (V2.left) -- +(0,0.6);
  \draw (2.5,11.2) node{generators};
    \node[draw,minimum width=1cm,minimum height=0.6cm,anchor=south west] at (3.4,11.9){storage};
    \draw (4,13) to (4,12.5);

  \draw [ultra thick] (-6,10) node[anchor=south]{transport} -- (-3,10);
  \draw (-5.5,10) -- +(0,-0.5) node[sground]{};
  \draw (-3.5,10) to [short,i_=${}$] (-3.5,13);
  \draw (-3.2,11.5)  node[rotate=90]{discharge};
  \draw (-4.5,13) to [short,i^=${}$] (-4.5,10);
  \draw (-4.2,11.5)  node[rotate=90]{charge};
  \node[draw,minimum width=1cm,minimum height=0.6cm,anchor=south west] at (-4.5,8.9){battery};
  \draw (-4,10) to (-4,9.5);

    \draw [ultra thick] (2,10) -- (6.5,10)  node[anchor=south]{heat};
  \draw (3.5,10) -- +(0,-0.5) node[sground]{};
  \draw (4.5,9.35) to [esource] (4.5,8.5);
  \draw (4.5,10) -- (4.5,9.35);
  \draw (4.5,8.3) node{solar thermal};
  \draw (5,13) to [short,i^=heat pump;] (5,10);
  \draw (6.2,11) node{resistive heater};
  \node[draw,minimum width=1cm,minimum height=0.6cm,anchor=south west] at (5.5,8.9){hot water tank};
  \draw (6,10) to (6,9.5);

  \draw [ultra thick] (-2,10)  -- (0.5,10) node[anchor=south]{hydrogen};
  \draw (-1.5,13) to [short,i_=${}$] (-1.5,10);
    \draw (-1.2,11.5)  node[rotate=90]{electrolysis};
  \draw (-0.5,10) to [short,i^=${}$] (-0.5,13);
  \draw (-0.2,11.5)  node[rotate=90]{fuel cell};
  \draw (-1,10) to (-1,9.5);
  \node[draw,minimum width=1cm,minimum height=0.6cm,anchor=south west] at (-1.5,8.9){store};

  \draw (0,10) to [short,i_=${}$] (0,8);
  \draw [ultra thick] (-1,8) node[anchor=south]{methane} -- (3,8);
  \draw (1.5,8) to [short,i_=${}$] (1.5,13);
  \draw (2.5,8) to [short,i_=${}$] (2.5,10);
  \node[draw,minimum width=1cm,minimum height=0.6cm,anchor=south west] at (0.5,6.9){store};
  \draw (1,8) to (1,7.5);
  \draw (0.3,9)  node[rotate=90]{methanation};
  \draw (1.8,9.2)  node[rotate=90]{generator/CHP};
  \draw (2.8,9)  node[rotate=90]{boiler/CHP};

  \end{circuitikz}

\end{adjustbox}
\caption{Energy flow at a single node. In this model, a node represents a whole European country. Within each node there is a bus (thick horizontal line) for each energy carrier (electric, transport, heat, hydrogen and methane), to which different loads (triangles), energy sources (circles), storage units (rectangles) and converters (lines connecting buses) are attached.}
\label{fig:flow}
\end{figure}
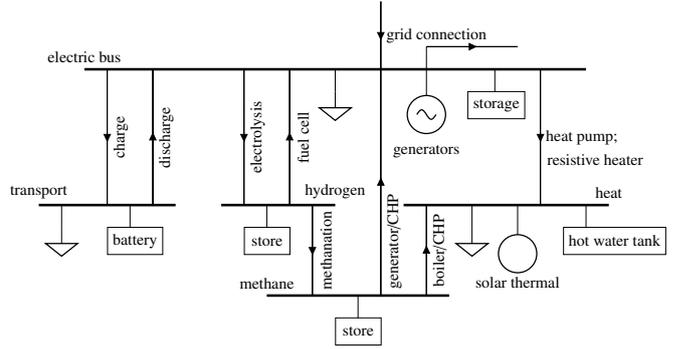

Each of the 30 European countries considered in the model is
aggregated to a single node, each of which consists of individual
`buses' (vertices to which energy assets are attached) for
electricity, heat, transport, hydrogen and methane. The electric buses
are connected together with transmission represented by High Voltage
Direct Current (HVDC) lines; the buses of the different sectors are
connected within a node with energy converters as shown in Figure
\ref{fig:flow}.

Generator capacities (for onshore wind, offshore wind, solar
photovoltaic (PV) and natural gas), storage capacities (for batteries,
hydrogen storage and conversion, methanisation and hot water tanks),
heating capacities (for heat pumps, resistive heaters, gas boilers,
combined heat and power (CHP) plants and solar thermal collector
units) and transmission capacities are all subject to optimisation, as
well as the operational dispatch of each unit in each hour. Demand
curves for the different sectors, the ratio of district heating to
decentralised heating, the number of electric vehicles, methane
storage and hydroelectricity capacities (for reservoir and
run-of-river generators and pumped hydro storage) are exogenous to the
model and not optimised.

Investment and operation are optimised over a full historical year of
hourly weather and demand data assuming perfect foresight, with 2011
chosen as the representative year.  In \cite{IWESextreme}, 2011 was
found to be ideal for scenario definition because of its average wind
conditions, slightly lower heating demand and higher PV feed-in than
average that can represent the (small) effects of global warming
expected by 2050, and the fact that it still contains a very cold
spell for dimensioning the supply of maximum heating demand. While it
would be desirable to model over multiple weather years in order to
capture inter-annual variability and more extreme weather events, as
well as to model forecast uncertainty, this is currently not
computationally feasible because of the large number of
variables. Results from other simulations for multiple years are
discussed in Section \ref{sec:limitations}.

If buses are labelled by $n$, generation and
storage technologies at the bus by $s$, hour of the year by $t$ and
bus connectors by $\ell$ (which includes transmission lines and energy converters such as heat
pumps and battery chargers), then the total annual system cost
consists of fixed annualised costs $c_{n,s}$ for generation and
storage power capacity $G_{n,s}$, fixed annualised costs
$\hat{c}_{n,s}$ for storage energy capacity $E_{n,s}$, fixed
annualised costs $c_\ell$ for bus connectors $F_{\ell}$, variable
costs $o_{n,s,t}$ for generation and storage dispatch $g_{n,s,t}$, as well as variable costs $o_{\ell,t}$ for power flow $f_{\ell,t}$ through connectors. The
objective function is then
\begin{eqnarray}
  &&  \min_{\substack{G_{n,s},E_{n,s},F_\ell,\\ g_{n,s,t},f_{\ell,t}}} \left[ \sum_{n,s} c_{n,s} \cdot G_{n,s} +\sum_{n,s} \hat{c}_{n,s} \cdot E_{n,s} + \sum_{\ell} c_{\ell} \cdot F_{\ell}\right. \nonumber \\
   & &\hspace{2cm} \left. + \sum_{n,s,t} o_{n,s,t} \cdot g_{n,s,t} + \sum_{\ell,t} o_{\ell,t} \cdot f_{\ell,t} \right]
\end{eqnarray}

The inelastic energy demand $d_{n,t}$ at each bus $n$ must be met at each time $t$ by either local generators and storage or by the flow $f_{\ell,t}$ from a connector $\ell$
\begin{equation}
\sum_{s} g_{n,s,t}+  \sum_{\ell} \alpha_{\ell, n,t}\cdot f_{\ell,t}  =  d_{n,t}  \hspace{.7cm} \leftrightarrow \hspace{0.5cm} \lambda_{n,t} \hspace{.3cm} \forall\, n,t \label{eq:energybalance}
\end{equation}
where $\alpha_{\ell,n,t}=-1$ if $\ell$ starts at $n$ and $\alpha_{\ell,n,t} = \eta_{\ell,t}$ if $\ell$ ends at $n$. $\eta_{\ell,t}$ is a factor for the efficiency of the energy conversion in $\ell$; it can be time-dependent for efficiency that, for example, depends on the outside temperature, like for a heat pump. The Karush-Kuhn-Tucker (KKT)/Lagrange multiplier $\lambda_{n,t}$ represents the market price of the energy carrier at this bus in this hour.

The dispatch $g_{n,s,t}$ of each generator and storage unit  is constrained by its
capacity $G_{n,s}$ and time-dependent availabilities
$\bar{g}_{n,s,t}$ and $\ubar{g}_{n,s,t}$, which are given per unit of
the capacity $G_{n,s}$:
\begin{equation}
 \ubar{g}_{n,s,t} \cdot G_{n,s} \leq  g_{n,s,t} \leq \bar{g}_{n,s,t} \cdot G_{n,s} \hspace{1cm} \forall\, n,s,t \label{eq:gen}
\end{equation}
For flexible conventional generators the availabilities are constant
$\ubar{g}_{n,s,t} = 0$ and $\bar{g}_{n,s,t} = 1$. For variable renewable
generators such as wind and solar, the time-varying $\bar{g}_{n,s,t}$
represents the weather-dependent power availability, and since curtailment is allowed, $\ubar{g}_{n,s,t}=0$. For battery
storage $\ubar{g}_{n,s,t} = -1$ and $\bar{g}_{n,s,t} = 1$.

The power capacity $G_{n,s}$ is optimised within minimum $\ubar{G}_{n,s}$ and maximum $\bar{G}_{n,s}$ installable potentials:
\begin{equation}
 \ubar{G}_{n,s} \leq G_{n,s}\leq \bar{G}_{n,s} \hspace{1cm} \forall\, n,s
\end{equation}

The energy levels $e_{n,s,t}$ of all storage units have to be consistent with the dispatch in all hours and are limited by the storage energy capacity $E_{n,s}$
\begin{eqnarray}
  e_{n,s,t} & = & \eta_0 \cdot e_{n,s,t-1} - \eta_{1} \left[g_{n,s,t}\right]^-- \eta_{2}^{-1} \left[g_{n,s,t}\right]^+ \nonumber \\
  & &+ g_{n,s,t,\textrm{inflow}} - g_{n,s,t,\textrm{spillage}} \nonumber \\
 \ubar{e}_{n,s,t}\cdot E_{n,s} & \leq &  e_{n,s,t}\,\, \leq \,\, \bar{e}_{n,s,t}\cdot E_{n,s}   \hspace{1cm} \forall\, n,s,t \label{eq:soc}
\end{eqnarray}
Positive and negative parts of a value are denoted as $[\cdot]^{+/-}=\max/\min(\cdot,0)$.
The storage units can have a standing leakage loss $\eta_0$, a charging efficiency $\eta_1$, a discharging efficiency $\eta_2$, inflow (e.g. river inflow in a reservoir) and spillage. The energy level can be set to be cyclic, i.e. $e_{n,s,t=0} = e_{n,s,t=T}$. The energy levels of the store can also be restricted by time series $\ubar{e}_{n,s,t}, \bar{e}_{n,s,t}$ given per unit of the energy capacity $E_{n,s}$. This is used to model the demand-side management of battery electric vehicles. The storage energy capacity $E_{n,s}$ can be optimised independently of the storage power capacity $G_{n,s}$, within installable potentials.

Flows on bus connectors are constrained by their capacities $F_{\ell}$
and time-dependent per unit availabilities $\ubar{f}_{\ell,t},
\bar{f}_{\ell,t}$
\begin{equation}
 \ubar{f}_{\ell,t} \cdot F_{\ell}  \leq f_{\ell,t} \leq \bar{f}_{\ell,t} \cdot F_{\ell} \hspace{1cm} \forall\,\ell,t
\end{equation}
For the HVDC links between electricity buses in different countries
$\ubar{f}_{\ell,t} = -1$ and $\bar{f}_{\ell,t}=1$; for a resistive
heater from an electricity bus to a heat bus in the same country the
connector is unidirectional $\ubar{f}_{\ell,t} = 0$ and
$\bar{f}_{\ell,t}=1$ since the heater cannot convert heat back into
electricity; the availabilities become time dependent for the charging
of electric vehicles.

In order to investigate the merits of international transmission,
the sum of transmission line capacities multiplied by their lengths $l_\ell$ can be
restricted by a line volume cap $\textrm{CAP}_{LV}$, which is then varied in different simulations:
\begin{equation}
  \sum_{\ell \in \textrm{HVDC}} l_\ell \cdot F_{\ell} \leq  \textrm{CAP}_{LV} \label{eq:lvcap}
\end{equation}
Line capacities are weighted by their lengths because the length
increases both the cost and potential public acceptance concerns for
overhead transmission lines.

\co{} emissions are also limited by a cap $\textrm{CAP}_{CO2}$, implemented using the
specific emissions $\varepsilon_{s}$ in \co{}-tonne-per-MWh\th{} of the fuel $s$, the efficiency $\eta_{n,s}$ and dispatch $g_{n,s,t}$ for generators, and the difference in energy level for non-cyclic stores (relevant for methane, which is depleted during the year):
\begin{equation}
  \sum_{n,s,t}  \varepsilon_{s} \frac{ g_{n,s,t} }{\eta_{n,s}} + \sum_{n,s} \varepsilon_s\, (e_{n,s,t=0} -e_{n,s,t=T}) \leq  \textrm{CAP}_{CO2} \hspace{.4cm} \leftrightarrow \hspace{0.3cm} \mu_{CO2} \label{eq:co2cap}
\end{equation}
The KKT multiplier $\mu_{CO2}$ indicates the carbon dioxide price
necessary to obtain this reduction in an open market.

The model was implemented in the free software energy modelling
framework `Python for Power System Analysis' (PyPSA)
\cite{PyPSA}. Each run took between 3 and 5 hours, depending on the
model parameters, using the commercial linear programming solver
Gurobi \cite{gurobi}. Gurobi was configured to use two threads on an
AMD Opteron 6274 machine with 64~GB of RAM and 1.4~GHz processing
speed per virtual core.

\section{Input Data}\label{sec:data}

In this section the input data for the model instance PyPSA-Eur-Sec-30
are described. Table \ref{tab:costs} summarises the different
investments the model can make, their costs, efficiencies and other
parameters. All energies and conversion efficiencies for methane and
hydrogen are given in terms of the higher heating value (HHV). For
power plants, all capacities refer to net generation capacities.

\ra{1.05}
\begin{table*}
\begin{threeparttable}
\caption{Input parameters based on 2030 value estimates} \label{tab:costs}
\centering
\begin{tabularx}{0.95\textwidth}{lrlrrrl}
\toprule
Technology                &Overnight   &Unit & FOM\tnote{a} & Lifetime & Efficiency & Source\\
& Cost [\euro] & & [\%/a] & [a] &  & \\
\midrule
Wind onshore    &1182   &kW\el &3 & 25  & 1 &  \cite{schroeder2013} \\
Wind offshore  &2506   &kW\el  &3& 25 & 1 &  \cite{schroeder2013} \\
Solar PV rooftop           &725   &kW\el &2 & 25  & 1 &  \cite{etip} \\
Solar PV utility           &425   &kW\el &3 & 25  & 1 &  \cite{etip} \\
Open cycle gas turbine (OCGT)             &400    &kW\el  &4& 30 & 0.39 &  \cite{schroeder2013,dea2016} \\
Pumped hydro storage\tnote{b} & 2000 &kW\el & 1 & 80 & $0.87\cdot 0.87$ & \cite{schroeder2013} \\
Hydro reservoir\tnote{b} & 2000 &kW\el & 1 & 80 & 0.9 & \cite{schroeder2013} \\
Run-of-river\tnote{b} & 3000 &kW\el & 2 & 80 & 0.9 & \cite{schroeder2013} \\
Battery inverter         &310   &kW\el  & 3 & 20 & $0.9\cdot0.9$ &  \cite{budischak2013} \\
Battery storage & 144.6 & kWh & 0 & 15 & 1 & \cite{budischak2013} \\
Hydrogen electrolysis        &350   &kW\el  & 4 & 18 & 0.8 & \cite{PalzerThesis} \\
Hydrogen fuel cell\tnote{c} & 339 & kW\el & 3 & 20 & 0.58 & \cite{NRELhydrogen,budischak2013} \\
Hydrogen storage\tnote{d}        & 8.4  &kWh  & 0 &20 & 1 & \cite{budischak2013} \\
Methanation\tnote{e} & 750 & kW$_{H_2}$ & 2.5 & 25 & 0.8 & \cite{PalzerThesis}\\
\co{} direct air capture (DAC)\tnote{e} & 228 &  t\co/a & 4 & 30 & see text & \cite{Fasihi2017}\\
Methanation+DAC\tnote{e} & 1000 & kW$_{H_2}$ & 3 & 25 & 0.6 & \cite{PalzerThesis,Fasihi2017}\\
Air-sourced heat pump decentral & 1050 & kW\th  & 3.5& 20 & variable & \cite{Henning20141003,PalzerThesis} \\
Air-sourced heat pump central & 700 & kW\th  & 3.5& 20 & variable & \cite{PalzerThesis} \\
Ground-sourced heat pump decentral & 1400 & kW\th & 3.5 &20& variable & \cite{PalzerThesis}\\
Resistive heater & 100 & kW\th & 2 & 20 & 0.9  & \cite{SchaberThesis} \\
Gas condensing boiler decentral & 175 & kW\th & 2 & 20 & 0.9& \cite{PalzerThesis} \\
Gas condensing boiler central & 63 & kW\th & 1 & 22 & 0.9& \cite{PalzerThesis} \\
Combined heat and power (CHP) central &600& kW\th & 3 & 25 & see text & \cite{Henning20141003} \\
Solar thermal collector decentral & 270 & m$^{2}$ & 1.3 & 20 & variable & \cite{Henning20141003} \\
Solar thermal collector central & 140 & m$^{2}$ & 1.4 & 20 & variable & \cite{Henning20141003} \\
Hot water tank decentral & 860 & m${}^3$  & 1& 20 & $\tau = $ 3 days & \cite{IEESWV,Henning20141003} \\
Hot water tank central & 30 & m${}^3$  & 1& 40 & $\tau = $ 180 days & \cite{IEESWV,Henning20141003} \\
Hot water tank (dis)charging & 0 &   & & & $0.9\cdot 0.9$ & \cite{Henning20141003} \\
High-density district heating network\tnote{f} & 220 & kW\th  & 1 & 40  & 1 & \cite{IEESWV} \\
Gas distribution network\tnote{f} & 387 & kW\th & 2 & 40 & 1 & based on \cite{bnetza2017} \\
Building retrofitting\tnote{f} & see text &  & 1 & 50 & 1 & \cite{Henning20141003,PalzerThesis} \\
HVDC transmission line       &400    &MWkm & 2 & 40 & 1 & \cite{Hagspiel} \\
HVDC converter pair & 150 & kW & 2 & 40 & 1 &  \cite{Hagspiel} \\
\bottomrule
\end{tabularx}

\begin{tablenotes}
\item [a] Fixed Operation and Maintenance (FOM) costs are give as a percentage of the overnight cost per year.
\item [b] Hydroelectric facilities are not expanded in this model and are considered to be fully amortized.
\item [c] The fuel cell technology is solid oxide, with partial (30\%)
  replacement after 10 years, following \cite{NRELhydrogen}. The more conservative estimate of efficiency has been taken, in line with other sources \cite{dea2016}.
\item [d] Hydrogen storage is in overground steel tanks following \cite{budischak2013}. The usage of existing underground caverns to store hydrogen could be more than 10 times cheaper \cite{Zakeri2015,ESYSTechnologiesteckbrief}, but a study of cavern potentials across Europe was not within the scope of this study.
\item [e] Investments in methanation and DAC are not allowed independently, only together as `Methanation+DAC', see text.
\item [f] The costs for distribution infrastructure and building retrofitting are approximate (see text) and they are therefore not optimised or included in the presented total system costs, but calculated retrospectively and analysed in the text.
\end{tablenotes}
\end{threeparttable}
\end{table*}

\subsection{Countries and network}

Following \cite{Schlachtberger2017}, there is one node in the model
for each country. The 30 countries consist of those in the major
synchronous zones of the European Network of Transmission System
Operators for Electricity (ENTSO-E), which includes the 28 European
Union member states as of 2018 minus Cyprus and Malta, plus Bosnia and
Herzegovina, Norway, Serbia and Switzerland. The nodes are connected
with a network based on existing and planned transmission line
interconnections between countries. The full network model was
presented and validated in \cite{pypsa-eur}.

\subsection{Electricity demand}

Hourly demand profiles are constructed that include current
electricity consumption and the electrification of fossil-fueled
cooking, but that exclude electricity consumption from space and water
heating; demand curves for transport and heating are considered
separately in Sections \ref{sec:transport_demand} and
\ref{sec:heating_demand} respectively. This allows the model to decide
independently how to meet demand from the different sectors.

The hourly electricity demand profiles for 2011 are based on those
from the Open Power System Data project \cite{OPSD}, which has
conveniently repackaged and unified data from the European Network of
Transmission System Operators for Electricity (ENTSO-E)
\cite{entsoe_load}. From these time series the time series for space
and water heating demand currently met by electricity in each country
is subtracted and added to the heating profiles (see Section \ref{sec:heating_demand}).
The remaining electricity time series are then scaled up linearly to
account for additional demand from the electrification of
fossil-fueled cooking demand in each country. These changes result in
a reduction of the original yearly electricity demand for the 30-node
model from 3153~TWh\el/a to 2970~TWh\el/a.

\subsection{Electricity supply}

The model for electricity generation and storage is fully documented
in \cite{Schlachtberger2017}, so only a summary is provided here;
differences with the model in \cite{Schlachtberger2017} are listed at
the end of this subsection. Electricity can be generated by solar
photovoltaics, wind onshore, wind offshore, hydro reservoirs,
run-of-river plants, open cycle gas turbines and combined heat and
power units.

The potential generation time series for wind generators are computed with the Aarhus University renewable energy atlas, described and validated in \cite{Andresen20151074}, based on hourly reanalysis wind data from 2011 with a spatial resolution of $40 \times 40 \textrm{km}^{2}$ \cite{Saha}. The time series for solar PV in 2011 are taken from the Renewables.ninja project, described and validated in \cite{Pfenninger2016}, based on the CM-SAF SARAH satellite-derived irradiance dataset \cite{rs70608067}.
The distribution of these generators is proportional to the quality of each site given by the local capacity factor.
However, protected sites as listed in Natura2000 \cite{natura2000} are excluded, as well as areas with certain land use types, as specified by \cite{Scholz} from the Corine Land Cover database \cite{corine2006}, to avoid, for example, placing  wind turbines in urban areas. The maximum water depth for offshore wind turbines is assumed to be 50~m.
The maximum installable capacity per country and generator type is
then determined by scaling these layouts until one site on the $40
\times 40 \textrm{km}^{2}$ lattice reaches the maximum installation
density. The theoretical maximum densities would be 10~MW/km$^2$ and
145~MW/km$^2$ for wind and solar respectively, but following
\cite{Schlachtberger2017} we take 20\% and 1\% of these values
respectively, in order to take account of competing land uses and
minimum-distance regulations in the case of onshore wind turbines.
Further validation of the renewable potentials was carried out in
\cite{pypsa-eur}.

The hydroelectricity generators in this model are fixed to their current size and are split into reservoir and run-of-river generators with river inflow, and pumped hydro storage as pure storage units.
Their respective power and energy storage capacities are based on country-aggregated data reported by \cite{kies2016,pfluger2011} and the inflow time series are provided by \cite{kies2016}. The power capacities and inflows of hydro reservoir and run-of-river are split in proportion to their respective national shares of installed capacity published by \cite{ENTSOEinstalledcapas}.

The model contains two extendable types of stationary electricity storage units:
batteries and hydrogen storage. Their charging and discharging
efficiencies, as well as cost assumptions for their power and energy
storage capacities are taken from \cite{budischak2013}.  For
batteries, it is assumed that the charging and discharging power
capacities of the inverter are equal; the energy storage capacity is
optimised independently. For hydrogen storage the capacities for the
production of hydrogen via electrolysis, the storage in steel tanks
and the generation of electricity with fuel cells can all be optimised
independently, since hydrogen is also used for non-electric purposes
such as methanation and for fuel cell vehicles. Any explicit standing
losses are neglected in the model.

In some scenarios synthetic methane can be produced from the hydrogen
using the Sabatier process. The methane can then be used in gas
turbines, CHPs or in gas boilers for heating. Carbon dioxide for the
methane production is sourced using Direct Air Capture (DAC). This
conservative assumption was chosen because other carbon sources could
not be guaranteed: biogenic sources are needed in sectors not covered
in the model, such as aviation, shipping and non-electric industrial
demand; capture of \co{} from fossil-burning industry and use in
synthetic fuels still results in net emissions; finally, carbon
captured from power plants could not be used, since the model does not
build enough centralised plants to generate the required carbon
dioxide.

DAC reduces the overall efficiency of the methanation because it is an
energy-intensive process. Based on figures from \cite{Fasihi2017}
(based in turn on private communications with the firm ClimeWorks),
0.23 kWh\el{} and 1.5 kWh\th{} are required for each kilogram of \co{}
extracted from the atmosphere, which reduces the overall energy
efficiency of the methanation from 80\% to 60\% (based on
0.19~kg\co{}/kWh\th{} for methane).


The differences with the model in \cite{Schlachtberger2017} are:
satellite data is used for the solar PV time series instead of
reanalysis data, since this was found to represent low generation in
winter more realistically; power and energy capacities for stationary
battery and hydrogen storage are optimised independently; PV is split
50-50\% between rooftop and utility installations, with cost changes
and a lower discount rate for rooftop PV (see Section
\ref{sec:costs}); and electricity can also be generated by CHP units
(see Section \ref{sec:heating_supply}).

\subsection{Transport demand}\label{sec:transport_demand}

For the transport final energy demand, only transport by road and rail
are considered in the model. Aviation, shipping and pipeline transport
are not considered. The mechanical drive for transport is assumed to
be provided in all cases by electric motors, since electric and fuel
cell electric vehicles are the most promising candidates for
fossil-free transport.

\begin{figure}[!t]
\centering
    \includegraphics[trim=0 0cm 0 0cm,width=\linewidth,clip=true]{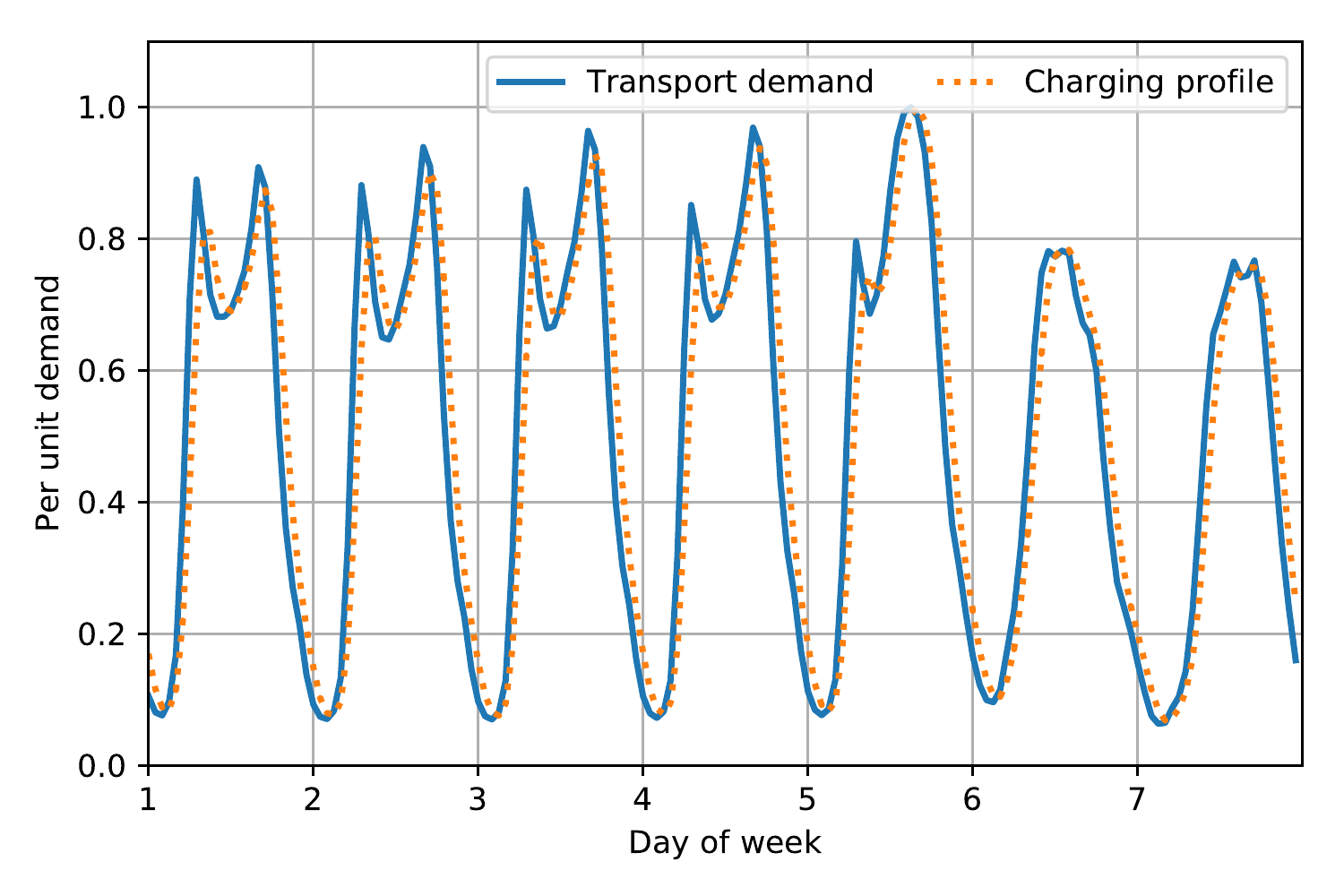}
\caption{Road transport demand based on statistics gathered by the German Federal Highway Research Institute (BASt) and derived BEV charging profile.}
\label{fig:transport}
\end{figure}

Transport demand time series are based on hourly vehicle counting
statistics from the German Federal Highway Research Institute (BASt)
\cite{BASt}, which the BASt has averaged to a weekly profile (see
Figure \ref{fig:transport}) based on the assumption that the profiles
change little from season to season. Given the lack of a unified
Europe-wide transport profile dataset, this German weekly profile is
assumed to be representative of transport demand for all countries in
all seasons and replicated for each country, taking account of time
zones and summer time. The profiles are scaled to the total road and
non-electric rail final energy demand for each country for 2011 taken
from the Odyssee database \cite{Odyssee}, corrected for the assumption
that all land-based transport is electrified and for the
heating and cooling requirements in electric vehicles.

To account for the fact that electric motors are significantly more
efficient in their consumption of electricity than internal combustion
engines are in their consumption of fossil fuels, the totals are
divided by a country-specific factor (averaging 3.5), giving a total
final electric energy transport demand in the model of 1075~TWh\el/a.
The country-specific efficiency factors are based on passenger car
final energy consumption per km in 2011 (averaging 0.70~kWh/km) from
the Odyssee database compared to the plug-to-wheels value of
0.20~kWh/km for the Tesla Model S, which was on the higher side for
the selection of electric cars tested by the US EPA in 2016 \cite{EPA}.

The profiles are then corrected with a temperature-dependent factor
for the heating and cooling demand in the vehicles. For both internal
combustion engine vehicles (ICEV) and electric vehicles (EV) it is
assumed that no climate control is required when the outside
temperature is between 15$^\circ$C and 20$^\circ$C, and that below or
above these temperatures the demand increases linearly with the
temperature. For EVs it is assumed that heating increases overall
demand by 0.98\%/$^\circ$C, while cooling increases it by
0.63\%/$^\circ$C, based on figures reported for the range of the Tesla
Model S in different conditions on the manufacturer's website \cite{tesla}.
For ICEVs the value for heating is 0.38\%/$^\circ$C and for cooling is
1.6\%/$^\circ$C, based on approximate figures from the US EPA
\cite{EPA}. The difference in heating demand is a reflection of the
fact that the internal combustion engine is a source of waste
heat, while for cooling, the driving of the compressor for air
conditioning has the same overall efficiency as the engine itself. To
correct the hourly profiles, the climate control demand for ICEVs is
first subtracted from the total transport demand, yearly profiles are
then extrapolated from the weekly profile, then finally the
temperature-dependent adjustment for the EV climate control is made to
the profiles. These corrections result in a final electric energy
transport demand of 1102~TWh\el/a.

In the basic transport scenario, charging profiles for battery
electric vehicles are constructed based on the simple assumption that
vehicles plug into the grid after travel and try to charge
immediately. This assumes that there is charging infrastructure
available at places of work, commerce and in homes. Given that the
average daily distance travelled is typically low (averaging around
40~km per day for passenger cars in Germany \cite{Odyssee},
corresponding to 8~kWh) and chargers are assumed to be at least 11~kW
for each vehicle, the charging profiles are spread one third
immediately after consumption, one third one hour after consumption
and one third two hours after consumption; the resulting charging
profile is plotted in Figure \ref{fig:transport}. The charging profile
peaks in the evening at 6-7~pm, raising the Europe-wide
electricity-only yearly peak demand from 459~GW to 659~GW.  Scenarios
with demand-side management (DSM) are also considered, as described in
the following section.

\subsection{Transport supply}\label{sec:transport_supply}

It is assumed that all road and rail transport demand is either met by
electric vehicles (EVs) or by hydrogen-consuming fuel cell electric
vehicles (FCEVs), depending on the scenario. The investment in
vehicles is not optimised in the model, but given exogenously. The
number of passenger vehicles is assumed to be the same as today
(246~million vehicles, 0.465 per population of 529~million
people). The effects of a reduced vehicle fleet are discussed below.

In the scenarios with EVs, all passenger cars are taken to be battery
electric vehicles (BEVs); whether road freight is electrified as BEVs,
or with electric roads (e.g. overhead pantographs) \cite{CONNOLLY2017235}, or on rail, is
left open. The BEVs are modelled in aggregate for each country,
following the approach in \cite{LUND20083578,KIVILUOMA20111758}.  In
the demand-side management (DSM) scenarios, a fixed fraction of cars
make a battery capacity of 50~kWh per car available to the model to shift the
BEV charging to times which reduce total system costs.  50~kWh
corresponds to today's mid-range capacity or today's top-range model
(the Tesla S100) with half its 100~kWh capacity reserved as a
buffer. The default fraction of cars participating in DSM is 50\%;
scenarios with 25\% and 100\% are also explored. Scenarios where the
cars can discharge into the grid (vehicle-to-grid, V2G) are also
examined. With 50\% of cars participating, there is 6.15 TWh of
storage available to the model, which corresponds to around three
quarters of the daily regular electricity consumption; this capacity
is ideal for smoothing out the diurnal variations of solar power.

In both DSM and V2G scenarios the BEV state of charge available to the
model is forced to be between 75\% and 100\% at 5~am every day (using
the variable $\ubar{e}_{n,s,t}$ from equation \eqref{eq:soc}), to meet
the expectation of consumers that the battery is reasonably full
in the morning before peak usage. The lower limit 75\% was chosen so
that the battery has room to be fully charged by PV during the day.
This restriction allows demand to be shifted within a day, but
prevents the wide-scale synoptic or seasonal shifting of BEV demand.

Each car can charge its battery with an efficiency of 90\% at a
maximum rate of 11~kW (i.e. with a three-phase 400~V 16~A connection);
discharging is also assumed to be 90\% efficient. The power
availability of the cars, i.e. the percentage connected to the grid at
any time, is assumed to be inversely proportional to the demand
profile. The profile is affinely transformed so that the average
availability is 80\% and the peak availability is 95\%. This results
in a minimum availability of around 62\%. These figures are
conservative compared to most of the literature: \cite{LUND20083578} uses a
minimum availability of 80\%; \cite{kempton2001} calculates that at least 83-92\% of the vehicle fleet in California is parked at any time; fields tests in the United States from 2007 (predating widespread charging infrastructure) \cite{markel2009} showed EVs were parked more than 90\% of the day and plugged in 60\% of the time; \cite{galus2013} reports that in Switzerland the minimum fraction of vehicles which are parked is just under 60\%.  The very high
charging power compared to average consumption (the total charging
power is theoretically 2700~GW, although the average transport
consumption is only 110~GW) means that reducing this power
availability has very little effect on the model; reducing the
availability by 50\% has no effect, and a 1\% change to system costs
is seen first at a 75\% reduction.\footnote{This also means that the results will not change if only single phase rather than three-phase connections are available at residential properties.} The availability assumptions would
only become significant if there were significant changes to consumer
behaviour such as a wide-scale (i.e. more than 75\%) shift to
car-sharing, so that the number of vehicles would be much lower and
the shared vehicles would be plugged in less often and for shorter periods. However, even the
most amibitious scenarios do not consider such a large shift to car
sharing; for example, \cite{rmi2017} foresees for the United States a
reduction in the total number of cars by 23\% by 2035 compared to
2015, of which half remain in personal ownership and half are
automated mobility service vehicles. The autonomous vehicles are still
plugged in much of the time outside peak hours \cite{rmi2017}.



The cost of the car charging infrastructure is calculated following
\cite{IEESWV}, which assumes 1.5 charging points per car, of which
90\% are private costing 200~\euro{} each and 10\% are public costing
667~\euro{} each. For 246 million cars in Europe this results in
annual costs of 6.2~billion~\euro/a. However, as discussed in Section
\ref{sec:limitations}, this does not include upstream upgrades to
electricity distribution networks that may be triggered by the
increase in electrical load.


Electric vehicles can be substituted with fuel cell electric vehicles
(FCEVs) which convert hydrogen to electricity with an efficiency of
58\%. Because there is cheap hydrogen storage in the model, the
hydrogen demand for transport represents a large source of flexible
demand to the model; this system benefit is offset by the lower
efficiency compared to electric vehicles. Based on cost assumptions
from \cite{IEESWV} of 2.8~\euro{} per GJ of hydrogen provided, the
cost of the hydrogen filling station infrastructure in Europe would be
17.2~billion~\euro/a for a land-based transport system based entirely
on FCEVs.



\subsection{Heating demand}\label{sec:heating_demand}

\begin{figure}[!t]
\centering
    \includegraphics[trim=0 0cm 0 0cm,width=\linewidth,clip=true]{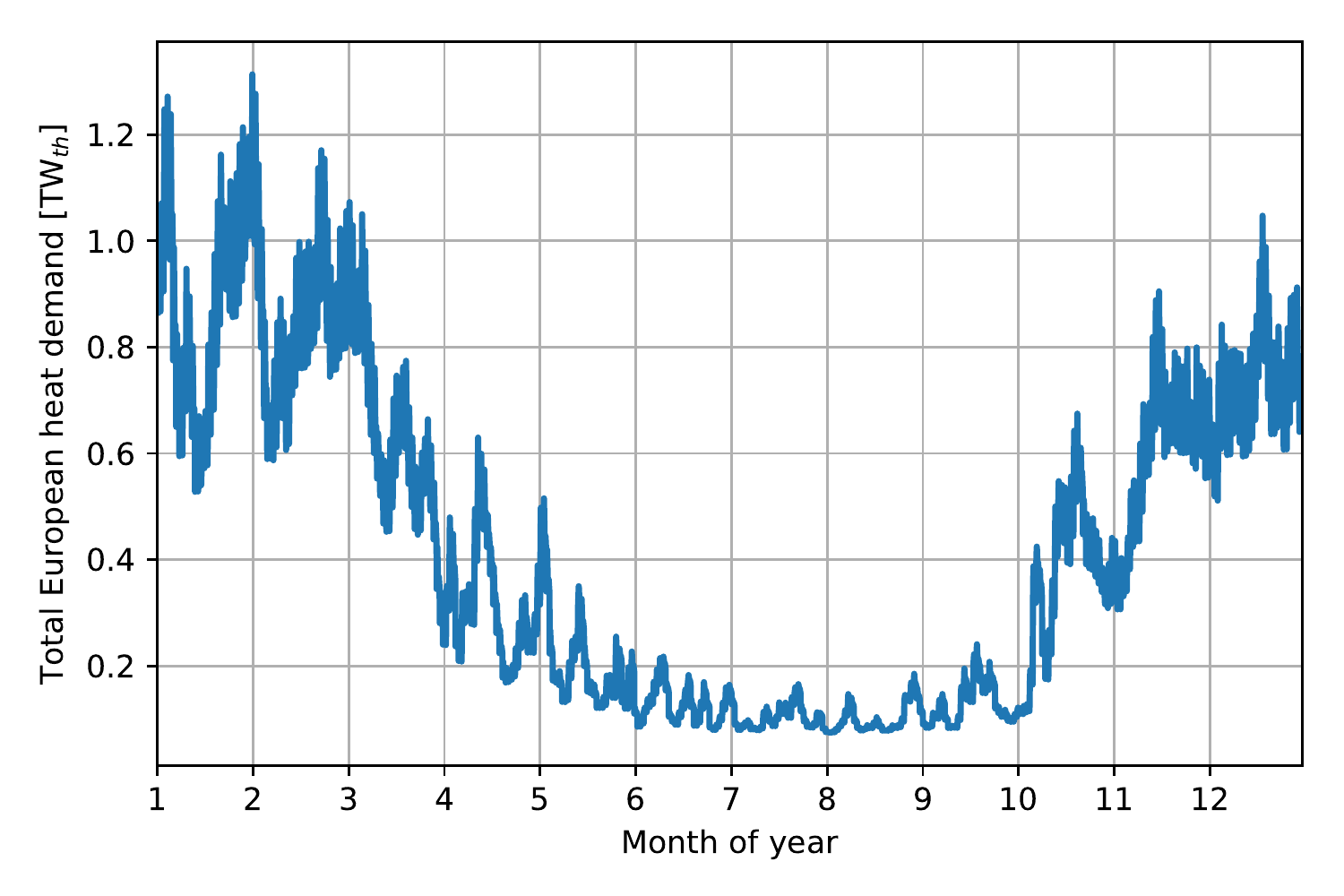}
\caption{Sum of heat demand profiles from 2011 for all 30 countries based on average daily temperatures in each country using the degree-day approximation.}
\label{fig:heat}
\end{figure}

For the heating final energy demand, only low-temperature space and
water heating in the residential and service sectors are
considered. Heating in the industrial sector is not included in the
model and cooking demand is included directly in the electricity
demand. Water heating demand is assumed to be constant over the year,
whereas profiles for the space heating demand are derived from
temperature time series using the degree-day approximation, assuming
that the heating demand rises linearly below an average daily
temperature of 15${}^\circ$C. Average daily temperature time series
for 2011 for each country are computed from the NCEP CFSR Reanalysis
air temperature dataset \cite{Saha}, using the NUTS3 population data
as a proxy for the geographical distribution of heat demand within
each country; see Figure \ref{fig:heat} for a graph of the total
European heat demand profile. Intraday profiles, which correspond to
typical consumer usage patterns, are based on weekday and weekend
profiles derived from heat demand data for Aarhus, Denmark in
\cite{Adam,ASHFAQ2018613}.

The water and space heating demand for the residential and service
sectors is scaled to energy totals for each country for 2011 taken
from the Odyssee database \cite{Odyssee}. Some data for the split
between water, space and cooking heating is missing for some
countries, particularly in the service sector, so here the total for
non-electric demand from the sectors was taken from the Eurostat
database \cite{EurostatEB} and split between space/water/cooking
according to the average ratios for the countries in the Odyssee
database. The average ratios were 79/15/6\% in the residential sector
and 78/14/4\% in the service sector (with 4\% remaining in services
for other heating applications).
For Switzerland a separate official data source was used \cite{SFOE}.
The total final energy heating
demand in the model is 3585~TWh\th/a.

In each country the heating demand is split between more rural areas
with low heating-per-area-density and more urban areas with high
heating-per-area-density. This distinction is made because it is
assumed that centralised district heating is only viable in
high-density areas, whereas ground-sourced heat pumps are only allowed
in low-density areas because of space restrictions \cite{Petrovic2016}.
High-density areas are defined as 60\% of all urban demand, since it
was determined to be cost-effective to use district heating for this
fraction in a selection of European countries in
\cite{Persson2011}. In Europe 74.4\% of the population lives in urban
areas, so according to this measure, 44.6\% of people live in
high-density areas. This fraction agrees with other assessments of the
potential penetration of district heating \cite{Gils}.

Heating efficiency measures are considered in the next section.



\subsection{Heating supply}\label{sec:heating_supply}

\begin{table*}[t]
  \centering
  \setlength{\tabcolsep}{6pt}
  \begin{tabular}{@{} lcll @{}}
    Low-density heat demand & & \multicolumn{2}{c}{High-density heat demand} \\
    \cmidrule{1-1}    \cmidrule{3-4}
    Individual & & Individual & Central (District Heating) \\
   \midrule
   Gas boiler & &    Gas boiler &    Gas boiler\\
   Resistive heater & & Resistive heater & Resistive heater \\
   Ground-sourced heat pump &  &  Air-sourced heat pump &    Air-sourced heat pump\\
   Solar thermal & & Solar thermal & Solar thermal \\
   Short-term TES & & Short-term TES & Long-term TES \\
   & & & Combined Heat and Power \\
  \end{tabular}
  \caption{Heating technologies allowed in the different density areas.}
  \label{tab:heating}
\end{table*}

Depending on the heat density of the area, different technologies are
allowed, as summarised in Table \ref{tab:heating}. In low-density
areas only decentralised individual heating units are allowed in the
model. For high-density areas, either individual heating units are
used or district heating can be turned on to provide heating
centrally, depending on the scenario. District heating is allowed in
all countries except those at southern latitudes (Portugal, Spain,
Italy, Greece and Bulgaria), where heating demand is low enough that
district heating is not considered economical.

District heating has the potential advantage that heating units can be
built more cheaply at scale and more easily interchanged; in addition,
large CHPs and long-term thermal energy storage (LTES) can be used
\cite{HeatRoadmapEurope}. LTES requires large, well-insulated hot
water tanks in pits containing tens of thousands of cubic metres of
hot water, which is only feasible for large heat demand.

District heating is costed with reference to the peak heating demand
at 220~\euro/kW, based on the cost for high-density, urban areas from
\cite{IEESWV}. This roughly agrees with the investment cost of
high-density district heating in \cite{Persson2011} of 208~\euro/kW
(converted from 2~\euro/GJ using the average European peak-to-average
ratio for heat demand of 3.57); the figure of 400~\euro/kW quoted in
\cite{Henning20141003} is more in line with the figure for low-density
heat demand of 370~\euro/kW from \cite{IEESWV}.  It must also be
considered that where district heating replaces individual gas
heating, costs are saved by no longer requiring a gas distribution
network.




It is assumed that gas distribution networks can be built for all
high-density areas and most low-density areas. There is not much
literature on the costs for gas distribution as a function of heat
demand density, so an approximation was made based on the residential
network charge of 15~\euro/MWh in Germany \cite{bnetza2017}. This
converts to an average investment cost of 387~\euro/kW per peak
demand. It is assumed that the peak demand is the main factor
when dimensioning the gas infrastructure.


Where gas distribution networks are not feasible because of low
demand density or distance from transmission pipelines, liquified gas must be
delivered in canisters. Liquified gas is considered preferable to oil
for remote areas, given the lower cost of gas and its lower \co{}
emissions (which also reduce the cost to consumers given the high \co{} prices seen in
the models).


Heat pumps are implemented with a coefficient of performance (COP)
that varies with the temperature. It is important to model the varying
COP because the COP drops low exactly when the heating demand is high
\cite{Petrovic2016}. The relationship between the COP and temperature
difference between the heat source and sink $\Delta T =
T_{\textrm{sink}} - T_{\textrm{source}}$ in degrees Celsius is taken
from a 2012 survey \cite{Staffell2012} (for air-sourced heat pumps:
$6.81 - 0.121\Delta T + 0.000630\Delta T^2$; for ground-sourced heat
pumps: $8.77 - 0.150\Delta T + 0.000734\Delta T^2$). The sink water
temperature was assumed to be 55$^{\circ}$C, following
\cite{Petrovic2016}, which is sufficient for domestic hot water, but
could be reduced for space heating with appropriate large-area
radiators. The source air and ground temperatures are taken from the
same dataset \cite{Saha} as for the heating demand. Ground-sourced
heat pumps (GSHP) are only allowed in low-density areas because of
land restrictions \cite{Petrovic2016}. In high-density areas only
air-sourced heat pumps (ASHP) are allowed, since their potentials are
not limited. However, noise regulations must be taken into account for
siting ASHPs. In cities other heat sources for heat pumps might be
available with higher temperatures than the air, such as boreholes or
sewage water, but this has not been considered in the limited scope of
this study.

\begin{figure}[!t]
\centering
    \includegraphics[trim=0 0cm 0 0cm,width=\linewidth,clip=true]{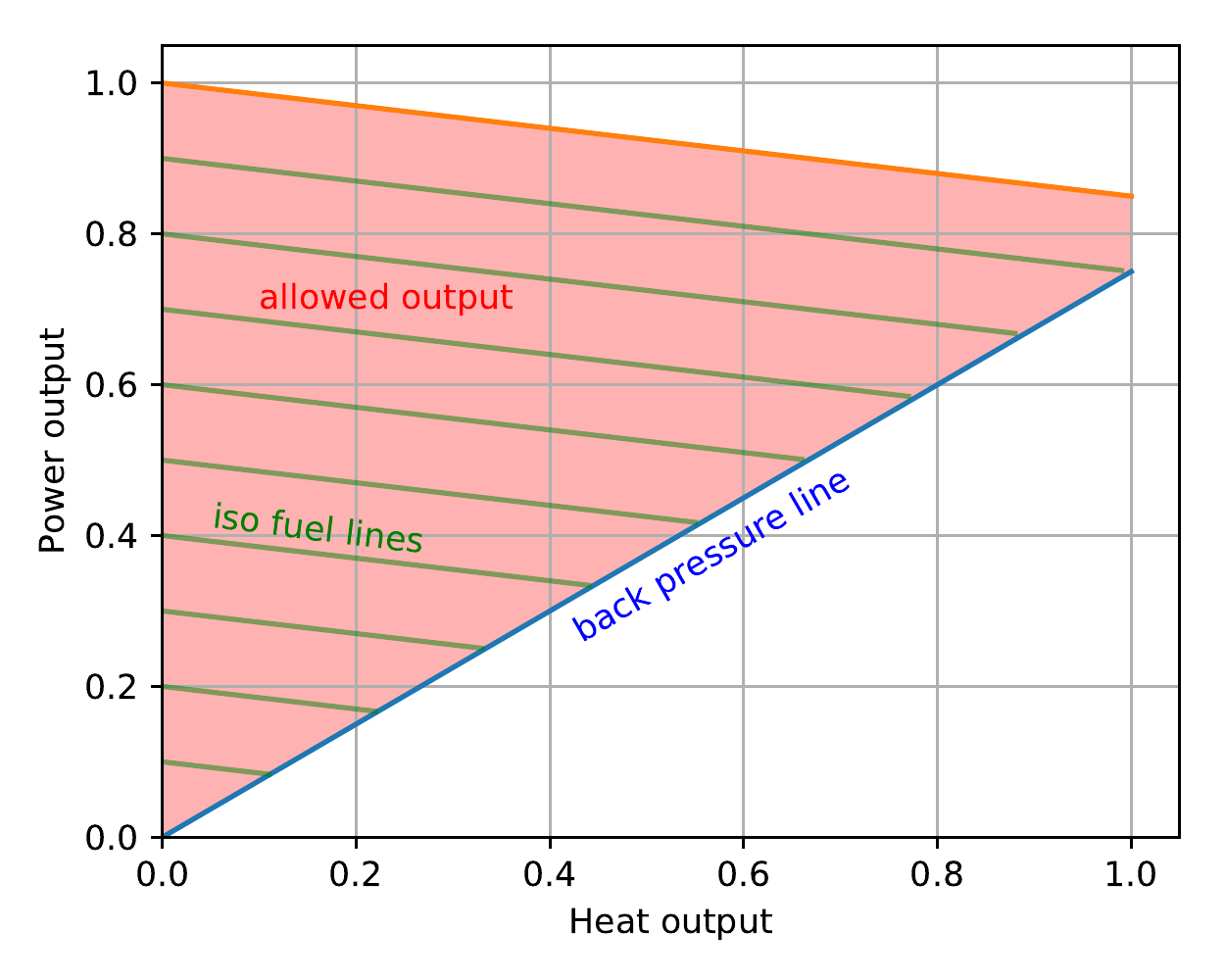}
\caption{The allowed area of heat and power production for a CHP unit.}
\label{fig:chp}
\end{figure}

The combined heat and power (CHP) model is based on the extraction
condensing unit described in \cite{Grohnheit1993}, which defines a
feasible operational area for power and heat production shown in
Figure \ref{fig:chp}. The feasible space is bounded at the bottom by
the back pressure line with slope 0.75 and bounded at the top by the
loss of power per unit of heat production with slope \num{-0.15}, whose
slope also defines the iso fuel lines. With no heat production
(i.e. in condensing mode), the electrical efficiency of the CHP is
46.8\%.

The solar thermal collector model uses the mathematical model from
\cite{Henning20141003}, which is based on the geometry of the
collectors in relation to the Sun, the downward shortwave radiation
flux $G$ (in W/m$^2$) on the collector, ambient temperature
$T_{\textrm{amb}}$ from \cite{Saha}, storage temperature of water
$T_{\textrm{stor}}$, assumed to be 80$^{\circ}$~C, the optical
efficiency $c_0$ and the heat loss coefficient $c_1$. The heat generated per
m$^2$ is then given by $Q = \eta_{\textrm{coll}} G$ where the
efficiency depends both on the irradiation and the ambient
temperature:
\begin{equation}
  \eta_{\textrm{coll}} = \left[c_0 - c_1 \left(\frac{T_{\textrm{stor}} - T_{\textrm{amb}}}{G}\right) \right]^+
\end{equation}
It was assumed that all solar collectors are tilted 45$^\circ$ to the
south which is close to the optimum position to maximize production in
winter in European countries. Following \cite{Henning20141003}, we
assume $c_0 = 0.8$ and $c_1 = 3$~W/m$^2$/K. As an example, German
collectors yield 532~kWh\th/m$^2$/a.

The model can also build thermal energy storage (TES), whose
parameters are based on insulated hot water tanks. The water tanks are
assumed to have a thermal energy density of 46.8~kWh\th/m${}^3$,
corresponding to a temperature difference of 40~K. The decay of thermal
energy is assumed to have a time constant of $\tau = $ 3 days for
short-term TES and $\tau = $ 180 days for long-term TES,
i.e. $1-\exp(-\frac{1}{24\tau})$ of the energy is lost per hour
regardless of the ambient temperature. Charging and discharging
efficiencies are 90\% due to pipe losses.

\begin{figure}[!t]
\centering
    \includegraphics[trim=0 0cm 0 0cm,width=\linewidth,clip=true]{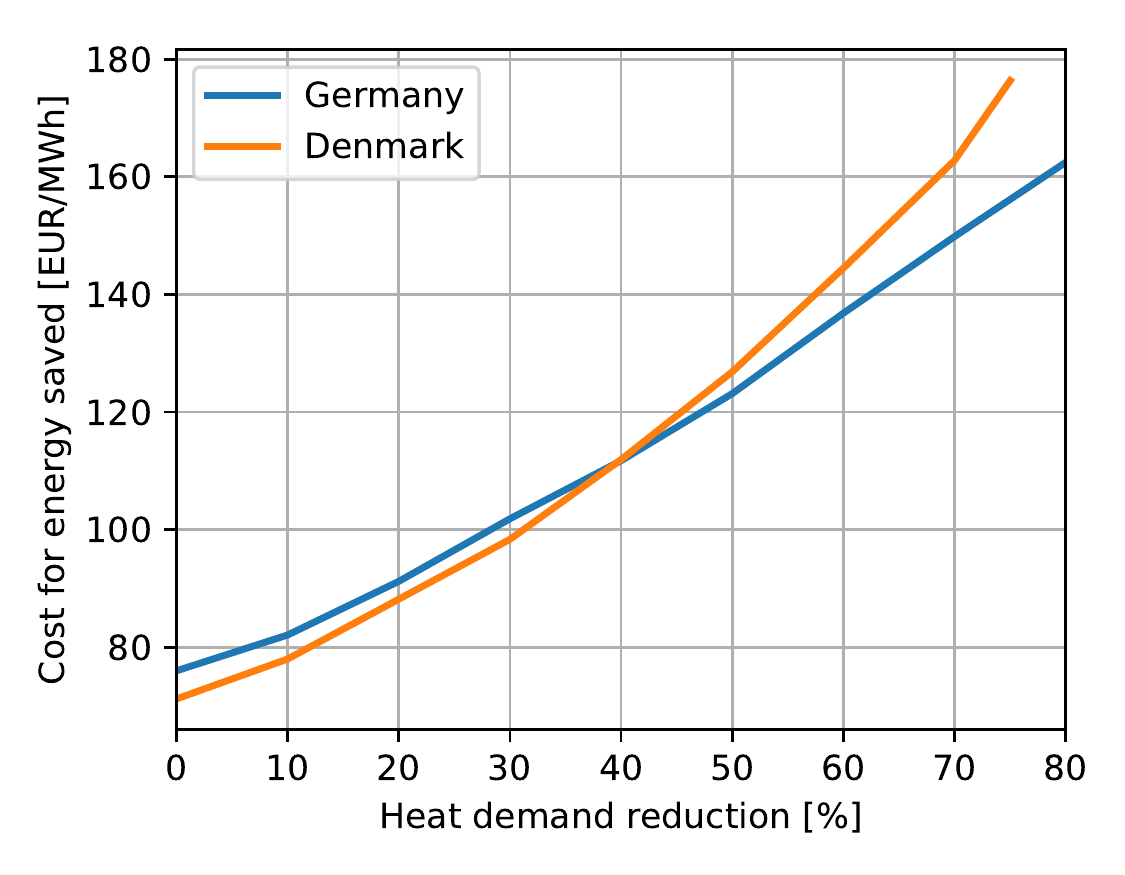}
\caption{Additional costs, averaged over each MWh\th{} of space heating consumption, for building retrofitting to reduce space heating demand by a specific fraction (x-axis). The German data is taken from \cite{Henning20141003,PalzerThesis}, with a lifetime of 50 years and with the reduction based on 2011 demand; the Danish data is taken from \cite{CONNOLLY2014475} with a lifetime of 30 years and based on 2010 demand; a discount rate of 4\% and FOM of 1\%/a have been used.}
\label{fig:retrofitting}
\end{figure}

Building retrofitting to reduce energy demand for space heating
requires a detailed database for each country of the building stock,
its current state of insulation and consumer heating behaviour. Since
adequate data was not available for each European state, investment in
retrofitting was not optimised directly in the model. Instead, a qualitative
analysis is given based on available retrofitting costs for Germany
\cite{Henning20141003,PalzerThesis} and Denmark \cite{CONNOLLY2014475}, plotted in Figure \ref{fig:retrofitting}. These
costs, averaged over each MWh\th{} of space heating demand, can then
be compared to the average marginal cost of space heating in each scenario (the
$\lambda_{n,t}$ for the heat bus from equation
\eqref{eq:energybalance}, weighted by the time series for space heating demand), to estimate what level of retrofitting is
efficient.


\subsection{Costs}\label{sec:costs}

Investment costs, fixed operation and maintenance (FOM) costs,
lifetimes, efficiencies and data sources for all assets are listed in
Table \ref{tab:costs}.  Natural gas fuel costs are 21.6~\euro/MWh\th, while
gas variable operation and maintenance (VOM) costs are
3~\euro/MWh\el{} \cite{schroeder2013}. Individual heating systems are
labelled `decentral', while heating systems that are connected to
district heating networks are labelled `central'.

Where possible, the costs are oriented towards predictions for 2030,
since this is a horizon within which cost projections might be
reliable, and also because this is the earliest point at which a 95\%
\co{} reduction might be plausible. The costs for generating assets are
mostly based on predictions for 2030 from DIW \cite{schroeder2013},
with the exception of solar PV, which has been updated with current
industry projections \cite{etip} given the fast changing costs; the
costs for battery and hydrogen electricity storage come from
\cite{budischak2013}; the costs of heating sector units are taken from
\cite{Henning20141003,IEESWV,PalzerThesis}. Costs are within the
ranges found in other databases
\cite{EnergyPLANDatabase,ESYSTechnologiesteckbrief,dea2016}.

For the annualisation of overnight costs a discount rate of 7\% is
used for large, utility and central assets, while a rate of 4\% is
used for decentral individual units (including rooftop solar PV and
building retrofitting), following the approach in
\cite{Henning20141003,PalzerThesis}. Solar PV units are split between
50\% for decentral rooftop and 50\% for utility-scale units.

\subsection{Carbon dioxide emissions}


The sectors in the model cover 72\% of the countries' final energy
consumption in 2011 (including from international marine bunkers)
\cite{EurostatEB}; the majority of the remaining final energy demand
comes from non-electric industrial demand (17\%) followed by shipping
(5\%) and aviation (4\%).


The sectors covered in the model emitted 3016 megatonnes of \co{}
(Mt\co{}) in 1990 \cite{Odyssee}, broken down into 1510~Mt\co{} from
electricity generation, 784~Mt\co{} for land-based transport and
723~Mt\co{} for heating in the residential and service sectors.  These
sectors comprised 68\% of the \co{} emissions in 1990 (76\% in 2011),
with almost all the rest coming from non-electric demand in industry,
shipping and aviation. A reduction in emissions of 95\% compared to
1990 thus corresponds to a limit of 151~Mt\co{}/a for the sectors
considered in the model.

The only net \co~emissions in the model come from the consumption of
natural gas in open-cycle gas turbines, combined heat and power plants
and gas boilers; \co{} is also captured from the air for methane
synthesis. Gas is assumed to have emissions of 0.19~t\co{}/MWh\th, so
the model can consume at most 795~TWh\th/a of natural gas.

\section{Results}\label{sec:results}

In this section different scenarios are presented, which successively
add demand and flexibility from the transport and heating sectors to
the model to assess the benefits of sector coupling. By adding
flexibility in stages, it is possible to understand how each
flexibility option impacts and interacts with the system, particularly
with respect to wind and solar generation, which dominate the system
costs and behaviour.

A \co{} reduction of 95\% compared to 1990 values is enforced for the
sum of the sectors considered in each scenario.

To weigh the benefits of sector coupling flexibility against the
expansion of cross-border inter-connectors, for each scenario
different levels of transmission are examined, including no
transmission, where every country is isolated ($\textrm{CAP}_{LV} = 0$ in equation \eqref{eq:lvcap}), and cost-optimal
transmission expansion using overhead lines ($\textrm{CAP}_{LV} = \infty$).

\begin{figure*}[!t]
\centering
\includegraphics[trim=0 0cm 0 0cm,width=\linewidth,clip=true]{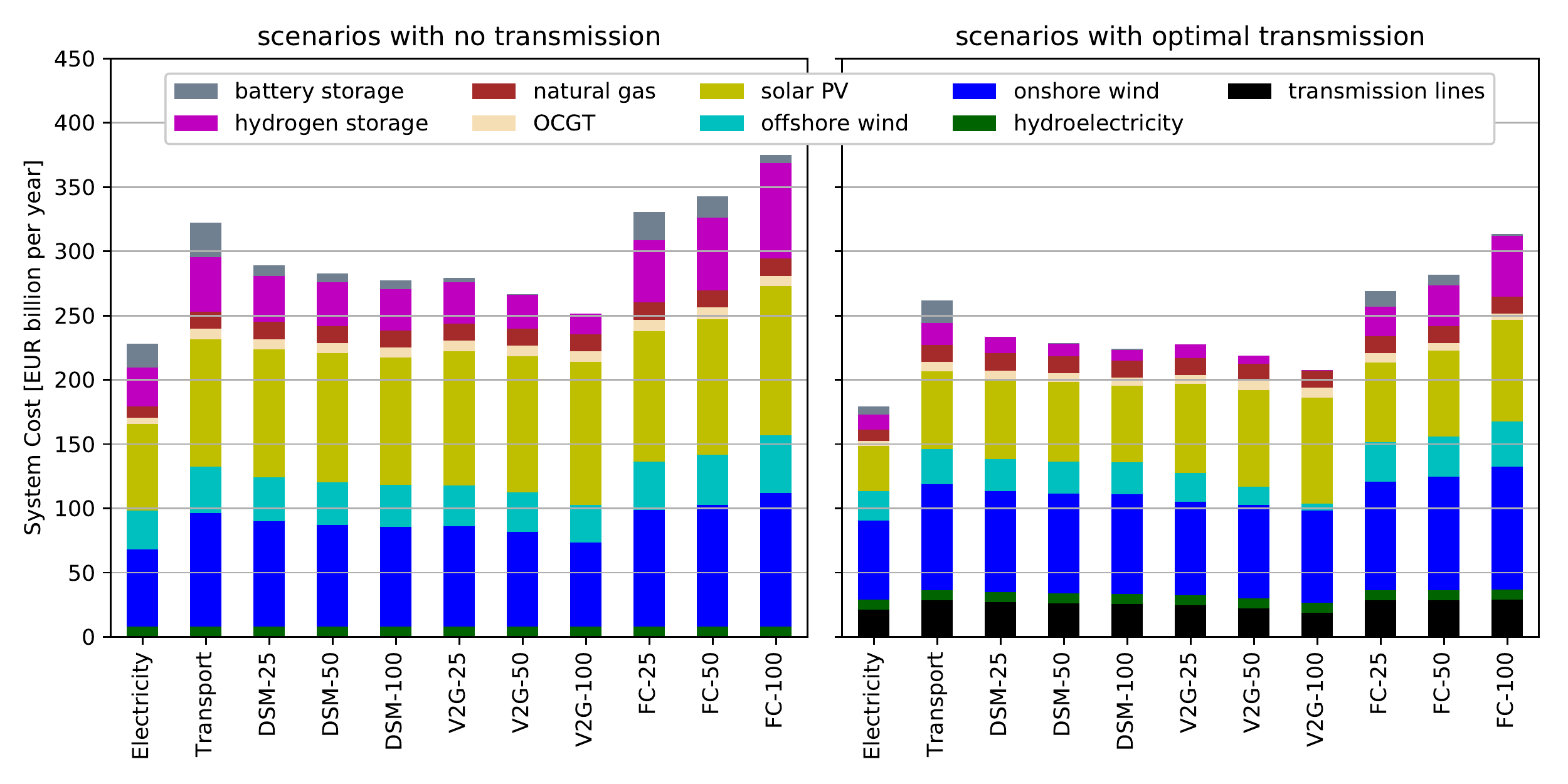}

\caption{Total annual system costs for the different scenarios with electricity and transport demand, with no interconnecting transmission (left) and optimal interconnecting transmission (right). Note that costs do not include distribution network costs. `Hydrogen storage' includes the costs of storage tanks, electrolysis and fuel cells.}
\label{fig:transport_scenarios}
\end{figure*}

\begin{figure*}[!t]
\centering
\includegraphics[trim=0 0cm 0 0cm,width=\linewidth,clip=true]{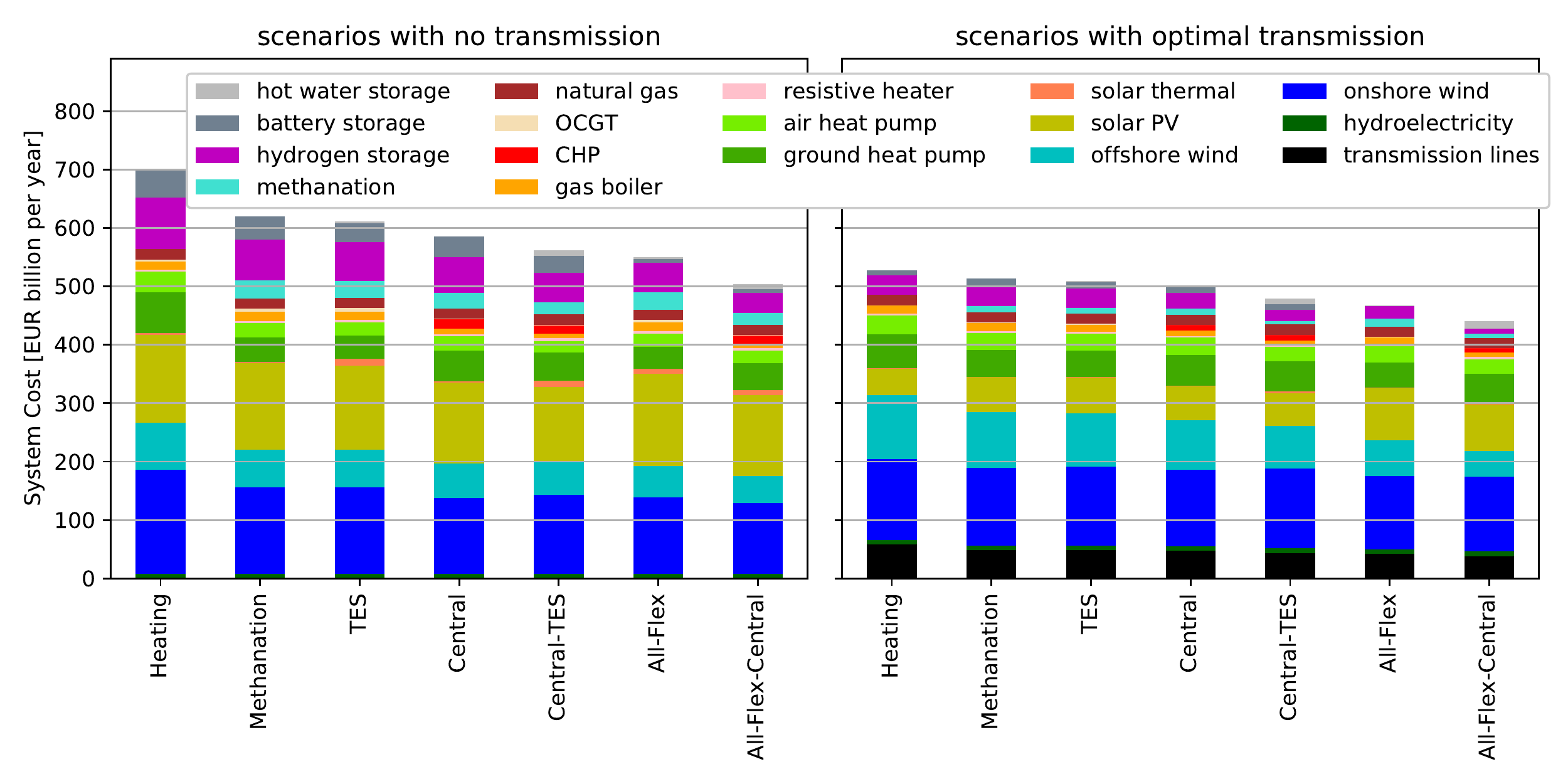}

\caption{Total annual system costs for the different scenarios with electricity, transport and heating demand, with no interconnecting transmission (left) and optimal interconnecting transmission (right). Note that costs do not include distribution network costs.}
\label{fig:heating_scenarios}
\end{figure*}

The options activated in each scenario are summarised in Table
\ref{tab:scenarios}, along with the main indicators for the results:
system costs, optimal transmission volume, the \co{} shadow price
($\mu_{CO2}$ from equation \eqref{eq:co2cap}) and the average
load-weighted marginal prices ($\lambda_{n,t}$ from equation
\eqref{eq:energybalance}) of electricity and low (L) and high (H)
density space heating demand. The breakdowns of the system costs into
individual technologies are plotted in Figures
\ref{fig:transport_scenarios} and \ref{fig:heating_scenarios}.

In the following subsections the results of each scenario are analysed
in detail.

\begin{table*}[t]
    \centering \small
  \setlength{\tabcolsep}{3pt}
  \begin{tabular}{@{} lccrrrccccc|rrrrrrrrrrrr @{}}
        &  \multicolumn{10}{c}{Scenario Definitions} &  \multicolumn{12}{c}{Results} \\[2ex]

Scenario & \rot{Electricity Demand} & \rot{Transport Demand} & \rot{BEV-DSM [\%]} & \rot{BEV-V2G [\%]} & \rot{FCEV [\%]} &\rot{Heating Demand} & \rot{Methanation} & \rot{Short-term TES} & \rot{Long-term TES} & \rot{District Heating} & \rot{\shortstack[l]{System costs w/o\\transmission [billion \euro/a]}}  & \rot{\shortstack[l]{System costs w/\\transmission [billion \euro/a]}}&\rot{\shortstack[l]{Ratio of costs w/o to\\w/ transmission}} & \rot{\shortstack[l]{Optimal transmission\\volume [TWkm]}} &\rot{\shortstack[l]{\co{} price w/o\\transmission [\euro/t\co]}}  &\rot{\shortstack[l]{\co{} price w/\\transmission [\euro/t\co]}} & \rot{\shortstack[l]{Electricity price w/o\\transmission [\euro/MWh\el]}} & \rot{\shortstack[l]{Electricity price w/\\ transmission [\euro/MWh\el]}} & \rot{\shortstack[l]{L space heating price w/o\\transmission  [\euro/MWh\th]}} & \rot{\shortstack[l]{L space heating price w/\\ transmission  [\euro/MWh\th]}}  & \rot{\shortstack[l]{H space heating price w/o\\transmission  [\euro/MWh\th]}} & \rot{\shortstack[l]{H space heating price w/\\transmission  [\euro/MWh\th]}}\\
   \midrule
Electricity  &  \OK &  &&&&&&&&  &  228 &  179 &   1.27 &          201 &    357 &      136 &      85 &        76 &        &          &         &          \\
Transport &  \OK &  \OK & &&&&&&&   &  322 &  262 &   1.23 &          267 &    371 &      145 &      84 &        74 &        &          &         &          \\
DSM-25 & \OK & \OK & 25&&&&&&&  &  289 &  233 &   1.24 &          253 &    345 &      133 &      84 &        75 &        &          &         &          \\
DSM-50 & \OK & \OK & 50&&&&&&&  &  283 &  229 &   1.24 &          248 &    335 &      130 &      84 &        75 &        &          &         &          \\
DSM-100 & \OK & \OK & 100&&&&&&&  &  277 &  224 &   1.24 &          243 &    324 &      127 &      84 &        76 &        &          &         &          \\
V2G-25 & \OK & \OK & 25 & 25&&&&&&  &  279 &  228 &   1.23 &          232 &    348 &      122 &      83 &        72 &        &          &         &          \\
V2G-50 & \OK & \OK & 50 & 50&&&&&&  &  267 &  219 &   1.22 &          210 &    345 &      114 &      82 &        71 &        &          &         &          \\
V2G-100 & \OK & \OK & 100 & 100  &&&&&& &  251 &  207 &   1.21 &          177 &    342 &      114 &      83 &        71 &        &          &         &          \\
FC-25 & \OK & \OK &  & &  25 &&&&& &  330 &  269 &   1.23 &          267 &    378 &      141 &      81 &        72 &        &          &         &          \\
FC-50 & \OK & \OK &  & &  50 &&&&& &  343 &  282 &   1.22 &          267 &    377 &      132 &      79 &        69 &        &          &         &          \\
FC-100 & \OK & \OK &  & &  100 &&&&& &  375 &  313 &   1.20 &          273 &    379 &      122 &      77 &        67 &        &          &         &          \\
\midrule
Heating &  \OK &  \OK &  & & & \OK &  &&& &  699 &  527 &   1.33 &          549 &   1184 &      682 &     118 &        85 &     153 &       112 &      161 &        114 \\
Methanation & \OK & \OK  &  & & & \OK & \OK&&&  &  620 &  514 &   1.21 &          457 &    509 &      434 &      77 &        75 &     106 &        94 &      108 &         94 \\
TES & \OK & \OK  &  & & & \OK & \OK & \OK &&  &  612 &  510 &   1.20 &          458 &    504 &      422 &      77 &        75 &     104 &        92 &      105 &         92 \\
Central & \OK & \OK  &  & & & \OK & \OK &  & &  \OK &  585 &  499 &   1.17 &          443 &    527 &      460 &      77 &        75 &     104 &        94 &       92 &         88 \\
Central-TES & \OK & \OK  &  & & & \OK & \OK &\OK  & \OK & \OK  &  562 &  479 &   1.17 &          411 &    497 &      413 &      75 &        73 &     101 &        90 &       79 &         76 \\
All-Flex & \OK & \OK  & 50 & 50 & & \OK & \OK &\OK &  & &  550 &  468 &   1.18 &          398 &    473 &      416 &      73 &        70 &     101 &        92 &      102 &         92 \\
All-Flex-Central & \OK & \OK  & 50 & 50 & & \OK & \OK &\OK  & \OK &\OK &  504 &  440 &   1.15 &          359 &    463 &      407 &      72 &        69 &      98 &        91 &       78 &         78 \\
   \end{tabular}
  \caption{Definition of scenarios in terms of activated options
    (left); major indicators for results (right). BEV-DSM corresponds
    to the fraction of passenger cars which are allowed to shift their
    charging to cheaper times; BEV-V2G is the fraction of
    passenger cars which are allowed to feed back into the grid if it
    is profitable. FCEV gives the fraction of transport demand which is met by
    fuel cell electric vehicles. Results are reported without transmission and with optimal transmission.}
  \label{tab:scenarios}
\end{table*}

\subsection{Electricity only scenario}

In the {\bf Electricity} scenario none of the transport or heating
demand is activated. This allows a comparison with electricity-only
scenarios in the literature, in particular with the recent results of
some of the authors \cite{Schlachtberger2017}. If no interconnecting
transmission is allowed, then countries must be electrically
self-sufficient at all times and balance the fluctuations of wind and
solar locally with hydro, gas and significant capacities of stationary
battery and hydrogen storage. This drives up the average price of
electricity to 85~\euro/MWh\el{} and favours generation from solar
PV. If cost-optimal interconnecting transmission of 201~TWkm is built,
corresponding to transmission volumes around six-and-a-half times today's capacity of 31~TWkm, then
cheaper renewables such as onshore wind can be shared between
countries, which brings down the average price by 11\%
to 76~\euro/MWh\el. It was shown in \cite{Schlachtberger2017} that the benefits of
transmission are highly non-linear: volumes of transmission
inter-connection only a few times bigger than today's can already
lock-in many of the benefits of international integration.

\subsection{Transport scenario}

In the {\bf Transport} scenario the electrified land transport demand
is added to the electricity-only demand without the potential for
demand-side management or for vehicles to feed electricity back into the grid.
Although the electrical demand increases by 37\%,
the total costs increase by 41\% in the case of no transmission.
This can be traced back to several effects: the transport load profile
exacerbates daytime and evening peak loads, increasing the need for
peak capacity, and the higher overall load means that renewable sites
with good load factors are already filled to potential, so that worse
sites must be exploited. The effect of the profile (high daytime
demand, very low night demand, see Figure \ref{fig:transport}) is also
visible in the stronger preference for solar PV compared to wind (see
Figure \ref{fig:transport_scenarios}), although PV cannot meet the
evening peak without storage.

\subsection{Transport with Demand-Side Management from BEVs}

In the Demand-Side Management ({\bf DSM}) scenarios, fractions of the
Battery Electric Vehicles (BEVs) are allowed to shift their charging
to the times when electricity is cheapest (which corresponds to the
charging times which minimise the total system costs), but do not
discharge back into the electricity grid. Each vehicle is assumed to
make a 50~kWh battery available to their system, so that, for example,
the {\bf DSM-25} scenario corresponds to 25\% of the vehicles
participating in DSM, or all vehicles participating, but only making
12.5~kWh available for DSM.

From Figure \ref{fig:transport_scenarios} it is clear that allowing
DSM significantly reduces the overall system costs compared to the
{\bf Transport} scenario, with a total reduction of 14\%
in the {\bf DSM-100} scenario. Much of the benefit is already accrued
in the {\bf DSM-25} scenario, which has 10\% lower costs than the {\bf
  Transport} scenario. Thus only 25\% of vehicles have to participate
in DSM to see the majority of the system benefit.

The cost reduction is seen primarily in the reduced investment in
stationary storage (both battery and hydrogen storage), which also
leads to lower efficiency losses and therefore lower investment in
renewable generators.  With optimal transmission capacity, the need
for stationary batteries is entirely eliminated. The use of
DSM also favours a slightly higher solar share, because the BEV
charging can easily be shifted to peak PV times.

\begin{figure}[!t]
\centering
\includegraphics[trim=0 0cm 0 0cm,width=\linewidth,clip=true]{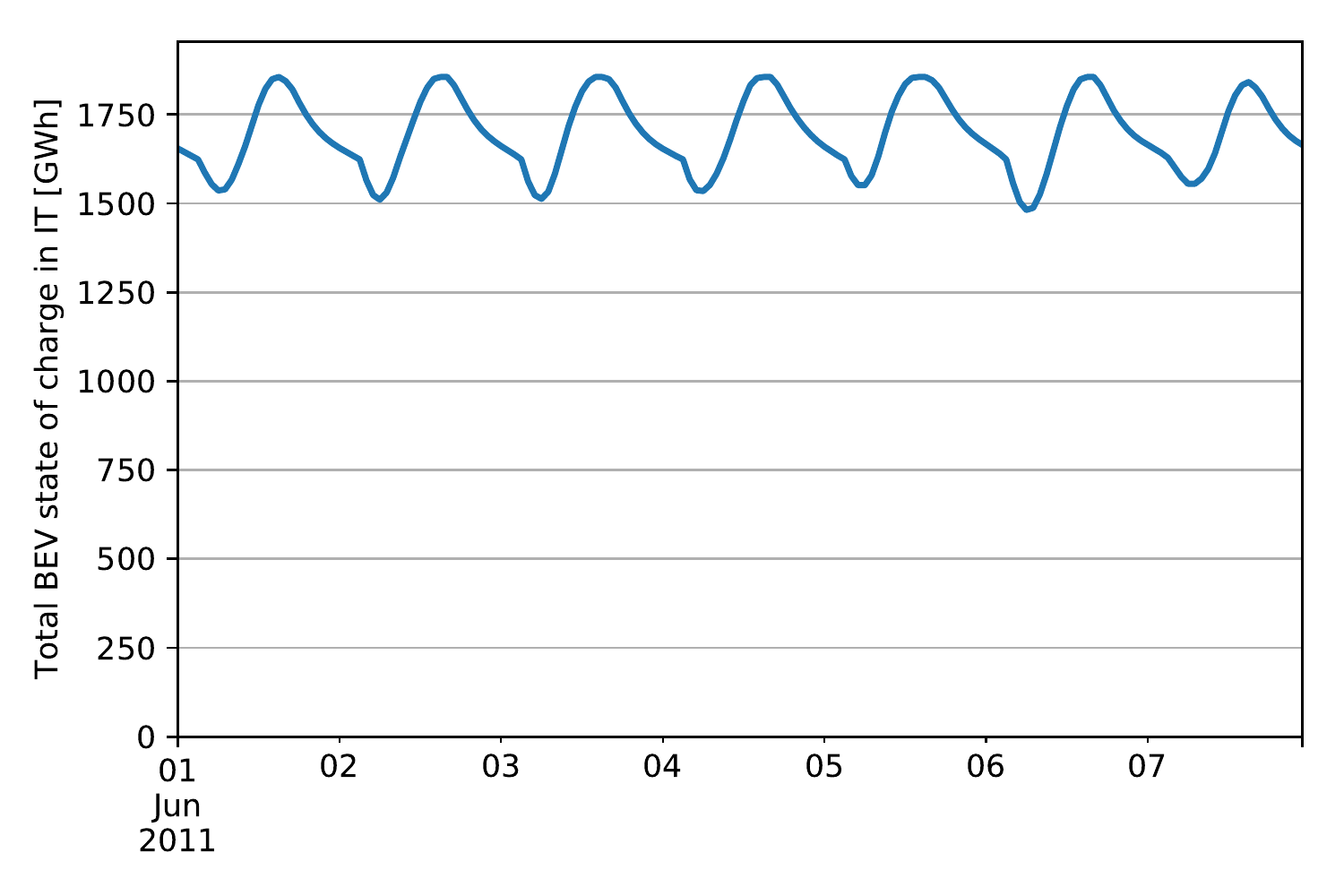}
\caption{The battery electric vehicle total state of charge in Italy
  for the scenario {\bf V2G-50} during a sunny two-week period. The
  total energy capacity of all BEVs is 1855~GWh.}
\label{fig:v2g}
\end{figure}

\subsection{Transport with Vehicle-To-Grid from BEVs}

In the Vehicle-To-Grid ({\bf V2G}) scenarios different fractions of
the BEVs are allowed to not only shift their charging time, but also
to discharge electricity back into the grid at times which are
profitable. This essentially makes battery capacity available to the
system without any additional investment (since the costs of the
vehicles are excluded from our consideration), but it is not free,
since the vehicle owners receive a payment from the system
corresponding to the price difference between the market price at
charging and discharging times.

With all vehicles participating in the {\bf V2G-100} scenario, the
total system costs are reduced a total of 22\%
compared to the {\bf Transport} scenario with no transmission, to a
level that is just 10\% above the cost of the {\bf Electricity}
scenario.
With each increase in V2G share, 25\% to 50\% to 100\%, there is a
substantial cost saving, although the saving is bigger with no
transmission than the case with optimal transmission. V2G leads to the
complete elimination of stationary battery storage and the successive
elimination of hydrogen storage. With all vehicles participating in
V2G and optimal transmission capacity there is no stationary storage
at all; in this case there is increased investment in solar PV and an
almost complete elimination of expensive offshore wind. The benefits
of V2G are simply due to the sheer volume of storage made available to
the system: 12.3~TWh, which is 1.5 days' worth of electricity demand.
This allows ample capacity to smooth out diurnal fluctuations, which
is reflected in the higher shares of solar PV in the energy mix.

The system benefits of V2G are also reflected in other
indicators. With optimal transmission the shadow price of \co{} drops
to 114~\euro/t\co{} in the {\bf V2G-100} scenario, which is 16\% less
than the value in the {\bf Electricity} scenario.
V2G also reduces
the need for interconnecting transmission, with the optimal capacity
dropping to 12\% below
the {\bf Electricity} scenario.

Such high levels of DSM and V2G may however be undesirable for other
reasons, such as inconvenience for consumers and the increased
wear-and-tear on battery components. Although vehicle owners are
compensated both for DSM and V2G according to market prices, this may
not be sufficient to cover their costs. However, as these results
demonstrate, there are already significant system benefits if only a
fraction of vehicles participate in DSM and V2G.

Furthermore, the way the BEV batteries are used in the V2G scenarios
does not involve the regular deep-discharge cycling that tends to
degrade battery performance (at least in the country-aggregated
profiles; individual consumption patterns may lead to deeper
discharging, but the consumers themselves are responsible for this).
Figure \ref{fig:v2g} shows the aggregated battery state of charge over
a two-week sunny period in Italy from the scenario {\bf V2G-50}. While
charging during the midday PV peak and discharging for the evening
electricity and transport demand peak is visible, the changes in
energy are small compared to the total vehicle energy capacity;
in addition, the requirement that the state of charge is above 75\% at
5~am every day keeps the overall level high and prevents the use of
BEVs for smoothing variable renewables over periods longer than a
day. If more than half of today's passenger car fleet were made
redundant by shared autonomous vehicles, this picture would change
because the available battery capacity would be lower, but such
dramatic changes in consumer behaviour are considered unlikely
\cite{rmi2017}.

\subsection{Transport with Fuel Cell Electric Vehicles}

In the Fuel Cell Electric Vehicle ({\bf FCEV}) scenarios fractions of
the electric vehicles are replaced with vehicles that use onboard fuel
cells consuming hydrogen. Since hydrogen is cheaper to store than
electricity, the demand for hydrogen represents a large time-shiftable
demand to the system that can be used to balance synoptic and seasonal
variations in solar and wind feed-in. On the other hand, the
efficiency of the electrolysis (80\%) and the fuel cell conversion of
hydrogen back to electricity (58\%) is much lower than for battery
charging and discharging (90\% and 90\% respectively).

In each of the {\bf FCEV} scenarios, with the fraction of FCEVs
ranging from 25\% to 100\%, the total system costs are higher than the
all-electric {\bf Transport} scenario, rising to 16\% higher in the
{\bf FCEV-100} scenario.
The higher costs are driven by higher investment in electrolysis
devices and hydrogen storage, and more investment in wind and solar to
supply the higher energy demand. These cost increases are not
offset by the lower investment in stationary battery storage.  These
results show that the higher energy demand resulting from the lower
round-trip efficiency of FCEVs increases costs more than they are
reduced by the large shiftable electrolysis demand. On the positive
side, the shiftable demand decreases the average electricity price
below the level in any of the other transport or electricity-only
scenarios.

To the additional costs of FCEVs must also be added the higher costs
of the vehicles themselves \cite{IEESWV} and the costs of the hydrogen
distribution system, which was calculated in
Section \ref{sec:transport_supply} to be around 11~billion~\euro/a
more for a 100\% FCEV scenario than the charging infrastructure for a
100\% BEVs scenario.


These results indicate that the FCEVs are not beneficial from a system
point of view and should be restricted to applications where the high
energy density of hydrogen is required, such as for long-range
journeys or for heavy duty vehicles (e.g. haulage trucks) on routes
where electrification of roads (e.g.  with overhead pantographs) is
not possible.

\begin{figure*}[!t]
\centering
\includegraphics[trim=0 0cm 0 0cm,width=\linewidth,clip=true]{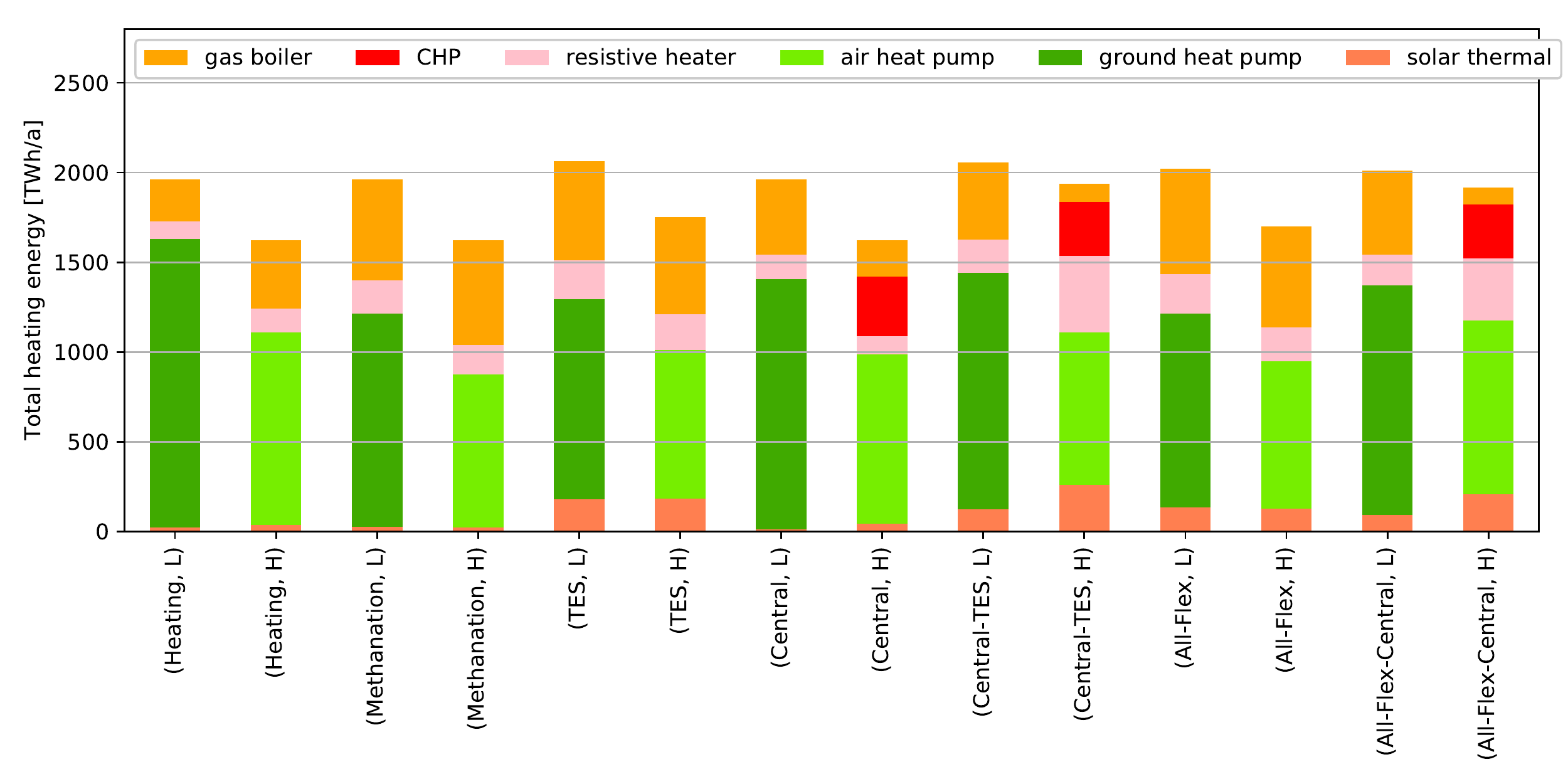}

\includegraphics[trim=0 0cm 0 0cm,width=\linewidth,clip=true]{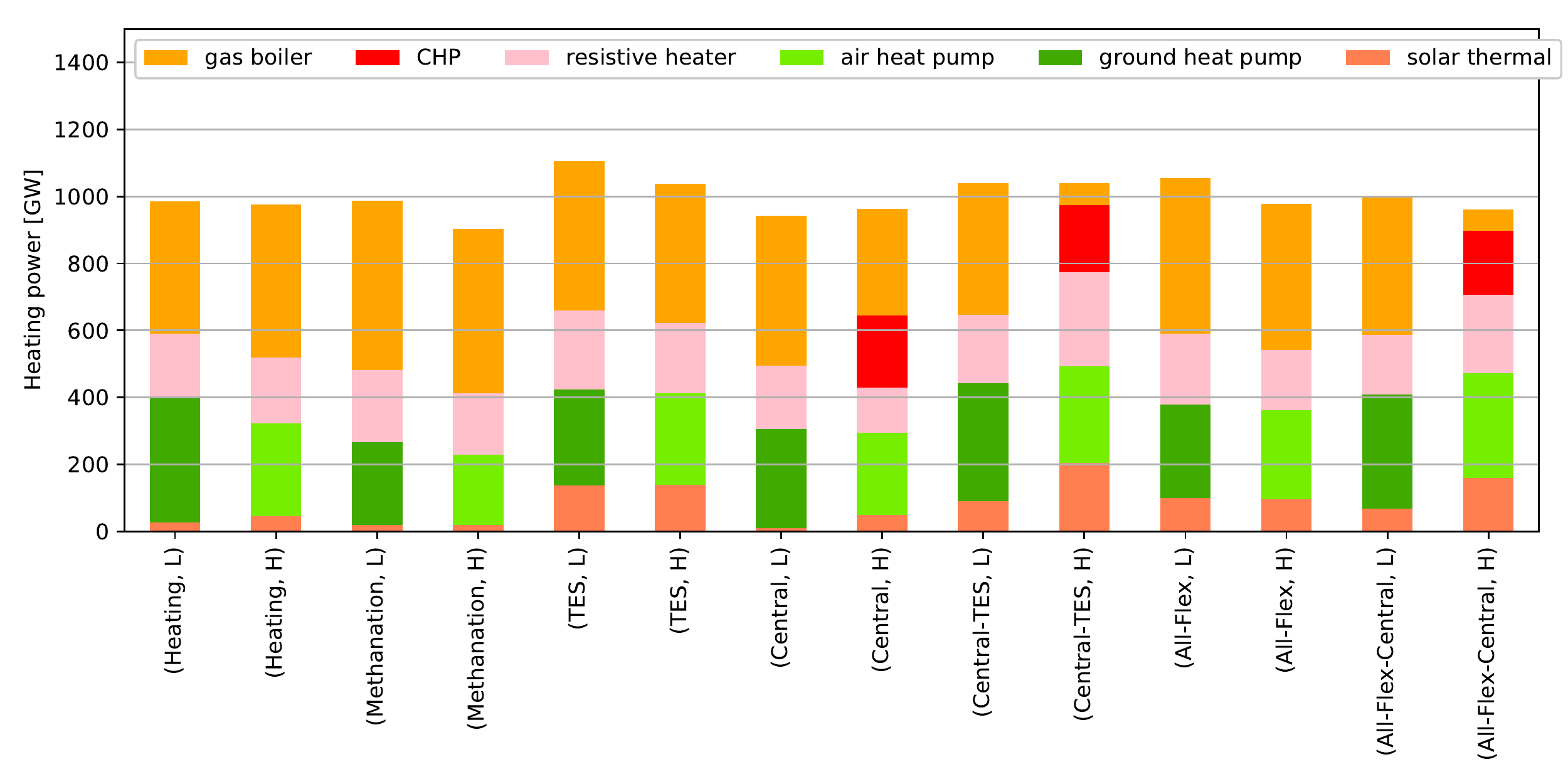}

  \caption{Heating total energy arriving at the heat buses (top) versus heating power capacity (bottom) for each heating scenario, split by the heating provision in low-density areas (L) versus high-density areas (H).  No electricity transmission is assumed for these results.}
\label{fig:supply}
\end{figure*}

\begin{figure*}[!t]
\centering
\includegraphics[trim=0 0cm 0 0cm,width=0.48\linewidth,clip=true]{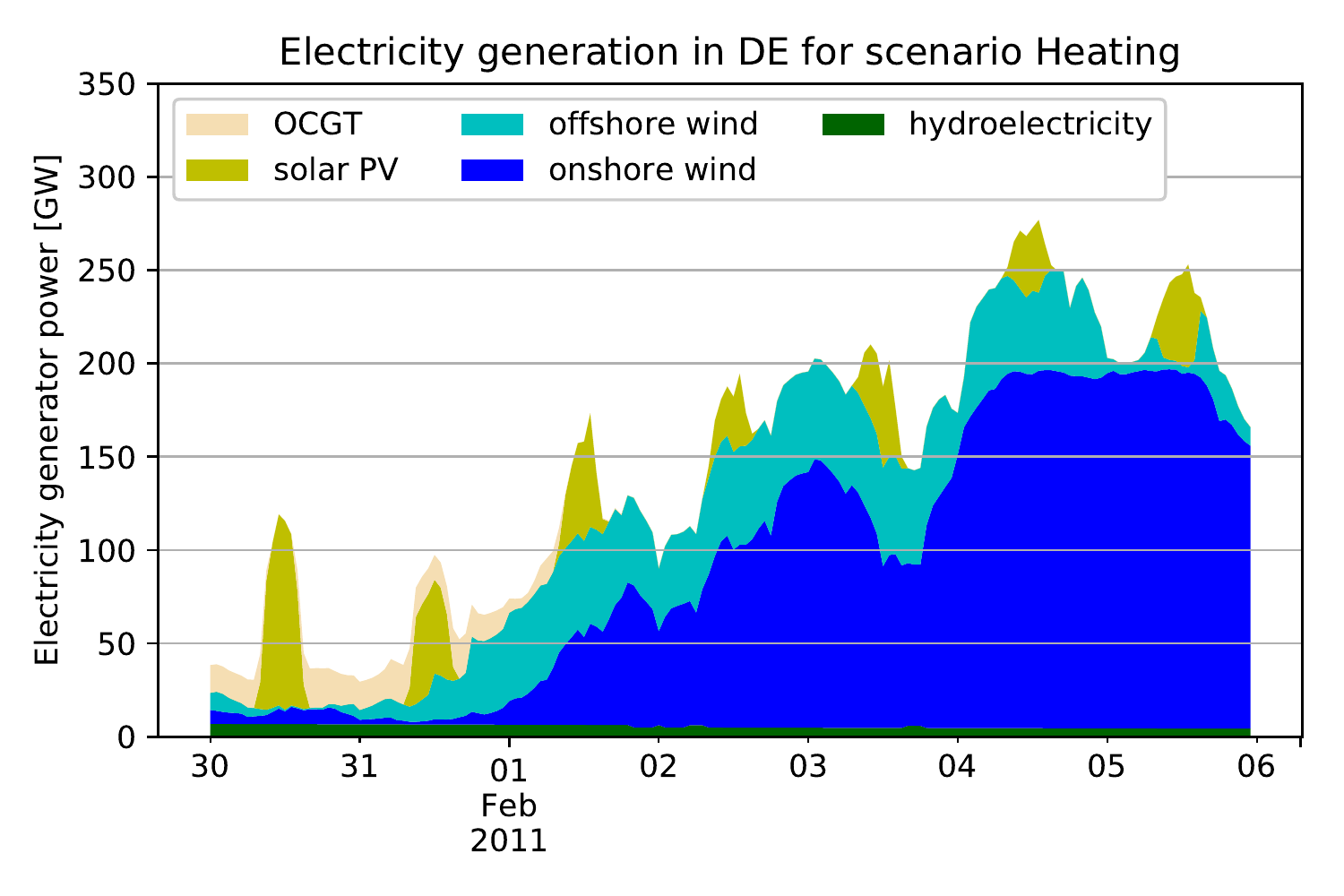}
\includegraphics[trim=0 0cm 0 0cm,width=0.48\linewidth,clip=true]{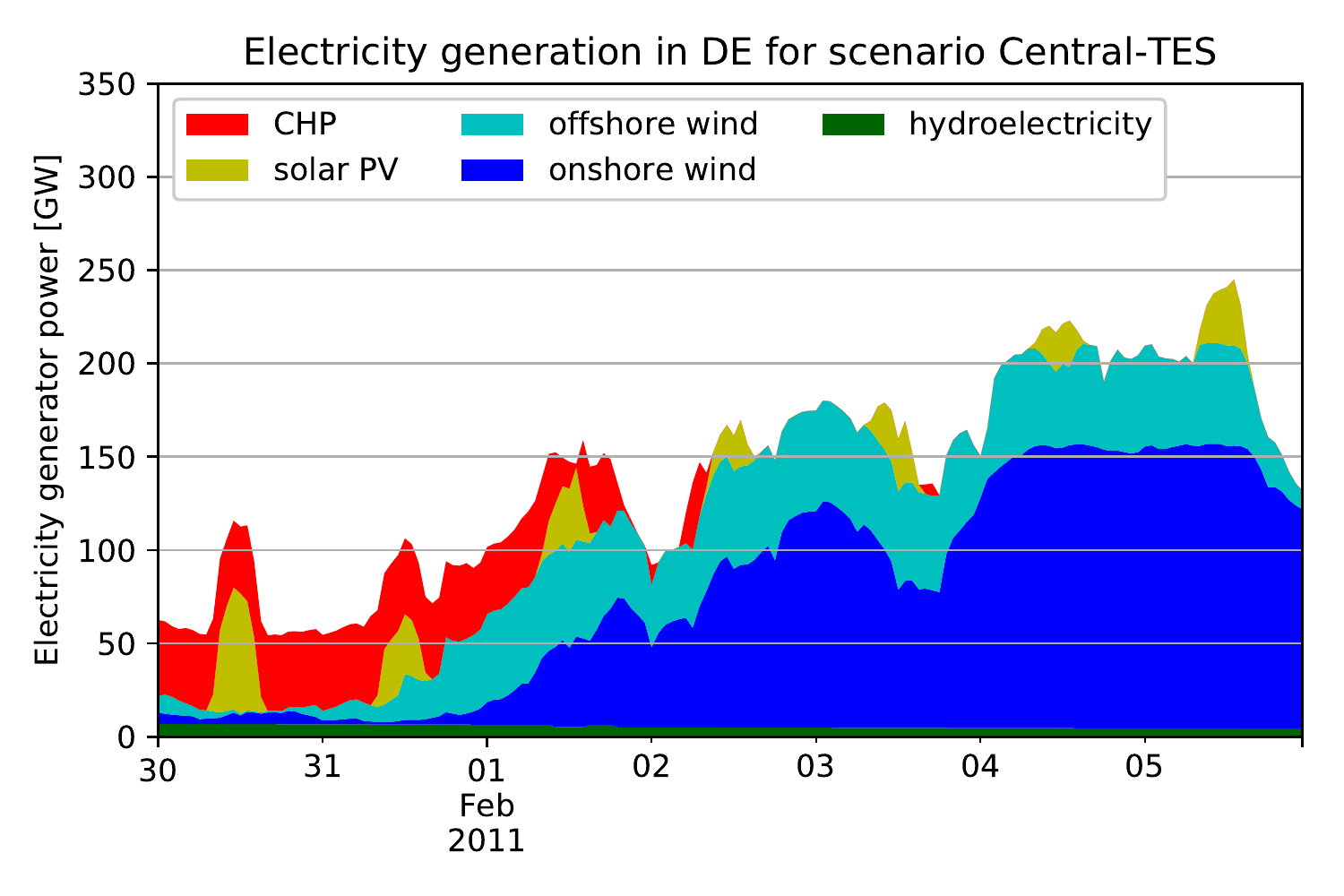}

\includegraphics[trim=0 0cm 0 0cm,width=0.48\linewidth,clip=true]{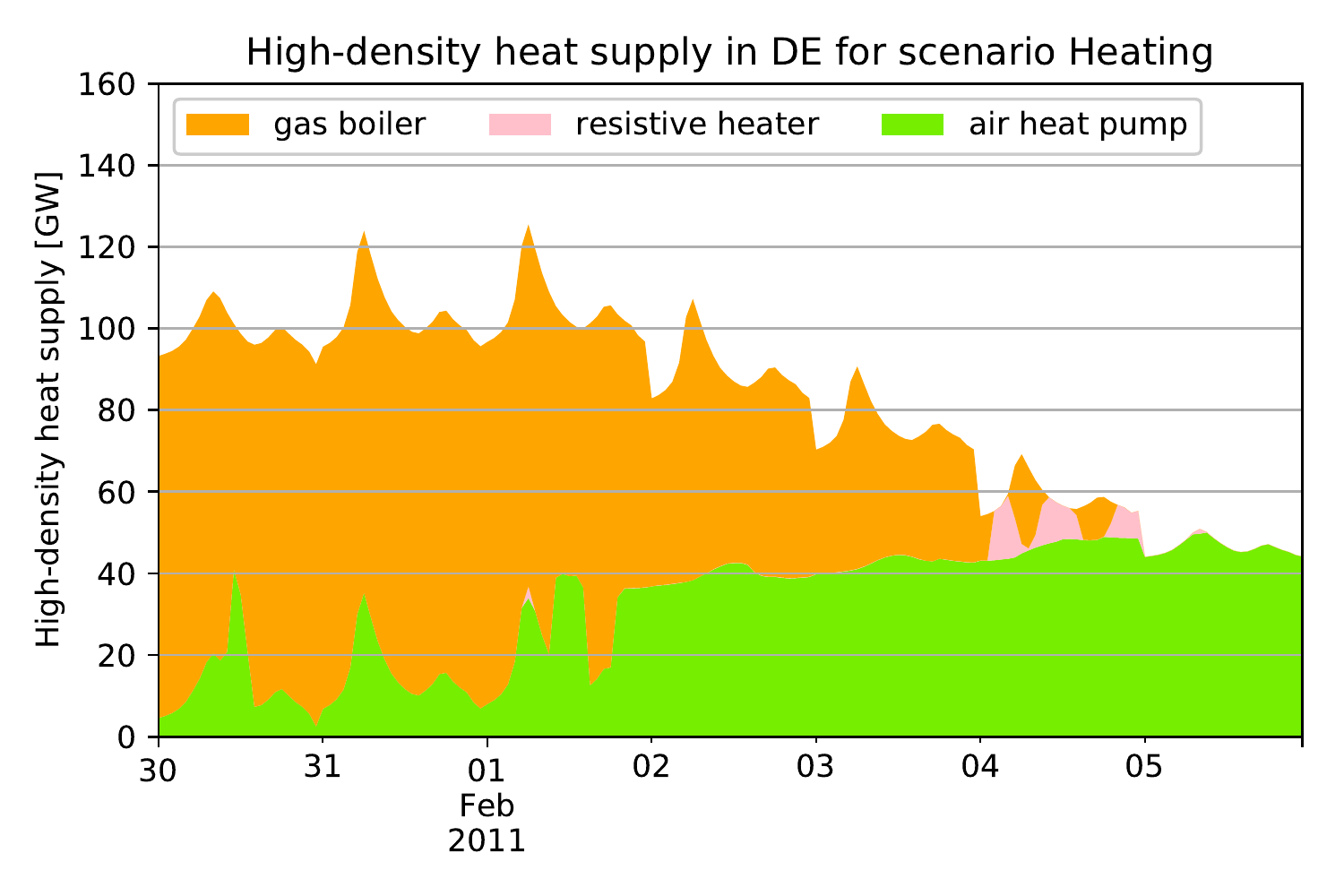}
\includegraphics[trim=0 0cm 0 0cm,width=0.48\linewidth,clip=true]{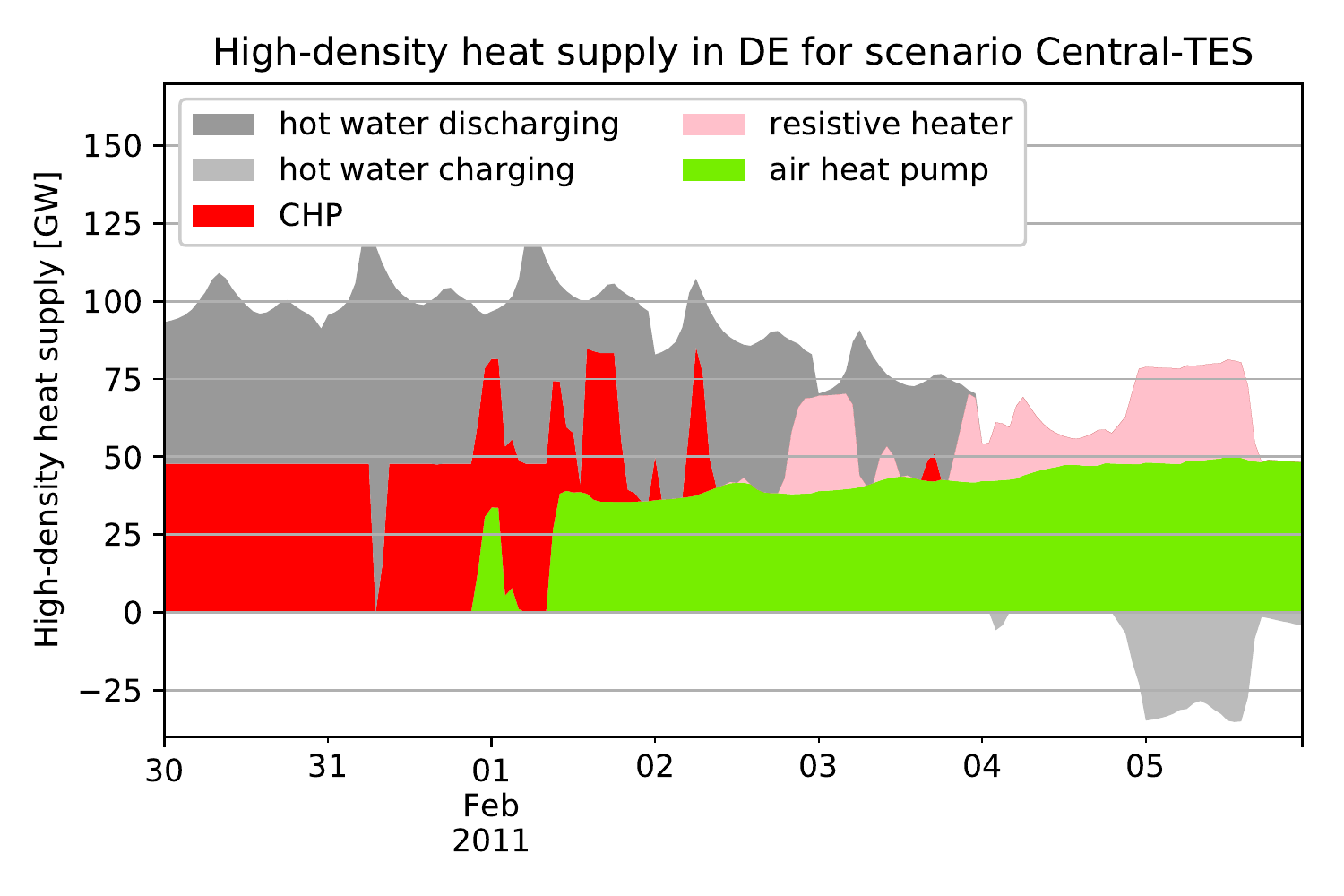}
\caption{Electricity supply ignoring storage (top) and heating supply in densely-population areas (bottom) for the {\bf Heating} scenario (left) and the {\bf Central-TES} scenario (right) in Germany during a week that includes the coldest days of the year. No transmission is assumed for these results.}
\label{fig:coldweek}
\end{figure*}

\begin{figure*}[!t]
  \centering
    \includegraphics[trim=0 0cm 0 0cm,width=\linewidth,clip=true]{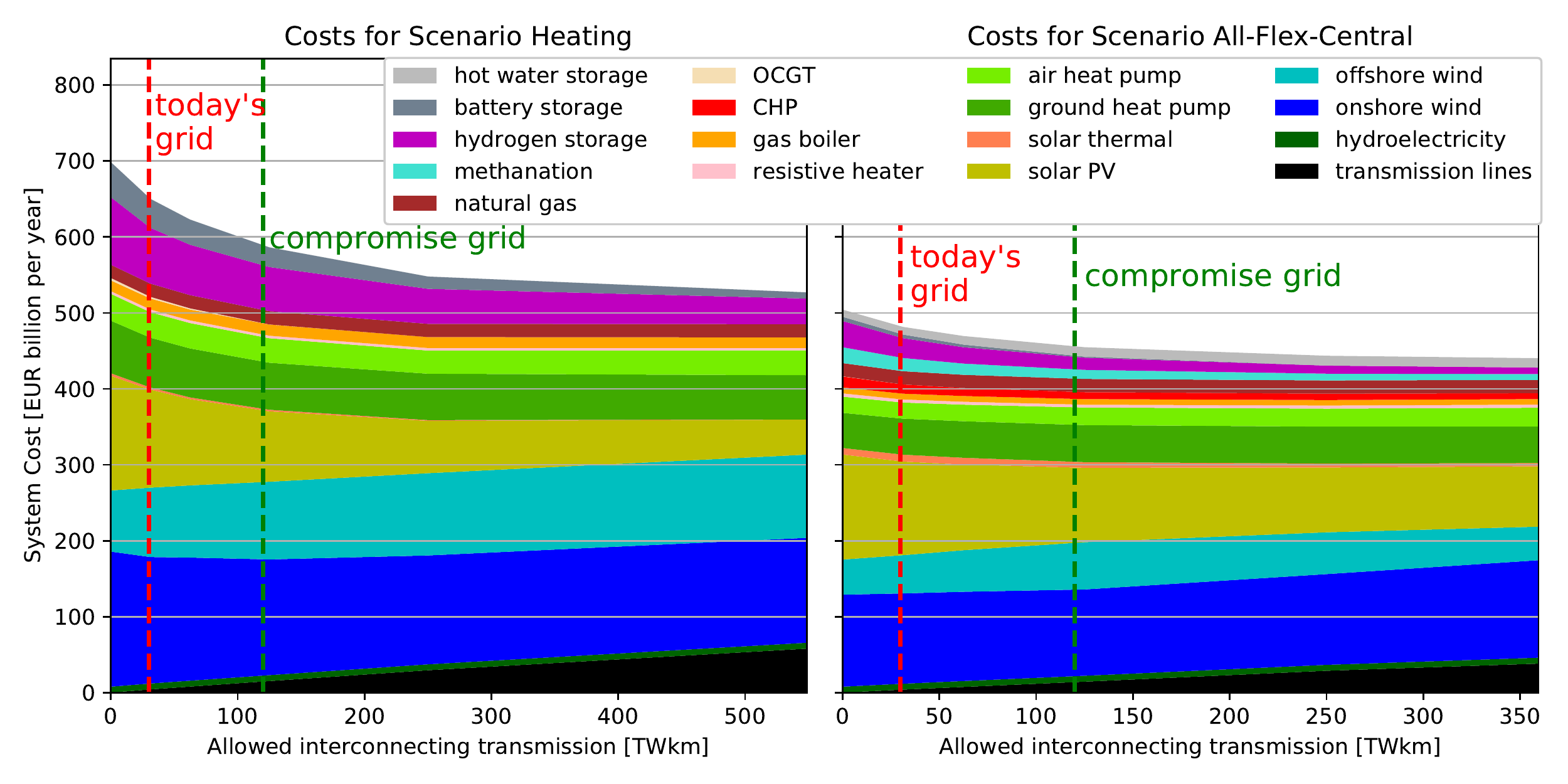}
    \caption{Total yearly system costs as a function of the allowed
      inter-connecting transmission capacity ($\textrm{CAP}_{LV}$ from
      equation \eqref{eq:lvcap}), assuming that transmission is costed
      as overhead lines. The left graphic shows the Heating scenario
      and the right graphic the All-Flex-Central scenario. The right
      axis of each graphic marks the optimal level of grid expansion, which is different in each scenario.}
\label{fig:costs}
\end{figure*}

\subsection{Heating scenario}

In the {\bf Heating} scenario, the heating demand is added to both the
electricity and transport demand without adding any extra flexibility
options, such as BEV DSM, V2G, thermal energy storage (TES),
power-to-gas (P2G) feeding into the natural gas network or district
heating in densely populated urban areas.

The addition of heating not only increases the energy demand in the
model (adding 3585~TWh\th/a to the transport and electricity demand of
4062~TWh\el/a), but it also requires new infrastructure to meet the
heating demand. Given the 95\% \co{} reduction target, much of the
heating demand has to be met by converting renewable electricity to
heating, primarily using heat pumps but also using resistive heaters
(i.e. electric boilers).

With no transmission, the total system costs increase by 117\%
compared to the {\bf Transport} scenario. Heat pumps (air-sourced in
densely-population areas, ground-sourced elsewhere) make up 15\% of
the total costs.

In Figure \ref{fig:supply} the heat supply for each scenario with no
transmission is plotted in terms of yearly energy contribution (top)
and in terms of the peak power capacity (bottom). The supply is split
between the low-density and high-density demand areas. In the {\bf
  Heating} scenario, the heating energy provision is dominated by heat
pumps, thanks to their efficient use of electricity, but gas
boilers provide the most heating power capacity (enough to cover 58\%
of the peak heating demand).


This discrepancy can be explained by examining, for example, the available
electricity generation and heat supply in Germany during a cold week
of the year; see the lefthand plots of Figure
\ref{fig:coldweek}. At the start of this week it is cold, so that the
heating demand is high, while the COP of heat pumps is low; at the
same time there is very little low-marginal cost wind and solar available. As a
result, the heat pumps are only used when there is a peak of solar PV,
and at other times gas boilers must step in to provide backup energy. At
these times resistive heaters would also be too expensive because of
the high price of electricity and their low efficiency compared to
heat pumps.

This means that for buildings supplied by individual heating units, a
cost-effective system is for heat pumps to provide the bulk of the
yearly heating demand and for gas boilers to provide backup capacity
for cold spells. This is more cost-effective than providing all
heating demand with heat pumps or resistive heaters, since this would
require a large backup OCGT fleet to meet the peak electricity demand,
which is less efficient. It does, however, require multiple heating
technologies for each building.

This reveals a significant difference between the economics driving
the electricity sector versus the heating sector: the heating demand
in Europe is much more strongly and seasonally peaked than the
electricity demand (refer back to Figure \ref{fig:heat} for the yearly
heating profile) making the balance between so-called `base load' and
`peaking' heat provision more skewed in favour of peaking plant.

The heating demand is also more strongly seasonal than the wind in
Europe (which also peaks in winter) and is anti-correlated with the
seasonality of solar energy in Europe. This mismatch helps to further
explain why the total system costs of the {\bf Heating} scenario are
disproportionately higher (given the change in energy demand) than the
{\bf Transport} scenario.

With no transmission, the high average marginal price of supplying
space heating demand (153~\euro/MWh\th{} in low-density and
161~\euro/MWh\th{} in high-density areas) is sufficiently high,
particularly when approximate gas distribution network costs of
15~\euro/MWh\th{} and/or taxes are added, to justify retrofitting
buildings to reduce heat demand by between 70\% and 80\%, based on
Figure \ref{fig:retrofitting} and assuming similar characteristics to
Germany and Denmark. This also assumes that the marginal price remains
constant as space heating demand is reduced; this assumption might be
warranted, given that the high price is caused by the shape of the
heating profile and the technology mix, but on the other hand the
price might go down because the \co{} limit is easier to meet with
lower energy demand, thus only justifying a slightly lower rate of
demand reduction. With optimal transmission, the marginal price for
space heating is 28\% lower, resulting in a lower rate of
retrofitting.



The \co{} shadow price, reflecting the marginal cost of further
reducing emissions, is high at 1184~\euro/t\co{} in this scenario with
no transmission. This high price is a direct reflection of the high
price paid for energy in the model (see Table \ref{tab:scenarios}),
versus the low fuel cost of natural gas (21.6~\euro/MWh\th), which
thus requires a high \co{} price to justify avoiding natural gas. As
energy becomes cheaper in the following scenarios, so the \co{} price
goes down.

As interconnecting transmission is expanded to its optimal level, costs reduce by 25\%
and there is a substantial shift of energy generation from solar PV to
wind, since the synoptic variability of wind can now be balanced in
space by the grid. The investments in stationary battery and hydrogen
storage are also significantly reduced. In the left graphic of Figure
\ref{fig:costs} the breakdown of the system costs is plotted as the
restriction on transmission ($\textrm{CAP}_{LV}$ from equation
\eqref{eq:lvcap}) is relaxed to its optimal level. As in the {\bf
  Electricity} scenario (see also \cite{Schlachtberger2017}), the cost
reduction is non-linear as transmission is expanded.
Thus, despite the fact that the optimal level of transmission is very
high (173\% higher than the {\bf Electricity} scenario), 66\% of cost
savings are already achieved with a compromise expansion of
inter-connecting capacity to 125~TWkm, which is four times today's net
transfer capacities (NTC). The European regulator ACER believes that
today's NTC could be doubled if congestion were managed more
effectively \cite{ACER2016}; a further doubling of cross-border
capacities through grid expansion is already foreseen by the official
planning process by 2030 \cite{TYNDP2016}, but not in exactly the same
places as seen in this model.


\subsection{Methanation scenario}

In the {\bf Methanation} scenario the conversion of hydrogen to
methane is allowed, which can then be fed into the natural gas network
for use both in the heating and electricity sectors. Since the carbon
dioxide required for the methanation is captured from the air, the
methanation has a low overall efficiency (60\%), but the resulting
methane is extremely valuable to meet the peak heating demand.

Despite the costs of the methanation equipment, total system costs
reduce by 11\% compared to the {\bf Heating} scenario.
In the heating sector, a substitution of heat pumps with gas heating
can be observed in Figure \ref{fig:supply}. Significantly reduced
\co{} prices and average marginal prices for electricity and heating
are also seen in Table \ref{tab:scenarios}. Furthermore, the benefit
of transmission reinforcement is weakened, since the methanation allows the use of
cheap gas storage to smooth synoptic and seasonal variations of
renewables. Optimal transmission reduces the total systems costs by
only 17\%, compared to 25\% in the {\bf Heating} scenario, and the
optimal transmission volume is also lower.

The total volume of synthetic methane produced with no transmission is
708~TWh\th, compared to 795~TWh\th{} from natural gas. With optimal
transmission the volume of synthetic methane reduces to 263~TWh\th{} as
transmission smoothes more synoptic variations of wind.



\subsection{Thermal energy storage scenario}

In the Thermal Energy Storage ({\bf TES}) scenario small hot water
tanks are added to the {\bf Methanation} scenario with a short
time constant of $\tau = $~3 days.

As can be seen from Figure \ref{fig:supply}, TES enables a higher
share of heating from solar thermal collectors, since the heat can be
shifted to hours of higher heat consumption. (Since most solar thermal
collectors are installed with TES already, the exclusion of TES from
the previous scenarios was somewhat contrived.) However, the thermal
losses of the TES mean that more heat must be provided. As a result,
the system costs are lowered by just 1.3\% compared to the {\bf
  Methanation} scenario.

In total 57 million cubic metres of TES is built in this scenario,
averaging 0.108 cubic metres per citizen.

\subsection{Central scenarios}

In the {\bf Central} scenario, heating demand in densely-populated
areas is served with district heating rather than individual heating
units. This enables large combined heat and power plants (CHPs) to be
deployed (see Table \ref{tab:heating}) and the larger scale of all
heating units reduces costs (see Table \ref{tab:costs}).  This leads
to a reduction in total costs of 6\% compared to the {\bf Methanation}
scenario, on which this scenario is based.
Figure \ref{fig:supply} shows that CHPs do indeed take
over some of the heating supply provided previously by gas boilers.

Another major advantage of district heating is seen when Long-term TES
(LTES) is allowed in the {\bf Central-TES} scenario. The large,
well-insulated water tanks used for LTES in the district heating
network have a heat decay time constant of $\tau = $~180 days, which
allows heat to be shifted seasonally. This results in a higher
share of solar thermal (see Figure \ref{fig:supply}), which is used to
charge the LTES in summer/autumn, and more usage of resistive heaters
when electricity prices are low. In the areas with district heating,
the total volume of hot water tanks in LTES is 3.1 billion cubic
metres, averaging to 13 cubic metres per citizen.

The LTES can then be used to supply heat during cold periods, as is
shown in the example for Germany's cold spell on the righthand side of
Figure \ref{fig:coldweek}. The majority of heat during the coldest
times comes from LTES, with the remainder covered by CHPs; gas boilers
have been almost totally eliminated.

The benefit of LTES can also be seen in the drop in the average space
heating price in high density areas from 92~\euro/MWh\th{} in the {\bf
  Central} scenario, to 79~\euro/MWh\th{} in the {\bf Central-TES}
scenario.

The annualised cost of building and maintaining the district heating
network to meet its total peak load of 548~GW\th{} is 10 billion~\euro/a
according to our cost assumptions. However, a gas distribution network
is no longer needed in areas with high-density heating demand, and this
reduces annual costs by 20 billion~\euro/a, more than offsetting the
cost of the district heating network and further contributing to the
attractiveness of district heating. The cost benefits of district
heating are also maintained with optimal transmission.

\begin{figure*}[!t]
\centering

    \includegraphics[trim=0 0cm 0 0cm,width=8.7cm,clip=true]{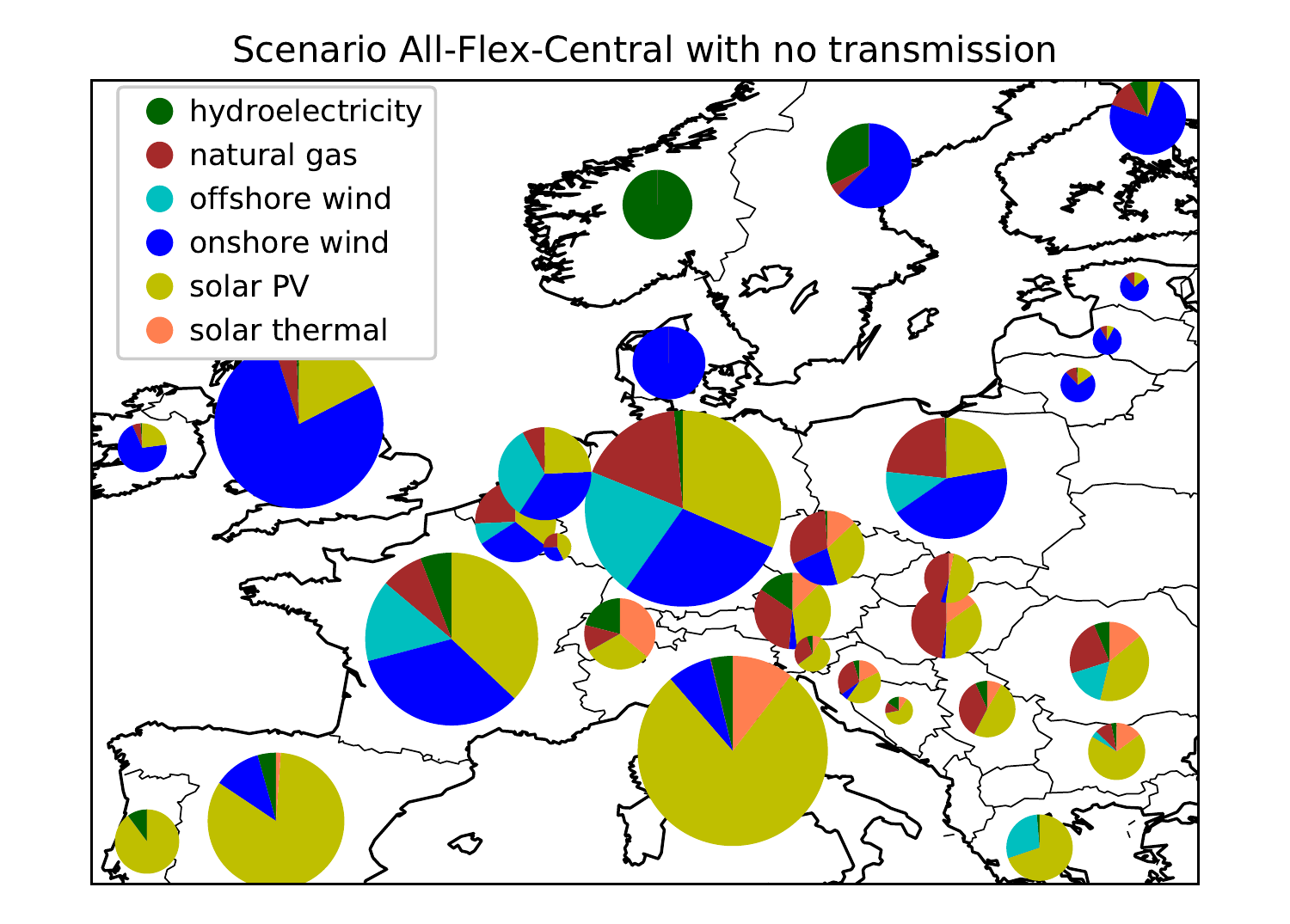}
    \includegraphics[trim=0 0cm 0 0cm,width=8.7cm,clip=true]{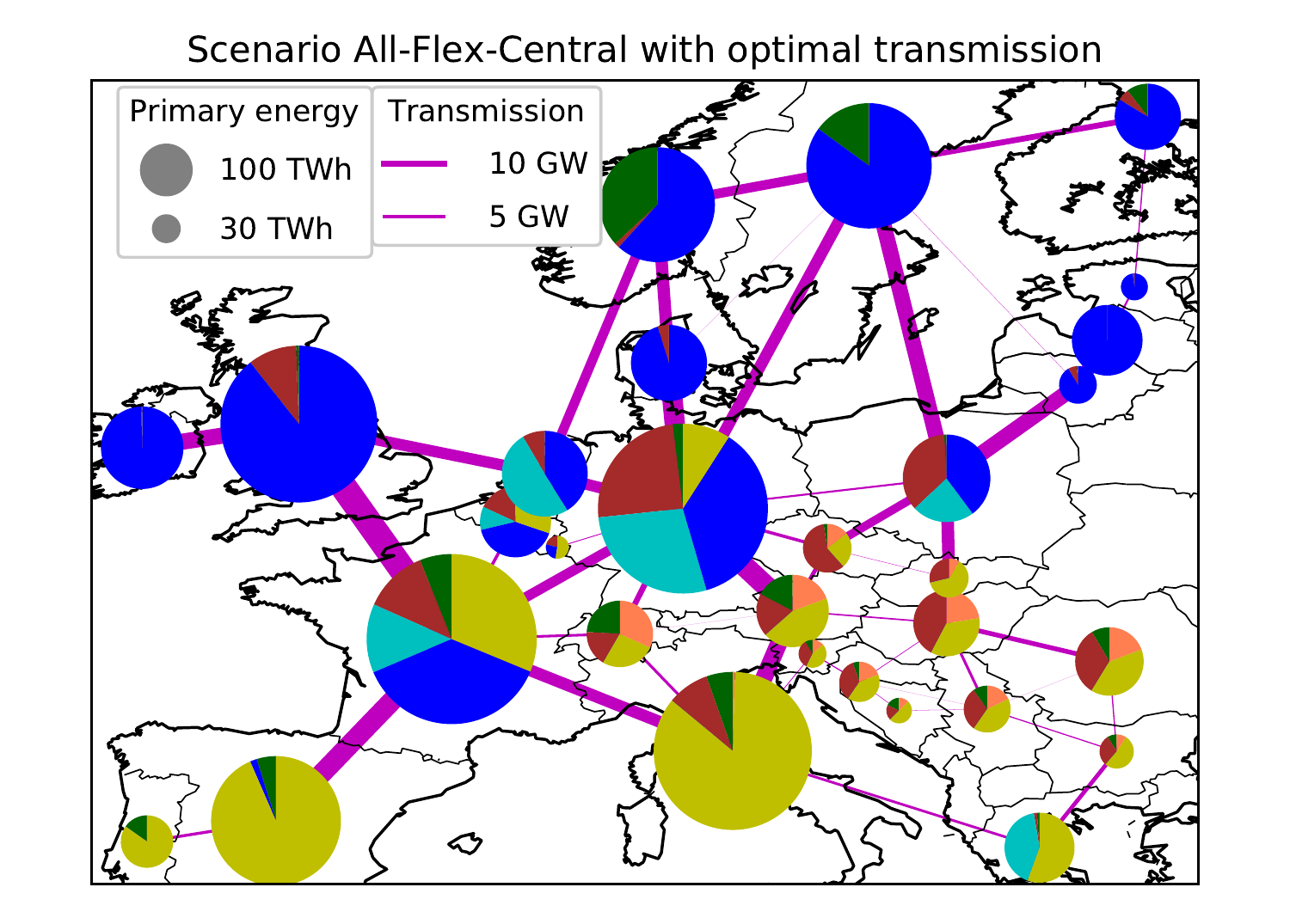}
    \caption{Primary energy consumption in the All-Flex-Central scenario with no transmission (left) and optimal cross-border transmission (right).  Given that each country is self-sufficient on the left, the difference in circle size on the right gives an indication of the heterogeneity of energy generation.}
\label{fig:primary}
\end{figure*}

\subsection{Scenarios with all flexibility options}

In the final scenarios {\bf All-Flex} and {\bf All-Flex-Central}, all
flexibility options are activated, including 50\% BEV-DSM, 50\%
BEV-V2G, methanation and TES.

The total costs in the scenario {\bf All-Flex-Central} with no
transmission are 28\% lower than the {\bf Heating} scenario, and 17\%
cheaper than the {\bf Heating} scenario with optimal transmission.
Much of the cost reduction in the {\bf All-Flex-Central} scenario
comes from a reduced need for stationary battery storage, hydrogen
storage and methanation, thanks to the availability of LTES and
BEV-V2G. The production of synthetic methane drops to 475~TWh/a
with no transmission and 184~TWh/a with optimal transmission, which is
around one third lower than the values in the {\bf Methanation}
scenarios. The daily smoothing provided by BEV-V2G also makes a larger
share of solar PV cost efficient in the {\bf All-Flex} scenarios.

The only difference between the {\bf Central-TES} and {\bf
  All-Flex-Central} scenarios is the introduction of 50\% BEV-DSM and
BEV-V2G. With no transmission, this reduces costs by 58
billion~\euro/a. This is almost identical to the cost reduction of 55
billion~\euro/a between the {\bf Transport} and {\bf V2G-50}
scenarios; similar changes in technology are also seen (more solar, less
electricity storage). This is an indication that the benefits of
transport flexibility (largely on daily time scales) are independent
from the benefits of heating flexibility (largely on synoptic and
seasonal time scales). This effect was also seen in \cite{Gils}.

Transmission is still beneficial in these scenarios, but the benefit
is much weaker: the ratio of the costs with and without optimal
transmission drops from 1.33 in {\bf Heating} to 1.15 in {\bf
  All-Flex-Central}. {\bf All-Flex-Central} with no transmission is
cheaper than {\bf Heating} with optimal transmission, but {\bf
  All-Flex-Central} with optimal transmission is the cheapest scenario
of all.

The optimal transmission volume of 359~TWkm is also much reduced.  As
can be seen from the righthand graphic in Figure \ref{fig:costs}, both the
drop in system costs and the change in system composition as
transmission is expanded are less dramatic than in the {\bf Heating}
scenario in the lefthand graphic. The compromise grid (four times today's
NTC) already captures 78\% of the cost benefit of optimal
transmission.
The technology choices also remain mostly stable as the transmission
volume is changed; increases in wind energy and reductions in hydrogen
storage reflect the availability of transmission for synoptic
smoothing.  This stability is also reflected in Table
\ref{tab:scenarios} in the barely-changing marginal prices; the
average marginal cost of heating in high density areas in fact remains
constant.

In Figure \ref{fig:primary} the spatial distribution of primary energy
consumption is plotted for the {\bf All-Flex-Central} scenario with
and without optimal transmission. The spatial distribution of
technologies remains broadly similar; optimal transmission allows more
onshore wind to be build around the North and Baltic Seas, while solar
PV is focussed on Southern Europe. Because export is possible with
transmission, some countries, such as Ireland, Norway and Sweden where
wind resource are good, generate more energy than they consume, while
others become net importers.

The \co{} price drops to 407~\euro/t\co{} because the cost of heating
is now lower compared to the fuel price. (If heating is the cheapest
place to displace \co{}, the relationship between the fuel cost
$o_{\textrm{gas}}$, \co{} price $\mu_{CO2}$, specific emissions
$\varepsilon_{\textrm{gas}}$ and heating price
$\lambda_{\textrm{heat}}$ at the cheapest hour and place where gas is
consumed is $o_{\textrm{gas}} +
\varepsilon_{\textrm{gas}}\cdot\mu_{CO2} = \lambda_{\textrm{heat}}$ using
the KKT relations.) Note that this price is high enough that other technologies for
carbon dioxide reduction that are not in the model, such as carbon
capture, might be attractive before this price is reached.


Finally, the reduced marginal costs of space heating would lead to a lower
optimal level of building retrofitting than seen, for example, in
the {\bf Heating scenario}. Taking account of the cost of distribution networks
and using Figure \ref{fig:retrofitting}, a reduction in heating demand
of around 20--35\% would be efficient in this scenario. Similar levels of
optimal retrofitting have also been seen in other studies
\cite{PALZER20141019,Connolly20161634}.

\subsection{Temporal scales}

\begin{figure}[!t]
\centering
  \includegraphics[trim=0 0cm 0 0cm,width=\linewidth,clip=true]{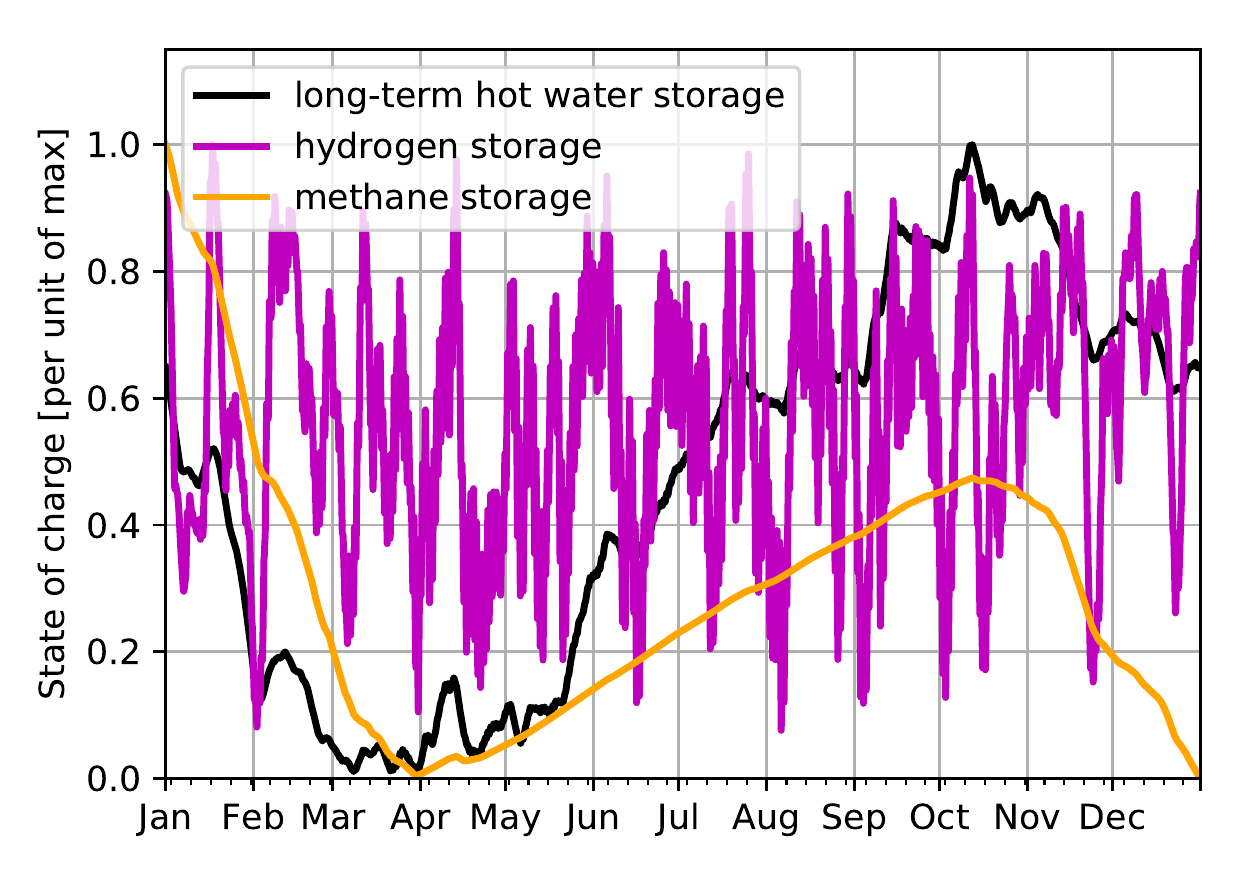}
  \caption{The state of charge for a selection of storage technologies in the {\bf All-Flex-Central} scenario without transmission, aggregated over all countries, normalised to the total energy capacity (141~TWh for long-term hot water, 9.4~TWh for hydrogen, 806~TWh for methane). Since the state of charge is aggregated, it does not necessarily drop to zero. While the other storage technologies are cyclic over the year, the methane storage finishes the year lower than at the start because of the depletion of natural gas reserves.}
\label{fig:scales}
\end{figure}

The time series of the states of charge of the different storage
technologies in the {\bf All-Flex-Central} scenario without transmission show the different
temporal scales on which each storage technology acts (see Figure
\ref{fig:scales}). The methane storage is depleted throughout the
winter when energy demand is highest, then replenished throughout the
summer with synthetic methane, mirroring seasonal imbalances in demand and renewable
supply. (It finishes the year lower than at the start to account for the depletion of natural gas reserves.) The hydrogen storage fluctuates on shorter, synoptic time
scales of 2--3 weeks, reflecting its role in balancing wind fluctuations. When spatial smoothing of synoptic variations is possible with the grid, investment in hydrogen storage drops.
The long-term hot water storage is dominated by a seasonal pattern
similar to that for methane storage superimposed with smaller
synoptic variations that match the variations in hydrogen storage.
Not plotted are the diurnally-varying storage technologies: the
short-term hot water storage that matches the solar thermal
collectors, and stationary and vehicular battery storage which follows
daily demand and solar PV fluctuations (see Figure \ref{fig:v2g}).

Recognising these different scales is critical to understanding the interaction between demand and renewable generation time series, and thus the resulting system composition.

\section{Discussion}\label{sec:discussion}

\subsection{Comparison of results to today's costs}

To approximate the cost of the current European energy system, some simplifications are made: we
assume an average cost of electricity of 50~\euro/MWh\el,
8~\euro/MWh\th{} for solid fuels, 40~\euro/MWh\th{} for oil and
22~\euro/MWh\th{} for gas, and assume that all non-electric heat load
is met by fossil fuel boilers priced like gas boilers, which are
dimensioned to meet the peak thermal load in each country. With these
assumptions and energy consumption figures from 2011
\cite{EurostatEB}, costs are 158 billion~\euro/a for electricity, 203
billion~\euro/a for fuel in the non-electric heating and transport
sectors considered here and 29 billion~\euro/a for the boilers,
resulting in total costs of 390 billion~\euro/a. The {\bf
  All-Flex-Central} scenario with optimal transmission costs just 13\%
more than today's system.
Furthermore, the estimation of today's costs excludes the external
costs due to greenhouse gases and airborne pollution, estimated by the
German Federal Environment Agency (UBA) to be 130~billion~\euro{} in
2014 in Germany alone \cite{uba}. Further analysis of the health
effects of air pollution from fossil-fuel-based energy systems can be
found in \cite{Jacobson2011b,ZVINGILAITE2011535}.

Note that the capital costs for vehicles have not been included in
this calculation. Although BEVs are currently more expensive than
vehicles with internal combustion engines, the upfront costs are
projected to be competitive already by the late 2020s \cite{ev2018}.

\subsection{Comparison of results to other similar studies}

Studies that focus only on the electricity sector typically find that
highly renewable systems are dominated by wind, which is most
cost-effectively integrated by expanding the pan-continental transmission
network \cite{Czisch,Schaber,Schaber2,Scholz,Rodriguez2013,OptHet};
without an expansion of the transmission network, solar energy is more
favoured, and expensive electricity storage solutions are needed to
balance variable renewables
\cite{HALLER2012282,GILS2017173,Schlachtberger2017,CEBULLA2017211}. The results
presented in this study still broadly support these conclusions, but
the cost benefit of transmission is weaker with sector-coupling than
with electricity alone thanks to the availability of cheap thermal
storage, BEV flexibility and power-to-gas facilities, which help to
replace much of the expensive stationary electricity storage.

The Smart Energy Europe study \cite{Connolly20161634} uses the
modelling tool EnergyPLAN to analyse a scenario with 100\% renewable
energy in all sectors in Europe by the year 2050, but with Europe
represented as a single node. The technology choices in that study
agree with many of the results found in the scenario {\bf
  All-Flex-Central}: optimal heat demand savings of 35\%, 80\%
electrification of private cars, heat pumps in rural areas, district
heating in urban areas and widespread use of synthetic
electrofuels. The annual system costs in that study are around
1400~billion~\euro/a (once the costs of vehicles have been excluded),
which is around three times the costs found here. This discrepancy is
largely due to the stricter renewable energy target and to the
inclusion of aviation, shipping and non-electric industrial demand,
which greatly increases the costs through the substitution of natural
gas, oil and coal with electrofuels. The discrepancy may also be due
to a lack of investment optimisation in that study. In the present
study, optimisation is used to find the most cost-effective energy
system given a fixed \co{} limit. Since \cite{Connolly20161634}
assumes perfect transmission within Europe, the present study improves
our understanding of the interaction between system characteristics
and cross-border transmission bottlenecks.

In \cite{PALZER20141019} a model of the German electricity and heating
sectors was optimised to meet a target of 100\% renewable energy. The
cost-optimal system (`REMax') sees a broadly similar selection of
technologies to the present study: a mix of solar PV, wind onshore and
offshore, CHPs, heat pumps, power-to-gas facilities and an optimal
level of space heating energy-saving of 31.8\% compared to 2010
values, which agrees with our analysis for the {\bf All-Flex-Central}
scenario. The total annual system costs were 111~billion~\euro/a,
which is within the range of our results once the size of Germany's
energy demand in relation to Europe's (around one fifth) is taken into
account. Nonetheless, that study misses some of the benefits of
interconnecting Germany within the European energy system identified
here.

The present study also confirms individual results of many other
studies: the higher benefit of cross-sector coupling compared to
cross-border coupling from the two-country study in
\cite{THELLUFSEN2017492}; the system benefits of district heating
\cite{Persson2011,Munster2012,CONNOLLY2014475,PENSINI2014530}; the
benefits of centralised heat storage in district heating networks
\cite{Nuytten2013,PENSINI2014530}; the independence of the benefits of
coupling BEV-DSM to the daily cycles of solar PV versus the seasonal
storage of energy for the heating sector \cite{Gils}; the importance
of modelling the temperature dependence of the heat pump COP
\cite{Petrovic2016}; and the advantages of power-to-gas and
methanation in particular in highly renewable energy systems
\cite{SternerPhD}.


\subsection{Limitations of the study}\label{sec:limitations}

Many of the limitations of this study arise from the simplifications
that are necessary to optimise the model in a reasonable amount of
time, such as aggregating each country to a single node or reducing
the range of available technologies. The impact of these
simplifications on the conclusions are now assessed.

Aggregating each country to a single node means that energy
distribution networks cannot be represented and that local resource
variations are not seen. In this study the costs of district heating
and gas distribution networks have been approximated based on their
peak loads. On the electrical side, the picture is more complicated,
since there is also distributed generation and storage that might
either relieve or exacerbate the pressure on distribution
networks. Thus, although the costs of inter-connecting transmission
lines have been taken into account, including the lengths of the
inter-connectors inside each country to reach each country's
mid-point, it has been assumed that the transmission and distribution
networks inside each country will be reinforced to relieve any
bottlenecks without attempting to assess the additional costs. The
additional costs are typically small compared to the total generation
costs (in the range of an addition 10-15\%) based on the results of
other studies (see \cite{burdenresponse} for a review). The grid costs
may also be offset by the ability to exploit good wind and solar sites better with a
finer-scaled model (see \cite{Hoersch2017} for an examination of these
trade-offs). More problematic than the costs is the potential for
public concerns about overhead transmission lines to delay or block
further grid extension. The costs of ancillary services for the grid have also
not been included, since their costs are negligible compared to the
total system costs \cite{burdenresponse}. On the positive side,
modelling one node per country accurately reflects the structure of
electricity markets in those parts of Europe without zonal pricing.

Constraints in the natural gas transmission network and storage
infrastructure are not represented, given that gas consumption in the
model is significantly lower than today's gas consumption. (Natural
gas consumption in the European Union (EU) was 2890~TWh in 2015
\cite{EurostatEB} and storage capacity in the EU was 1075~TWh in mid
2017 \cite{gie}; total yearly consumption
of gas in our model was at most 1500~TWh including synthetic methane.)

Many conservative assumptions regarding technology choices have been
made: no biomass has been considered for energy use, given concerns
about the sustainability of fuel crops \cite{GCBB:GCBB12205} and given
that sustainable second-generation biofuels \cite{SIMS20101570} will
be needed for the hard-to-defossilise sectors not considered in the
model, i.e. process heat in industry and energy-dense fuels for
aviation and shipping; the potential to use low-cost second-hand
batteries from BEVs as stationary batteries for the grid has been
ignored; the use of waste heat from methanation or other industrial
processes for electrolysis, direct air capture (DAC) or in district
heating networks has been neglected; \co{} for methanation has been
conservatively assumed to come from DAC, whereas it would be cheaper
to derive it from biogenic sources, power stations or industry;
heating directly with hydrogen \cite{DODDS20152065} rather than
natural gas was not considered, since it would require a more detailed
consideration of changes to delivery infrastructure; carbon capture
and storage (CCS) has not been included, although the high prices of
\co{} in the model might make it an attractive option; the
exploitation of thermal stratification in water tanks, which could
improve the efficiency of TES, has been ignored; the efficiency of the
natural gas CHP could be improved by using a combined cycle gas
turbine for the CHP \cite{LUND2015389}; although building retrofitting
was analysed qualitatively, no further energy efficiency measures were
considered, even though these might be desirable beyond the economic
optimum in order to reduce the need for energy infrastructure; other
synthetic electrofuels for transport, such as methane, dimethyl ether,
(m)ethanol or hydrocarbons produced from the Fischer-Tropf process have
not been considered, because efficiency losses in their production
will be higher than hydrogen and the analysis will be therefore
analogous; demand-side management has only been considered for BEV and
heat demand, whereas other electric load shifting measures could
further decrease the system cost.

A more detailed subdivision of transport demand by vehicle type and
usage would allow a more accurate assessment of transport
technologies, but given the uncertainty around some of the
technologies (such as electric roads, overhead pantographs or the distribution of
charging points), it may not necessarily be useful. Delineation of
each country's building stock would enable a finer analysis of the
potential for building retrofitting for energy saving, the thermal
inertia of buildings, district heating and the potential for
low-temperature heating provision \cite{LUND20141}.

Upstream emissions from manufacturing renewable generators have not
been considered, although these are small (the energy required
for manufacture as a fraction of lifetime energy generation is just
2.3\% for wind and 3.8\% for PV \cite{pehl2017}); on the other hand, it
was recently estimated that 12.6\% of all end-use energy worldwide is
used to mine, transport and refine fossil fuels and uranium
\cite{Jacobson2017}. These concerns can be addressed once all
non-electric industrial demand is included in the model.

On the modelling side, only a single historical weather year (2011)
has been modelled with perfect foresight, which may mean the model is
over-tuned to this year and ignores the future effects of global
warming. In an upcoming paper \cite{Schlachtberger2018} by some of the
authors, the sensitivity of the electricity-only model from
\cite{Schlachtberger2017} to different years or multiple years was
examined and found to be negligible; a similar low sensitivity to the
year was found in the sector-coupled model of most of Europe in
\cite{IWESextreme} based on 7 historical weather
years. \cite{Schlachtberger2018} also analyses the sensitivity of the
model to changing cost assumptions for generation and storage, wherein
some sensitivity is seen to the ratio of solar PV to wind costs.

Finally the availability of the model online \cite{zenodo,zenodo-full} facilitates further
analysis and experimentation by other researchers.

\section{Conclusions}\label{sec:conclusions}

In this paper the model PyPSA-Eur-Sec-30 has been
presented. PyPSA-Eur-Sec-30 is the first open, spatially-resolved,
temporally-resolved and sector-coupled energy model covering the whole
of Europe.  The coupling of the heating and transport sectors to
electricity in a European context enables both the consideration of a
higher share (75\%) of the total final energy usage in the model and
an assessment of the benefits of cross-border transmission versus
enhanced flexibility from sector-coupling in highly renewable
scenarios.

In scenarios where \co{} emissions are reduced by 95\% compared to
1990 levels, the cost-optimal use of battery electric vehicles,
synthetic electrofuels, heat pumps, district heating and long-term
thermal energy storage removes the economic case for almost all
stationary electricity storage and can reduce total system costs by up
to 28\%. These flexibility options work on different time scales
(diurnal, synoptic and seasonal) to help balance the variability of
demand and that of solar and wind generation, which provide the bulk
of primary energy in these scenarios and comprise the majority of the
system costs.

The cost benefit of these flexibility options (28\%) is greater than
the benefit of cross-border transmission on its own
(25\%). Transmission helps to smooth renewables, particularly wind, in
space across the continent, rather than in time.  However, if used
together, sector-coupling flexibility and transmission can reduce
total system costs by 37\% compared to a scenario with no
inter-connection and inflexible sector-coupling. This leads to
scenarios with 95\% lower emissions that are only marginally more
expensive than today's energy system. If the damage from greenhouse
gas emissions and air pollution is taken into account, the highly
renewable systems presented here are of considerable benefit to
society compared to today's system.


Based on the results of this study, the following policy conclusions
can be drawn: in cost-optimal energy systems with low emissions, wind
and solar dominate primary energy generation, while heat pumps
dominate heat provision; increasing cross-border transmission capacity
by a few multiples of today's capacity, particularly around the North
and Baltic Seas, is robustly cost-efficient across a wide range of
scenarios; electrification of transport is more cost-effective than
using synthetic fuels in transport because of efficiency losses when
producing the fuels; the algorithms for managing battery electric
vehicle charging should be exposed to dynamic electricity market
prices; district heating in high-density, urban areas with long-term
thermal energy storage can significantly reduce costs (as long as it
is carefully regulated in view of the potential for monopoly
exploitation); for heating individual buildings in rural areas,
heating systems with multiple technologies (heat pumps, resistive
heating, solar thermal collectors and backup gas boilers for cold
periods) can be efficient; converting power to hydrogen and methane is
advantageous in highly renewable systems, and the technologies for
methanation and carbon dioxide capture should be developed further in
view of this; finally, there are a variety of different possible paths
to a highly renewable energy system, and no significant technical or
economic barriers could be identified.


\section*{Acknowledgments}

The authors thank Gorm Andresen, Tobias Bischof-Niemz, Christian
Breyer, Tom Brown Senior, Magnus Dahl, Thomas Grube, Veit Hagenmeyer,
Heidi Heinrichs, Jonas Hörsch, Robbie Morrison, Andreas Palzer, Marta
Victoria Pérez, Martin Robinius, Mirko Schäfer and Kun Zhu for helpful
discussions and suggestions.  This research was conducted as part of
the CoNDyNet project, which was supported by the German Federal
Ministry of Education and Research under grant number
03SF0472C. T.B. and M.G. were partially funded by the RE-INVEST
project, which is supported by the Innovation Fund Denmark under grant
number 6154-00022B. T.B. also acknowledges funding from the Helmholtz
Association under grant no.~VH-NG-1352. The responsibility for the
contents lies solely with the authors.

\bibliographystyle{elsarticle-num}

\biboptions{sort&compress}
\bibliography{sector}

\begin{thebibliography}{100}
\expandafter\ifx\csname url\endcsname\relax
  \def\url#1{\texttt{#1}}\fi
\expandafter\ifx\csname urlprefix\endcsname\relax\def\urlprefix{URL }\fi
\expandafter\ifx\csname href\endcsname\relax
  \def\href#1#2{#2} \def\path#1{#1}\fi

\bibitem{Czisch}
G.~Czisch, Szenarien zur zuk\"unftigen {S}tromversorgung, Ph.D. thesis,
  Universit\"at Kassel (2005).

\bibitem{Schaber}
{Schaber, K.}, {Steinke, F.}, {Hamacher, T.},
  \href{https://doi.org/10.1016/j.enpol.2011.12.040}{Transmission grid
  extensions for the integration of variable renewable energies in {E}urope:
  Who benefits where?}, Energy Policy 43 (2012) 123 -- 135.
\newblock \href {http://dx.doi.org/10.1016/j.enpol.2011.12.040}
  {\path{doi:10.1016/j.enpol.2011.12.040}}.
\newline\urlprefix\url{https://doi.org/10.1016/j.enpol.2011.12.040}

\bibitem{Schaber2}
{Schaber, K.}, {Steinke, F.}, {M{\"u}hlich, P.}, {Hamacher, T.},
  \href{https://doi.org/10.1016/j.enpol.2011.12.016}{Parametric study of
  variable renewable energy integration in {E}urope: Advantages and costs of
  transmission grid extensions}, Energy Policy 42 (2012) 498--508.
\newblock \href {http://dx.doi.org/10.1016/j.enpol.2011.12.016}
  {\path{doi:10.1016/j.enpol.2011.12.016}}.
\newline\urlprefix\url{https://doi.org/10.1016/j.enpol.2011.12.016}

\bibitem{Scholz}
Y.~Scholz, \href{https://doi.org/10.18419/opus-2015}{{Renewable energy based
  electricity supply at low costs - Development of the {REM}ix model and
  application for Europe}}, Ph.D. thesis, Universit\"at Stuttgart (2012).
\newline\urlprefix\url{https://doi.org/10.18419/opus-2015}

\bibitem{Rodriguez2013}
{Rodriguez, R.A.}, {Becker, S.}, {Andresen, G.}, {Heide, D.}, {Greiner, M.},
  \href{https://doi.org/10.1016/j.renene.2013.10.005}{Transmission needs across
  a fully renewable {E}uropean power system}, Renewable Energy 63 (2014)
  467--476.
\newblock \href {http://dx.doi.org/10.1016/j.renene.2013.10.005}
  {\path{doi:10.1016/j.renene.2013.10.005}}.
\newline\urlprefix\url{https://doi.org/10.1016/j.renene.2013.10.005}

\bibitem{Bussar201440}
C.~Bussar, M.~Moos, R.~Alvarez, P.~Wolf, T.~Thien, H.~Chen, Z.~Cai,
  M.~Leuthold, D.~U. Sauer, A.~Moser, {Optimal Allocation and Capacity of
  Energy Storage Systems in a Future European Power System with 100\% Renewable
  Energy Generation}, Energy Procedia 46 (2014) 40 -- 47, 8th International
  Renewable Energy Storage Conference and Exhibition (IRES 2013).
\newblock \href {http://dx.doi.org/10.1016/j.egypro.2014.01.156}
  {\path{doi:10.1016/j.egypro.2014.01.156}}.

\bibitem{OptHet}
E.~H. Eriksen, L.~J. Schwenk-Nebbe, B.~Tranberg, T.~Brown, M.~Greiner,
  \href{https://doi.org/10.1016/j.energy.2017.05.170}{{Optimal heterogeneity in
  a simplified highly renewable European electricity system}}, Energy
  133~(Supplement C) (2017) 913 -- 928.
\newblock \href {http://dx.doi.org/10.1016/j.energy.2017.05.170}
  {\path{doi:10.1016/j.energy.2017.05.170}}.
\newline\urlprefix\url{https://doi.org/10.1016/j.energy.2017.05.170}

\bibitem{Breyer2017}
C.~Breyer, D.~Bogdanov, A.~Aghahosseini, A.~Gulagi, M.~Child, A.~S. Oyewo,
  J.~Farfan, K.~Sadovskaia, P.~Vainikka,
  \href{http://dx.doi.org/10.1002/pip.2950}{Solar photovoltaics demand for the
  global energy transition in the power sector}, Progress in Photovoltaics:
  Research and Applications (2017) n/a--n/aPIP-17-137.R1.
\newblock \href {http://dx.doi.org/10.1002/pip.2950}
  {\path{doi:10.1002/pip.2950}}.
\newline\urlprefix\url{http://dx.doi.org/10.1002/pip.2950}

\bibitem{burdenresponse}
T.~Brown, T.~Bischof-Niemz, K.~Blok, C.~Breyer, H.~Lund, B.~Mathiesen,
  \href{https://doi.org/10.1016/j.rser.2018.04.113}{Response to `burden of
  proof: A comprehensive review of the feasibility of 100\%
  renewable-electricity systems'}, Renewable and Sustainable Energy Reviews 92
  (2018) 834 -- 847.
\newblock \href {http://dx.doi.org/10.1016/j.rser.2018.04.113}
  {\path{doi:10.1016/j.rser.2018.04.113}}.
\newline\urlprefix\url{https://doi.org/10.1016/j.rser.2018.04.113}

\bibitem{HALLER2012282}
M.~Haller, S.~Ludig, N.~Bauer,
  \href{https://doi.org/10.1016/j.enpol.2012.04.069}{{Decarbonization scenarios
  for the EU and MENA power system: Considering spatial distribution and short
  term dynamics of renewable generation}}, Energy Policy 47 (2012) 282 -- 290.
\newblock \href {http://dx.doi.org/10.1016/j.enpol.2012.04.069}
  {\path{doi:10.1016/j.enpol.2012.04.069}}.
\newline\urlprefix\url{https://doi.org/10.1016/j.enpol.2012.04.069}

\bibitem{Hagspiel}
S.~Hagspiel, C.~J\"agemann, D.~Lindenburger, T.~Brown, S.~Cherevatskiy,
  E.~Tr\"oster,
  \href{https://doi.org/10.1016/j.energy.2014.01.025}{Cost-optimal power system
  extension under flow-based market coupling}, Energy 66 (2014) 654--666.
\newblock \href {http://dx.doi.org/10.1016/j.energy.2014.01.025}
  {\path{doi:10.1016/j.energy.2014.01.025}}.
\newline\urlprefix\url{https://doi.org/10.1016/j.energy.2014.01.025}

\bibitem{GILS2017173}
H.~C. Gils, Y.~Scholz, T.~Pregger, D.~L. de~Tena, D.~Heide,
  \href{10.1016/j.energy.2017.01.115}{{Integrated modelling of variable
  renewable energy-based power supply in Europe}}, Energy 123 (2017) 173 --
  188.
\newblock \href
  {http://dx.doi.org/https://doi.org/10.1016/j.energy.2017.01.115}
  {\path{doi:https://doi.org/10.1016/j.energy.2017.01.115}}.
\newline\urlprefix\url{10.1016/j.energy.2017.01.115}

\bibitem{Schlachtberger2017}
D.~Schlachtberger, T.~Brown, S.~Schramm, M.~Greiner,
  \href{https://doi.org/10.1016/j.energy.2017.06.004}{The benefits of
  cooperation in a highly renewable {E}uropean electricity network}, Energy 134
  (2017) 469 -- 481.
\newblock \href {http://dx.doi.org/10.1016/j.energy.2017.06.004}
  {\path{doi:10.1016/j.energy.2017.06.004}}.
\newline\urlprefix\url{https://doi.org/10.1016/j.energy.2017.06.004}

\bibitem{CEBULLA2017211}
F.~Cebulla, T.~Naegler, M.~Pohl,
  \href{https://doi.org/10.1016/j.est.2017.10.004}{{Electrical energy storage
  in highly renewable European energy systems: Capacity requirements, spatial
  distribution, and storage dispatch}}, Journal of Energy Storage 14~(Part 1)
  (2017) 211 -- 223.
\newblock \href {http://dx.doi.org/10.1016/j.est.2017.10.004}
  {\path{doi:10.1016/j.est.2017.10.004}}.
\newline\urlprefix\url{https://doi.org/10.1016/j.est.2017.10.004}

\bibitem{LUND2017556}
H.~Lund, P.~A. Østergaard, D.~Connolly, B.~V. Mathiesen,
  \href{https://doi.org/10.1016/j.energy.2017.05.123}{Smart energy and smart
  energy systems}, Energy 137~(Supplement C) (2017) 556 -- 565.
\newblock \href {http://dx.doi.org/10.1016/j.energy.2017.05.123}
  {\path{doi:10.1016/j.energy.2017.05.123}}.
\newline\urlprefix\url{https://doi.org/10.1016/j.energy.2017.05.123}

\bibitem{Lund2010}
H.~Lund (Ed.), The Choice and Modelling of 100\% Renewable Solutions, Academic
  Press, 2010.

\bibitem{Meibom2007}
P.~Meibom, J.~Kiviluoma, R.~Barth, H.~Brand, C.~Weber, H.~V. Larsen,
  \href{http://dx.doi.org/10.1002/we.224}{Value of electric heat boilers and
  heat pumps for wind power integration}, Wind Energy 10~(4) (2007) 321--337.
\newblock \href {http://dx.doi.org/10.1002/we.224} {\path{doi:10.1002/we.224}}.
\newline\urlprefix\url{http://dx.doi.org/10.1002/we.224}

\bibitem{PENSINI2014530}
A.~Pensini, C.~N. Rasmussen, W.~Kempton,
  \href{https://doi.org/10.1016/j.apenergy.2014.04.111}{Economic analysis of
  using excess renewable electricity to displace heating fuels}, Applied Energy
  131~(Supplement C) (2014) 530 -- 543.
\newblock \href {http://dx.doi.org/10.1016/j.apenergy.2014.04.111}
  {\path{doi:10.1016/j.apenergy.2014.04.111}}.
\newline\urlprefix\url{https://doi.org/10.1016/j.apenergy.2014.04.111}

\bibitem{ASHFAQ2017363}
A.~Ashfaq, Z.~H. Kamali, M.~H. Agha, H.~Arshid,
  \href{https://doi.org/10.1016/j.energy.2017.01.084}{{Heat coupling of the
  pan-European vs. regional electrical grid with excess renewable energy}},
  Energy 122~(Supplement C) (2017) 363 -- 377.
\newblock \href {http://dx.doi.org/10.1016/j.energy.2017.01.084}
  {\path{doi:10.1016/j.energy.2017.01.084}}.
\newline\urlprefix\url{https://doi.org/10.1016/j.energy.2017.01.084}

\bibitem{KEMPTON1997157}
W.~Kempton, S.~E. Letendre,
  \href{https://doi.org/10.1016/S1361-9209(97)00001-1}{Electric vehicles as a
  new power source for electric utilities}, Transportation Research Part D:
  Transport and Environment 2~(3) (1997) 157 -- 175.
\newblock \href {http://dx.doi.org/10.1016/S1361-9209(97)00001-1}
  {\path{doi:10.1016/S1361-9209(97)00001-1}}.
\newline\urlprefix\url{https://doi.org/10.1016/S1361-9209(97)00001-1}

\bibitem{KIVILUOMA20111758}
J.~Kiviluoma, P.~Meibom,
  \href{https://doi.org/10.1016/j.energy.2010.12.053}{Methodology for modelling
  plug-in electric vehicles in the power system and cost estimates for a system
  with either smart or dumb electric vehicles}, Energy 36~(3) (2011) 1758 --
  1767.
\newblock \href {http://dx.doi.org/10.1016/j.energy.2010.12.053}
  {\path{doi:10.1016/j.energy.2010.12.053}}.
\newline\urlprefix\url{https://doi.org/10.1016/j.energy.2010.12.053}

\bibitem{SCHILL2015185}
W.-P. Schill, C.~Gerbaulet,
  \href{https://doi.org/10.1016/j.apenergy.2015.07.012}{{Power system impacts
  of electric vehicles in Germany: Charging with coal or renewables?}}, Applied
  Energy 156 (2015) 185 -- 196.
\newblock \href {http://dx.doi.org/10.1016/j.apenergy.2015.07.012}
  {\path{doi:10.1016/j.apenergy.2015.07.012}}.
\newline\urlprefix\url{https://doi.org/10.1016/j.apenergy.2015.07.012}

\bibitem{en10070956}
M.~Robinius, A.~Otto, P.~Heuser, L.~Welder, K.~Syranidis, D.~S. Ryberg,
  T.~Grube, P.~Markewitz, R.~Peters, D.~Stolten,
  \href{https://doi.org/10.3390/en10070956}{{Linking the Power and Transport
  Sectors—Part 1: The Principle of Sector Coupling}}, Energies 10~(7).
\newblock \href {http://dx.doi.org/10.3390/en10070956}
  {\path{doi:10.3390/en10070956}}.
\newline\urlprefix\url{https://doi.org/10.3390/en10070956}

\bibitem{en10070957}
M.~Robinius, A.~Otto, K.~Syranidis, D.~S. Ryberg, P.~Heuser, L.~Welder,
  T.~Grube, P.~Markewitz, V.~Tietze, D.~Stolten,
  \href{http://doi.org/10.3390/en10070957}{{Linking the Power and Transport
  Sectors—Part 2: Modelling a Sector Coupling Scenario for Germany}},
  Energies 10~(7).
\newblock \href {http://dx.doi.org/10.3390/en10070957}
  {\path{doi:10.3390/en10070957}}.
\newline\urlprefix\url{http://doi.org/10.3390/en10070957}

\bibitem{SCHIEBAHN20154285}
S.~Schiebahn, T.~Grube, M.~Robinius, V.~Tietze, B.~Kumar, D.~Stolten,
  \href{https://doi.org/10.1016/j.ijhydene.2015.01.123}{{Power to gas:
  Technological overview, systems analysis and economic assessment for a case
  study in Germany}}, International Journal of Hydrogen Energy 40~(12) (2015)
  4285 -- 4294.
\newblock \href {http://dx.doi.org/10.1016/j.ijhydene.2015.01.123}
  {\path{doi:10.1016/j.ijhydene.2015.01.123}}.
\newline\urlprefix\url{https://doi.org/10.1016/j.ijhydene.2015.01.123}

\bibitem{ROBINIUS2018182}
M.~Robinius, T.~Raje, S.~Nykamp, T.~Rott, M.~Müller, T.~Grube, B.~Katzenbach,
  S.~Küppers, D.~Stolten,
  \href{https://doi.org/10.1016/j.apenergy.2017.10.117}{{Power-to-Gas:
  Electrolyzers as an alternative to network expansion – An example from a
  distribution system operator}}, Applied Energy 210 (2018) 182 -- 197.
\newblock \href {http://dx.doi.org/10.1016/j.apenergy.2017.10.117}
  {\path{doi:10.1016/j.apenergy.2017.10.117}}.
\newline\urlprefix\url{https://doi.org/10.1016/j.apenergy.2017.10.117}

\bibitem{Henning20141003}
H.-M. Henning, A.~Palzer, \href{https://doi.org/10.1016/j.rser.2013.09.012}{{A
  comprehensive model for the German electricity and heat sector in a future
  energy system with a dominant contribution from renewable energy
  technologies—Part I: Methodology}}, Renewable and Sustainable Energy
  Reviews 30 (2014) 1003 -- 1018.
\newline\urlprefix\url{https://doi.org/10.1016/j.rser.2013.09.012}

\bibitem{PALZER20141019}
A.~Palzer, H.-M. Henning, \href{https://doi.org/10.1016/j.rser.2013.11.032}{{A
  comprehensive model for the German electricity and heat sector in a future
  energy system with a dominant contribution from renewable energy technologies
  – Part II: Results}}, Renewable and Sustainable Energy Reviews
  30~(Supplement C) (2014) 1019 -- 1034.
\newblock \href {http://dx.doi.org/10.1016/j.rser.2013.11.032}
  {\path{doi:10.1016/j.rser.2013.11.032}}.
\newline\urlprefix\url{https://doi.org/10.1016/j.rser.2013.11.032}

\bibitem{IEESWV}
N.~Gerhardt, A.~Scholz, F.~Sandau, H.~Hahn,
  \href{{http://www.energiesystemtechnik.iwes.fraunhofer.de/de/projekte/suche/2015/interaktion_strom_waerme_verkehr.html}}{{Interaktion
  EE-Strom, Wärme und Verkehr}}, Tech. rep., Fraunhofer IWES (2015).
\newline\urlprefix\url{{http://www.energiesystemtechnik.iwes.fraunhofer.de/de/projekte/suche/2015/interaktion_strom_waerme_verkehr.html}}

\bibitem{Quaschning}
V.~Quaschning, {Sektorkopplung durch die Energiewende}, Tech. rep., HTW Berlin
  (2016).

\bibitem{LUND2009524}
H.~Lund, B.~Mathiesen,
  \href{https://doi.org/10.1016/j.energy.2008.04.003}{{Energy system analysis
  of 100\% renewable energy systems - The case of Denmark in years 2030 and
  2050}}, Energy 34~(5) (2009) 524 -- 531, 4th Dubrovnik Conference.
\newblock \href {http://dx.doi.org/10.1016/j.energy.2008.04.003}
  {\path{doi:10.1016/j.energy.2008.04.003}}.
\newline\urlprefix\url{https://doi.org/10.1016/j.energy.2008.04.003}

\bibitem{mathiesen2014smart}
B.~V. Mathiesen, H.~Lund, D.~Conolly, H.~Wenzel, P.~{\O}stergaard,
  B.~M{\"o}ller, S.~Nielsen, I.~Ridjan, P.~Karn{\o}e, K.~Sperling,
  F.~Hvelplund, \href{https://doi.org/10.1016/j.apenergy.2015.01.075}{Smart
  energy systems for coherent 100\% renewable energy and transport solutions},
  Applied Energy 145 (2015) 139--154.
\newblock \href {http://dx.doi.org/10.1016/j.apenergy.2015.01.075}
  {\path{doi:10.1016/j.apenergy.2015.01.075}}.
\newline\urlprefix\url{https://doi.org/10.1016/j.apenergy.2015.01.075}

\bibitem{Lund201296}
H.~Lund, A.~N. Andersen, P.~A. Østergaard, B.~V. Mathiesen, D.~Connolly,
  \href{https://doi.org/10.1016/j.energy.2012.04.003}{From electricity smart
  grids to smart energy systems – a market operation based approach and
  understanding}, Energy 42~(1) (2012) 96 -- 102, 8th World Energy System
  Conference, {WESC} 2010.
\newblock \href {http://dx.doi.org/10.1016/j.energy.2012.04.003}
  {\path{doi:10.1016/j.energy.2012.04.003}}.
\newline\urlprefix\url{https://doi.org/10.1016/j.energy.2012.04.003}

\bibitem{CONNOLLY2011502}
D.~Connolly, H.~Lund, B.~Mathiesen, M.~Leahy,
  \href{https://doi.org/10.1016/j.apenergy.2010.03.006}{{The first step towards
  a 100\% renewable energy-system for Ireland}}, Applied Energy 88~(2) (2011)
  502 -- 507, the 5th Dubrovnik Conference on Sustainable Development of
  Energy, Water and Environment Systems, held in Dubrovnik September/October
  2009.
\newblock \href {http://dx.doi.org/10.1016/j.apenergy.2010.03.006}
  {\path{doi:10.1016/j.apenergy.2010.03.006}}.
\newline\urlprefix\url{https://doi.org/10.1016/j.apenergy.2010.03.006}

\bibitem{Deane2012303}
J.~Deane, A.~Chiodi, M.~Gargiulo, B.~P.~O. Gallachoir,
  \href{https://doi.org/10.1016/j.energy.2012.03.052}{Soft-linking of a power
  systems model to an energy systems model}, Energy 42~(1) (2012) 303 -- 312,
  8th World Energy System Conference, \{WESC\} 2010.
\newblock \href {http://dx.doi.org/10.1016/j.energy.2012.03.052}
  {\path{doi:10.1016/j.energy.2012.03.052}}.
\newline\urlprefix\url{https://doi.org/10.1016/j.energy.2012.03.052}

\bibitem{Connolly20161634}
D.~Connolly, H.~Lund, B.~Mathiesen,
  \href{https://doi.org/10.1016/j.rser.2016.02.025}{{Smart Energy Europe: The
  technical and economic impact of one potential 100\% renewable energy
  scenario for the European Union}}, Renewable and Sustainable Energy Reviews
  60 (2016) 1634 -- 1653.
\newblock \href {http://dx.doi.org/10.1016/j.rser.2016.02.025}
  {\path{doi:10.1016/j.rser.2016.02.025}}.
\newline\urlprefix\url{https://doi.org/10.1016/j.rser.2016.02.025}

\bibitem{THELLUFSEN2017492}
J.~Z. Thellufsen, H.~Lund,
  \href{https://doi.org/10.1016/j.energy.2017.02.112}{Cross-border versus
  cross-sector interconnectivity in renewable energy systems}, Energy
  124~(Supplement C) (2017) 492 -- 501.
\newblock \href {http://dx.doi.org/10.1016/j.energy.2017.02.112}
  {\path{doi:10.1016/j.energy.2017.02.112}}.
\newline\urlprefix\url{https://doi.org/10.1016/j.energy.2017.02.112}

\bibitem{Primes}
{The PRIMES Model}, Tech. rep., NTUA (2009).

\bibitem{REMIND}
M.~Leimbach, N.~Bauer, L.~Baumstark, M.~Luken, O.~Edenhofer,
  \href{https://doi.org/10.5547/ISSN0195-6574-EJ-Vol31-NoSI-5}{{Technological
  Change and International Trade - Insights from REMIND-R}}, {The Energy
  Journal} 31.
\newblock \href {http://dx.doi.org/10.5547/ISSN0195-6574-EJ-Vol31-NoSI-5}
  {\path{doi:10.5547/ISSN0195-6574-EJ-Vol31-NoSI-5}}.
\newline\urlprefix\url{https://doi.org/10.5547/ISSN0195-6574-EJ-Vol31-NoSI-5}

\bibitem{CAPROS2014231}
P.~Capros, L.~Paroussos, P.~Fragkos, S.~Tsani, B.~Boitier, F.~Wagner, S.~Busch,
  G.~Resch, M.~Blesl, J.~Bollen,
  \href{https://doi.org/10.1016/j.esr.2013.12.007}{European decarbonisation
  pathways under alternative technological and policy choices: A multi-model
  analysis}, Energy Strategy Reviews 2~(3) (2014) 231 -- 245, sustainable
  Energy System Changes.
\newblock \href {http://dx.doi.org/10.1016/j.esr.2013.12.007}
  {\path{doi:10.1016/j.esr.2013.12.007}}.
\newline\urlprefix\url{https://doi.org/10.1016/j.esr.2013.12.007}

\bibitem{SIMOES2017353}
S.~Simoes, W.~Nijs, P.~Ruiz, A.~Sgobbi, C.~Thiel,
  \href{https://doi.org/10.1016/j.enpol.2016.10.006}{{Comparing policy routes
  for low-carbon power technology deployment in EU – an energy system
  analysis}}, Energy Policy 101 (2017) 353 -- 365.
\newblock \href {http://dx.doi.org/10.1016/j.enpol.2016.10.006}
  {\path{doi:10.1016/j.enpol.2016.10.006}}.
\newline\urlprefix\url{https://doi.org/10.1016/j.enpol.2016.10.006}

\bibitem{en10101468}
K.~Löffler, K.~Hainsch, T.~Burandt, P.-Y. Oei, C.~Kemfert, C.~von
  Hirschhausen, \href{https://doi.org/10.3390/en10101468}{{Designing a Model
  for the Global Energy System—GENeSYS-MOD: An Application of the Open-Source
  Energy Modeling System (OSeMOSYS)}}, Energies 10~(10).
\newblock \href {http://dx.doi.org/10.3390/en10101468}
  {\path{doi:10.3390/en10101468}}.
\newline\urlprefix\url{https://doi.org/10.3390/en10101468}

\bibitem{LUDIG20116674}
S.~Ludig, M.~Haller, E.~Schmid, N.~Bauer,
  \href{https://doi.org/10.1016/j.energy.2011.08.021}{Fluctuating renewables in
  a long-term climate change mitigation strategy}, Energy 36~(11) (2011) 6674
  -- 6685.
\newblock \href {http://dx.doi.org/10.1016/j.energy.2011.08.021}
  {\path{doi:10.1016/j.energy.2011.08.021}}.
\newline\urlprefix\url{https://doi.org/10.1016/j.energy.2011.08.021}

\bibitem{KOTZUR2018474}
L.~Kotzur, P.~Markewitz, M.~Robinius, D.~Stolten,
  \href{https://doi.org/10.1016/j.renene.2017.10.017}{Impact of different time
  series aggregation methods on optimal energy system design}, Renewable Energy
  117 (2018) 474 -- 487.
\newblock \href {http://dx.doi.org/10.1016/j.renene.2017.10.017}
  {\path{doi:10.1016/j.renene.2017.10.017}}.
\newline\urlprefix\url{https://doi.org/10.1016/j.renene.2017.10.017}

\bibitem{Council2009}
{European Council}, {Presidency Conclusions of Meeting of European Council on
  29/30 October 2009},
  \url{http://www.consilium.europa.eu/uedocs/cms_data/docs/pressdata/en/ec/110889.pdf},
  {Online, retrieved August 2016} (2009).

\bibitem{kyoto}
{Kyoto Protocol: Reference Manual}, Tech. rep., United Nations Framework
  Convention on Climate Change (2008).

\bibitem{zenodo}
T.~Brown, D.~Schlachtberger,
  \href{https://doi.org/10.5281/zenodo.1146665}{{Supplementary Data: Code,
  Input Data and Result Summaries: Synergies of sector coupling and
  transmission extension in a cost-optimised, highly renewable European energy
  system}} (2018).
\newblock \href {http://dx.doi.org/10.5281/zenodo.1146665}
  {\path{doi:10.5281/zenodo.1146665}}.
\newline\urlprefix\url{https://doi.org/10.5281/zenodo.1146665}

\bibitem{zenodo-full}
T.~Brown, D.~Schlachtberger,
  \href{https://doi.org/10.5281/zenodo.1146649}{{Supplementary Data: Full
  Results: Synergies of sector coupling and transmission extension in a
  cost-optimised, highly renewable European energy system}} (2018).
\newblock \href {http://dx.doi.org/10.5281/zenodo.1146649}
  {\path{doi:10.5281/zenodo.1146649}}.
\newline\urlprefix\url{https://doi.org/10.5281/zenodo.1146649}

\bibitem{PFENNINGER2017211}
S.~Pfenninger, J.~DeCarolis, L.~Hirth, S.~Quoilin, I.~Staffell,
  \href{https://doi.org/10.1016/j.enpol.2016.11.046}{{The importance of open
  data and software: Is energy research lagging behind?}}, Energy Policy 101
  (2017) 211 -- 215.
\newblock \href {http://dx.doi.org/10.1016/j.enpol.2016.11.046}
  {\path{doi:10.1016/j.enpol.2016.11.046}}.
\newline\urlprefix\url{https://doi.org/10.1016/j.enpol.2016.11.046}

\bibitem{PFENNINGER201863}
S.~Pfenninger, L.~Hirth, I.~Schlecht, E.~Schmid, F.~Wiese, T.~Brown, C.~Davis,
  M.~Gidden, H.~Heinrichs, C.~Heuberger, S.~Hilpert, U.~Krien, C.~Matke,
  A.~Nebel, R.~Morrison, B.~Müller, G.~Pleßmann, M.~Reeg, J.~C. Richstein,
  A.~Shivakumar, I.~Staffell, T.~Tröndle, C.~Wingenbach,
  \href{https://doi.org/10.1016/j.esr.2017.12.002}{{Opening the black box of
  energy modelling: Strategies and lessons learned}}, Energy Strategy Reviews
  19 (2018) 63 -- 71.
\newblock \href {http://dx.doi.org/10.1016/j.esr.2017.12.002}
  {\path{doi:10.1016/j.esr.2017.12.002}}.
\newline\urlprefix\url{https://doi.org/10.1016/j.esr.2017.12.002}

\bibitem{PyPSA}
T.~Brown, J.~H\"orsch, D.~Schlachtberger,
  \href{https://doi.org/10.5334/jors.188}{{PyPSA: Python for Power System
  Analysis}}, Journal of Open Research Software 6~(4).
\newblock \href {http://arxiv.org/abs/1707.09913} {\path{arXiv:1707.09913}},
  \href {http://dx.doi.org/10.5334/jors.188} {\path{doi:10.5334/jors.188}}.
\newline\urlprefix\url{https://doi.org/10.5334/jors.188}

\bibitem{IWESextreme}
N.~Gerhardt, D.~Böttger, T.~Trost, A.~Scholz, C.~Pape, A.-K. Gerlach,
  P.~Härtel, I.~Ganal,
  \href{{http://www.energieversorgung-elektromobilitaet.de/includes/reports/Auswertung_7Wetterjahre_95Prozent_FraunhoferIWES.pdf}}{{Analyse
  eines europäischen 95\%-Klimazielszenarios über mehrere Wetterjahre}},
  Tech. rep., Fraunhofer IWES (2017).
\newline\urlprefix\url{{http://www.energieversorgung-elektromobilitaet.de/includes/reports/Auswertung_7Wetterjahre_95Prozent_FraunhoferIWES.pdf}}

\bibitem{gurobi}
{Gurobi Optimization Inc.}, \href{http://www.gurobi.com}{Gurobi optimizer
  reference manual} (2017).
\newline\urlprefix\url{http://www.gurobi.com}

\bibitem{schroeder2013}
A.~Schr\"{o}der, F.~Kunz, J.~Meiss, R.~Mendelevitch, C.~von Hirschhausen,
  \href{http://hdl.handle.net/10419/80348}{Current and prospective costs of
  electricity generation until 2050}, Data Documentation, DIW~68, Deutsches
  Institut f\"{u}r Wirtschaftsforschung (DIW), Berlin (2013).
\newline\urlprefix\url{http://hdl.handle.net/10419/80348}

\bibitem{etip}
E.~Vartiainen, G.~Masson, C.~Breyer,
  \href{http://www.etip-pv.eu/fileadmin/Documents/ETIP_PV_Publications_2017-2018/LCOE_Report_March_2017.pdf}{{The
  True Competitiveness of Solar PV: A European Case Study}}, Tech. rep.,
  European Technology and Innovation Platform for Photovoltaics (2017).
\newline\urlprefix\url{http://www.etip-pv.eu/fileadmin/Documents/ETIP_PV_Publications_2017-2018/LCOE_Report_March_2017.pdf}

\bibitem{dea2016}
\href{https://ens.dk/en/our-services/projections-and-models/technology-data}{Technology
  data for generation of electricity and district heating, energy storage and
  energy carrier generation and conversion}, Tech. rep., Danish Energy Agency
  and Energinet.dk (2016).
\newline\urlprefix\url{https://ens.dk/en/our-services/projections-and-models/technology-data}

\bibitem{budischak2013}
C.~Budischak, D.~Sewell, H.~Thomson, L.~Mach, D.~E. Veron, W.~Kempton,
  \href{https://doi.org/10.1016/j.jpowsour.2012.09.054}{Cost-minimized
  combinations of wind power, solar power and electrochemical storage, powering
  the grid up to 99.9\% of the time}, Journal of Power Sources 225 (2013) 60 --
  74.
\newblock \href {http://dx.doi.org/10.1016/j.jpowsour.2012.09.054}
  {\path{doi:10.1016/j.jpowsour.2012.09.054}}.
\newline\urlprefix\url{https://doi.org/10.1016/j.jpowsour.2012.09.054}

\bibitem{PalzerThesis}
A.~Palzer, {Sektorübergreifende Modellierung und Optimierung eines
  zukünftigen deutschen Energiesystems unter Berücksichtigung von
  Energieeffizienzmaßnahmen im Gebäudesektor}, Ph.D. thesis, KIT (2016).

\bibitem{NRELhydrogen}
D.~M. Steward, Scenario development and analysis of hydrogen as a large-scale
  energy storage medium, Tech. rep. (2009).

\bibitem{Fasihi2017}
M.~Fasihi, D.~Bogdanov, C.~Breyer,
  \href{https://doi.org/10.3390/su9020306}{{Long-Term Hydrocarbon Trade Options
  for the Maghreb Region and Europe—Renewable Energy Based Synthetic Fuels
  for a Net Zero Emissions World}}, Sustainability 9~(2).
\newblock \href {http://dx.doi.org/10.3390/su9020306}
  {\path{doi:10.3390/su9020306}}.
\newline\urlprefix\url{https://doi.org/10.3390/su9020306}

\bibitem{SchaberThesis}
K.~Schaber, {Integration of Variable Renewable Energies in the European power
  system: a model-based analysis of transmission grid extensions and energy
  sector coupling}, Ph.D. thesis, TU München (2013).

\bibitem{bnetza2017}
\href{https://www.bundesnetzagentur.de/DE/Sachgebiete/ElektrizitaetundGas/Unternehmen_Institutionen/DatenaustauschundMonitoring/Monitoring/Monitoringberichte/Monitoring_Berichte.html}{{Monitoringbericht
  2017}}, Tech. rep., Bundesnetzagentur (2017).
\newline\urlprefix\url{https://www.bundesnetzagentur.de/DE/Sachgebiete/ElektrizitaetundGas/Unternehmen_Institutionen/DatenaustauschundMonitoring/Monitoring/Monitoringberichte/Monitoring_Berichte.html}

\bibitem{Zakeri2015}
B.~Zakeri, S.~Syri,
  \href{https://doi.org/10.1016/j.rser.2014.10.011}{Electrical energy storage
  systems: A comparative life cycle cost analysis}, Renewable and Sustainable
  Energy Reviews 42~(Supplement C) (2015) 569 -- 596.
\newblock \href {http://dx.doi.org/10.1016/j.rser.2014.10.011}
  {\path{doi:10.1016/j.rser.2014.10.011}}.
\newline\urlprefix\url{https://doi.org/10.1016/j.rser.2014.10.011}

\bibitem{ESYSTechnologiesteckbrief}
P.~Elsner, D.~U. Sauer,
  \href{http://www.acatech.de/fileadmin/user_upload/Baumstruktur_nach_Website/Acatech/root/de/Publikationen/Materialien/ESYS_Technologiesteckbrief_Energiespeicher.pdf}{{Energiespeicher:
  Technologiesteckbrief zur Analyse ,,Flexibilitätskonzepte für die
  Stromversorgung 2050``}}, Tech. rep. (2015).
\newline\urlprefix\url{http://www.acatech.de/fileadmin/user_upload/Baumstruktur_nach_Website/Acatech/root/de/Publikationen/Materialien/ESYS_Technologiesteckbrief_Energiespeicher.pdf}

\bibitem{pypsa-eur}
J.~Hörsch, F.~Hofmann, D.~Schlachtberger, T.~Brown,
  \href{https://arxiv.org/abs/1806.01613}{{PyPSA-Eur: An Open Optimisation
  Model of the European Transmission System}}.
\newline\urlprefix\url{https://arxiv.org/abs/1806.01613}

\bibitem{OPSD}
{Open Power System Data}, {Data Package Time series. Version 2017-07-09},
  \url{https://data.open-power-system-data.org/time_series/2017-07-09/} (July
  2017).

\bibitem{entsoe_load}
{European Transmission System Operators}, {Country-specific hourly load data},
  \url{https://www.entsoe.eu/data/data-portal/consumption/} (2011).

\bibitem{Andresen20151074}
G.~B. Andresen, A.~A. Søndergaard, M.~Greiner,
  \href{https://doi.org/10.1016/j.energy.2015.09.071}{{Validation of Danish
  wind time series from a new global renewable energy atlas for energy system
  analysis}}, Energy 93, Part 1 (2015) 1074 -- 1088.
\newblock \href {http://dx.doi.org/10.1016/j.energy.2015.09.071}
  {\path{doi:10.1016/j.energy.2015.09.071}}.
\newline\urlprefix\url{https://doi.org/10.1016/j.energy.2015.09.071}

\bibitem{Saha}
S.~Saha, et~al., {The NCEP Climate Forecast System Reanalysis}, Bulletin of the
  American Meteorological Society 91~(8) (2010) 1015--1057.
\newblock \href {http://dx.doi.org/10.1175/2010BAMS3001.1}
  {\path{doi:10.1175/2010BAMS3001.1}}.

\bibitem{Pfenninger2016}
S.~Pfenninger, I.~Staffell,
  \href{https://doi.org/10.1016/j.energy.2016.08.060}{{Long-term patterns of
  European PV output using 30 years of validated hourly reanalysis and
  satellite data}}, Energy 114~(Supplement C) (2016) 1251 -- 1265.
\newblock \href {http://dx.doi.org/10.1016/j.energy.2016.08.060}
  {\path{doi:10.1016/j.energy.2016.08.060}}.
\newline\urlprefix\url{https://doi.org/10.1016/j.energy.2016.08.060}

\bibitem{rs70608067}
R.~M\"uller, U.~Pfeifroth, C.~Tr\"ager-Chatterjee, J.~Trentmann, R.~Cremer,
  \href{https://doi.org/10.3390/rs70608067}{Digging the meteosat treasure—3
  decades of solar surface radiation}, Remote Sensing 7~(6) (2015) 8067--8101.
\newblock \href {http://dx.doi.org/10.3390/rs70608067}
  {\path{doi:10.3390/rs70608067}}.
\newline\urlprefix\url{https://doi.org/10.3390/rs70608067}

\bibitem{natura2000}
EEA, {Natura 2000 data - the European network of protected sites},
  \url{http://www.eea.europa.eu/data-and-maps/data/natura-7} (2016).

\bibitem{corine2006}
EEA, Corine land cover 2006 (2014).

\bibitem{kies2016}
{Kies, A.}, {Chattopadhyay, K.}, {von Bremen, L.}, {Lorenz, E.}, {Heinemann,
  D.}, {RESTORE 2050 Work Package Report D12: Simulation of renewable feed-in
  for power system studies.}, Tech. rep., RESTORE 2050, in preparation (2016).

\bibitem{pfluger2011}
B.~Pfluger, F.~Sensfu{\ss}, G.~Schubert, J.~Leisentritt, {Tangible ways towards
  climate protection in the European Union (EU Long-term scenarios 2050)},
  Fraunhofer ISI.

\bibitem{ENTSOEinstalledcapas}
{Installed Capacity per Production Type in 2015}, Tech. rep., ENTSO-E (2016).

\bibitem{BASt}
{Verkehrszählung - Stundenwerte}, Tech. rep., Bundesanstalt für
  Straßenwesen.

\bibitem{Odyssee}
{ODYSSEE database on energy efficiency data \& indicators}, Tech. rep.,
  Enerdata (2016).

\bibitem{EPA}
{US Environmental Protection Agency}, {2017 Car Fuel Economy Estimates},
  \url{https://www.fueleconomy.gov/feg/} (2017).

\bibitem{tesla}
\href{https://www.tesla.com/models}{Tesla model s range estimator} (2017).
\newline\urlprefix\url{https://www.tesla.com/models}

\bibitem{CONNOLLY2017235}
D.~Connolly, \href{https://doi.org/10.1016/j.esr.2017.09.005}{Economic
  viability of electric roads compared to oil and batteries for all forms of
  road transport}, Energy Strategy Reviews 18~(Supplement C) (2017) 235 -- 249.
\newblock \href {http://dx.doi.org/10.1016/j.esr.2017.09.005}
  {\path{doi:10.1016/j.esr.2017.09.005}}.
\newline\urlprefix\url{https://doi.org/10.1016/j.esr.2017.09.005}

\bibitem{LUND20083578}
H.~Lund, W.~Kempton,
  \href{https://doi.org/10.1016/j.enpol.2008.06.007}{{Integration of renewable
  energy into the transport and electricity sectors through V2G}}, Energy
  Policy 36~(9) (2008) 3578 -- 3587.
\newblock \href {http://dx.doi.org/10.1016/j.enpol.2008.06.007}
  {\path{doi:10.1016/j.enpol.2008.06.007}}.
\newline\urlprefix\url{https://doi.org/10.1016/j.enpol.2008.06.007}

\bibitem{kempton2001}
W.~Kempton, J.~Tomi\'c, S.~Letendre, A.~Brooks, T.~Lipman,
  \href{http://www1.udel.edu/V2G/docs/V2G-Cal-2001.pdf}{{Vehicle-to-Grid Power:
  Battery, Hybrid, and Fuel Cell Vehicles as Resources for Distributed Electric
  Power in California}}, Tech. rep. (2001).
\newline\urlprefix\url{http://www1.udel.edu/V2G/docs/V2G-Cal-2001.pdf}

\bibitem{markel2009}
T.~Markel, K.~Bennion, W.~Kramer, J.~Bryan, J.~Giedd,
  \href{http://citeseerx.ist.psu.edu/viewdoc/download?doi=10.1.1.504.1653&rep=rep1&type=pdf}{{Field
  Testing Plug-in Hybrid Electric Vehicles with Charge Control Technology in
  the Xcel Energy Territory}}, Tech. rep., NREL (2009).
\newline\urlprefix\url{http://citeseerx.ist.psu.edu/viewdoc/download?doi=10.1.1.504.1653&rep=rep1&type=pdf}

\bibitem{galus2013}
M.~D. Galus, M.~G. Vayá, T.~Krause, G.~Andersson,
  \href{https://doi.org/10.1002/wene.56}{The role of electric vehicles in smart
  grids}, Wiley Interdisciplinary Reviews: Energy and Environment 2~(4) (2013)
  384--400.
\newblock \href {http://dx.doi.org/10.1002/wene.56}
  {\path{doi:10.1002/wene.56}}.
\newline\urlprefix\url{https://doi.org/10.1002/wene.56}

\bibitem{rmi2017}
J.~Walker, C.~Johnson, \href{http://www.rmi.org/peak_car_ownership}{Peak car
  ownership: The market opportunity of electric automated mobility services},
  Tech. rep., Rocky Mountain Institute (2016).
\newline\urlprefix\url{http://www.rmi.org/peak_car_ownership}

\bibitem{Adam}
A.~R. Jensen, Coupling of a highly renewable electricity system to the heating
  sector, Master's thesis, Aarhus University (2016).

\bibitem{ASHFAQ2018613}
A.~Ashfaq, A.~Ianakiev,
  \href{https://doi.org/10.1016/j.energy.2018.03.155}{Cost-minimised design of
  a highly renewable heating network for fossil-free future}, Energy 152 (2018)
  613 -- 626.
\newblock \href {http://dx.doi.org/10.1016/j.energy.2018.03.155}
  {\path{doi:10.1016/j.energy.2018.03.155}}.
\newline\urlprefix\url{https://doi.org/10.1016/j.energy.2018.03.155}

\bibitem{EurostatEB}
{Energy Balances 1900 -- 2014}, Tech. rep., Eurostat (2016).

\bibitem{SFOE}
\href{http://www.bfe.admin.ch/themen/00526/00541/00542/02167/index.html?dossier_id=02169}{{Analyse
  des schweizerischen Energieverbrauchs 2000-2016 nach Verwendungszwecken}},
  Tech. rep., Swiss Federal Office of Energy (2017).
\newline\urlprefix\url{http://www.bfe.admin.ch/themen/00526/00541/00542/02167/index.html?dossier_id=02169}

\bibitem{Petrovic2016}
S.~N. Petrović, K.~B. Karlsson,
  \href{https://doi.org/10.1016/j.energy.2016.08.007}{{Residential heat pumps
  in the future Danish energy system}}, Energy 114~(Supplement C) (2016) 787 --
  797.
\newblock \href {http://dx.doi.org/10.1016/j.energy.2016.08.007}
  {\path{doi:10.1016/j.energy.2016.08.007}}.
\newline\urlprefix\url{https://doi.org/10.1016/j.energy.2016.08.007}

\bibitem{Persson2011}
U.~Persson, S.~Werner,
  \href{https://doi.org/10.1016/j.apenergy.2010.09.020}{Heat distribution and
  the future competitiveness of district heating}, Applied Energy 88~(3) (2011)
  568 -- 576.
\newblock \href {http://dx.doi.org/10.1016/j.apenergy.2010.09.020}
  {\path{doi:10.1016/j.apenergy.2010.09.020}}.
\newline\urlprefix\url{https://doi.org/10.1016/j.apenergy.2010.09.020}

\bibitem{Gils}
H.~C. Gils, Balancing of intermittent renewable power generation by demand
  response and thermal energy storage, Ph.D. thesis, University of Stuttgart
  (2015).

\bibitem{HeatRoadmapEurope}
{Heat Roadmap Europe}, Tech. rep. (2016).

\bibitem{Staffell2012}
I.~Staffell, D.~Brett, N.~Brandon, A.~Hawkes,
  \href{https://doi.org/10.1039/C2EE22653G}{A review of domestic heat pumps},
  Energy Environ. Sci. 5 (2012) 9291--9306.
\newblock \href {http://dx.doi.org/10.1039/C2EE22653G}
  {\path{doi:10.1039/C2EE22653G}}.
\newline\urlprefix\url{https://doi.org/10.1039/C2EE22653G}

\bibitem{Grohnheit1993}
P.~E. Grohnheit, \href{https://doi.org/10.1016/0301-4215(93)90282-K}{{Modelling
  CHP within a national power system}}, Energy Policy 21~(4) (1993) 418 -- 429.
\newblock \href {http://dx.doi.org/10.1016/0301-4215(93)90282-K}
  {\path{doi:10.1016/0301-4215(93)90282-K}}.
\newline\urlprefix\url{https://doi.org/10.1016/0301-4215(93)90282-K}

\bibitem{CONNOLLY2014475}
D.~Connolly, H.~Lund, B.~Mathiesen, S.~Werner, B.~Möller, U.~Persson,
  T.~Boermans, D.~Trier, P.~Østergaard, S.~Nielsen,
  \href{10.1016/j.enpol.2013.10.035}{{Heat Roadmap Europe: Combining district
  heating with heat savings to decarbonise the EU energy system}}, Energy
  Policy 65~(Supplement C) (2014) 475 -- 489.
\newblock \href {http://dx.doi.org/https://doi.org/10.1016/j.enpol.2013.10.035}
  {\path{doi:https://doi.org/10.1016/j.enpol.2013.10.035}}.
\newline\urlprefix\url{10.1016/j.enpol.2013.10.035}

\bibitem{EnergyPLANDatabase}
D.~Connolly, \href{http://www.energyplan.eu/costdatabase/}{{EnergyPLAN Cost
  Database 3.0}}, Tech. rep. (2015).
\newline\urlprefix\url{http://www.energyplan.eu/costdatabase/}

\bibitem{ACER2016}
{ACER},
  \href{http://www.acer.europa.eu/Official_documents/Acts_of_the_Agency/Publication/ACER%20Market%20Monitoring%20Report%202016%20-%20ELECTRICITY.pdf}{{ACER
  Market Monitoring Report 2016}}, Tech. rep. (2017).
\newline\urlprefix\url{http://www.acer.europa.eu/Official_documents/Acts_of_the_Agency/Publication/ACER%20Market%20Monitoring%20Report%202016%20-%20ELECTRICITY.pdf}

\bibitem{TYNDP2016}
{European Network of Transmission System Operators for Electricity},
  \href{http://tyndp.entsoe.eu/}{{Ten-Year Network Development Plan (TYNDP)
  2016}}, Tech. rep., ENTSO-E (2016).
\newline\urlprefix\url{http://tyndp.entsoe.eu/}

\bibitem{uba}
\href{https://www.umweltbundesamt.de/indikator-umweltkosten-von-energie-strassenverkehr}{{Indikator:
  Umweltkosten von Energie und Straßenverkehr}} (2017).
\newline\urlprefix\url{https://www.umweltbundesamt.de/indikator-umweltkosten-von-energie-strassenverkehr}

\bibitem{Jacobson2011b}
{Delucchi, M.A.}, {Jacobson, M.Z.},
  \href{https://doi.org/10.1016/j.enpol.2010.11.045}{Providing all global
  energy with wind, water, and solar power, {Part II}: Reliability, system and
  transmission costs, and policies}, Energy Policy 39~(3) (2011) 1170--1190.
\newblock \href {http://dx.doi.org/10.1016/j.enpol.2010.11.045}
  {\path{doi:10.1016/j.enpol.2010.11.045}}.
\newline\urlprefix\url{https://doi.org/10.1016/j.enpol.2010.11.045}

\bibitem{ZVINGILAITE2011535}
E.~Zvingilaite, \href{https://doi.org/10.1016/j.apenergy.2010.08.007}{{Human
  health-related externalities in energy system modelling the case of the
  Danish heat and power sector}}, Applied Energy 88~(2) (2011) 535 -- 544, the
  5th Dubrovnik Conference on Sustainable Development of Energy, Water and
  Environment Systems, held in Dubrovnik September/October 2009.
\newblock \href {http://dx.doi.org/10.1016/j.apenergy.2010.08.007}
  {\path{doi:10.1016/j.apenergy.2010.08.007}}.
\newline\urlprefix\url{https://doi.org/10.1016/j.apenergy.2010.08.007}

\bibitem{ev2018}
Electric vehicle outlook 2018, Tech. rep., {Bloomberg New Energy Finance}
  (2018).

\bibitem{Munster2012}
M.~Münster, P.~E. Morthorst, H.~V. Larsen, L.~Bregnbæk, J.~Werling, H.~H.
  Lindboe, H.~Ravn, \href{https://doi.org/10.1016/j.energy.2012.06.011}{{The
  role of district heating in the future Danish energy system}}, Energy 48~(1)
  (2012) 47 -- 55, 6th Dubrovnik Conference on Sustainable Development of
  Energy Water and Environmental Systems, SDEWES 2011.
\newblock \href {http://dx.doi.org/10.1016/j.energy.2012.06.011}
  {\path{doi:10.1016/j.energy.2012.06.011}}.
\newline\urlprefix\url{https://doi.org/10.1016/j.energy.2012.06.011}

\bibitem{Nuytten2013}
T.~Nuytten, B.~Claessens, K.~Paredis, J.~V. Bael, D.~Six,
  \href{https://doi.org/10.1016/j.apenergy.2012.11.029}{Flexibility of a
  combined heat and power system with thermal energy storage for district
  heating}, Applied Energy 104~(Supplement C) (2013) 583 -- 591.
\newblock \href {http://dx.doi.org/10.1016/j.apenergy.2012.11.029}
  {\path{doi:10.1016/j.apenergy.2012.11.029}}.
\newline\urlprefix\url{https://doi.org/10.1016/j.apenergy.2012.11.029}

\bibitem{SternerPhD}
M.~Sterner,
  \href{http://www.upress.uni-kassel.de/katalog/abstract.php?978-3-89958-798-2}{Bioenergy
  and renewable power methane in integrated 100\% renewable energy systems},
  Ph.D. thesis, Kassel University (2009).
\newline\urlprefix\url{http://www.upress.uni-kassel.de/katalog/abstract.php?978-3-89958-798-2}

\bibitem{Hoersch2017}
J.~H\"orsch, T.~Brown, \href{https://arxiv.org/abs/1705.07617}{The role of
  spatial scale in joint optimisations of generation and transmission for
  {E}uropean highly renewable scenarios}, in: Proceedings of 14th International
  Conference on the European Energy Market (EEM 2017), 2017.
\newblock \href {http://dx.doi.org/10.1109/EEM.2017.7982024}
  {\path{doi:10.1109/EEM.2017.7982024}}.
\newline\urlprefix\url{https://arxiv.org/abs/1705.07617}

\bibitem{gie}
G.~I. Europe, Gas storage data, \url{https://agsi.gie.eu/},
  \url{https://agsi.gie.eu/}.

\bibitem{GCBB:GCBB12205}
F.~Creutzig, N.~H. Ravindranath, G.~Berndes, S.~Bolwig, R.~Bright,
  F.~Cherubini, H.~Chum, E.~Corbera, M.~Delucchi, A.~Faaij, J.~Fargione,
  H.~Haberl, G.~Heath, O.~Lucon, R.~Plevin, A.~Popp, C.~Robledo-Abad, S.~Rose,
  P.~Smith, A.~Stromman, S.~Suh, O.~Masera,
  \href{http://dx.doi.org/10.1111/gcbb.12205}{Bioenergy and climate change
  mitigation: an assessment}, GCB Bioenergy 7~(5) (2015) 916--944.
\newblock \href {http://dx.doi.org/10.1111/gcbb.12205}
  {\path{doi:10.1111/gcbb.12205}}.
\newline\urlprefix\url{http://dx.doi.org/10.1111/gcbb.12205}

\bibitem{SIMS20101570}
R.~E. Sims, W.~Mabee, J.~N. Saddler, M.~Taylor,
  \href{https://doi.org/10.1016/j.biortech.2009.11.046}{An overview of second
  generation biofuel technologies}, Bioresource Technology 101~(6) (2010) 1570
  -- 1580.
\newblock \href {http://dx.doi.org/10.1016/j.biortech.2009.11.046}
  {\path{doi:10.1016/j.biortech.2009.11.046}}.
\newline\urlprefix\url{https://doi.org/10.1016/j.biortech.2009.11.046}

\bibitem{DODDS20152065}
P.~E. Dodds, I.~Staffell, A.~D. Hawkes, F.~Li, P.~Grünewald, W.~McDowall,
  P.~Ekins, \href{https://doi.org/10.1016/j.ijhydene.2014.11.059}{Hydrogen and
  fuel cell technologies for heating: A review}, International Journal of
  Hydrogen Energy 40~(5) (2015) 2065 -- 2083.
\newblock \href {http://dx.doi.org/10.1016/j.ijhydene.2014.11.059}
  {\path{doi:10.1016/j.ijhydene.2014.11.059}}.
\newline\urlprefix\url{https://doi.org/10.1016/j.ijhydene.2014.11.059}

\bibitem{LUND2015389}
R.~Lund, B.~V. Mathiesen,
  \href{https://doi.org/10.1016/j.apenergy.2015.01.013}{Large combined heat and
  power plants in sustainable energy systems}, Applied Energy 142 (2015) 389 --
  395.
\newblock \href {http://dx.doi.org/10.1016/j.apenergy.2015.01.013}
  {\path{doi:10.1016/j.apenergy.2015.01.013}}.
\newline\urlprefix\url{https://doi.org/10.1016/j.apenergy.2015.01.013}

\bibitem{LUND20141}
H.~Lund, S.~Werner, R.~Wiltshire, S.~Svendsen, J.~E. Thorsen, F.~Hvelplund,
  B.~V. Mathiesen, \href{https://doi.org/10.1016/j.energy.2014.02.089}{{4th
  Generation District Heating (4GDH)}}, Energy 68 (2014) 1 -- 11.
\newblock \href {http://dx.doi.org/10.1016/j.energy.2014.02.089}
  {\path{doi:10.1016/j.energy.2014.02.089}}.
\newline\urlprefix\url{https://doi.org/10.1016/j.energy.2014.02.089}

\bibitem{pehl2017}
M.~Pehl, A.~Arvesen, F.~Humpen\"oder, A.~Popp, E.~G. Hertwich, G.~Luderer,
  \href{https://doi.org/10.1038/s41560-017-0032-9}{Understanding future
  emissions from low-carbon power systems by integration of life-cycle
  assessment and integrated energy modelling}, Nature Energy 2 (2017) 939--945.
\newblock \href {http://dx.doi.org/10.1038/s41560-017-0032-9}
  {\path{doi:10.1038/s41560-017-0032-9}}.
\newline\urlprefix\url{https://doi.org/10.1038/s41560-017-0032-9}

\bibitem{Jacobson2017}
M.~Z. Jacobson, M.~A. Delucchi, Z.~A. Bauer, S.~C. Goodman, W.~E. Chapman,
  M.~A. Cameron, C.~Bozonnat, L.~Chobadi, H.~A. Clonts, P.~Enevoldsen, J.~R.
  Erwin, S.~N. Fobi, O.~K. Goldstrom, E.~M. Hennessy, J.~Liu, J.~Lo, C.~B.
  Meyer, S.~B. Morris, K.~R. Moy, P.~L. O'Neill, I.~Petkov, S.~Redfern,
  R.~Schucker, M.~A. Sontag, J.~Wang, E.~Weiner, A.~S. Yachanin,
  \href{https://doi.org/10.1016/j.joule.2017.07.005}{{100\% Clean and Renewable
  Wind, Water, and Sunlight All-Sector Energy Roadmaps for 139 Countries of the
  World}}, Joule 1 (2017) 1--14.
\newblock \href {http://dx.doi.org/10.1016/j.joule.2017.07.005}
  {\path{doi:10.1016/j.joule.2017.07.005}}.
\newline\urlprefix\url{https://doi.org/10.1016/j.joule.2017.07.005}

\bibitem{Schlachtberger2018}
D.~Schlachtberger, T.~Brown, M.~Schäfer, S.~Schramm, M.~Greiner,
  \href{https://arxiv.org/abs/1803.09711}{{Cost optimal scenarios of a future
  highly renewable European electricity system: Exploring the influence of
  weather data, cost parameters and policy constraints}}.
\newline\urlprefix\url{https://arxiv.org/abs/1803.09711}

\end{thebibliography}

\end{document}